\documentclass[12pt]{article}
\usepackage{amsmath,amssymb,amsthm,color}
\usepackage{graphicx}
\usepackage{cite}
\usepackage{epsfig}
\usepackage{lineno}
\renewcommand{\baselinestretch}{1.66}
\begin{document}

\title {Electronic and optical properties of graphite-related systems}
\author{
\small Chiun-Yan Lin$^{a}$, Rong-Bin Chen$^{b}$, Yen-Hung Ho$^{c}$, Ming-Fa Lin$^{a,*}$ $$\\
\small  $^a$Department of Physics, National Cheng Kung University, Taiwan\\
\small  $^b$Center of General Studies, National Kaohsiung Marine University, Taiwan\\
\small  $^c$Department of Physics, National Tsing Hua University, Taiwan\\
 }
\renewcommand{\baselinestretch}{1.66}
\maketitle

\renewcommand{\baselinestretch}{1.66}
\begin{abstract}

A systematic review is made for the AA-, AB- and ABC-stacked
graphites. The generalized tight-binding model, accompanied with the
effective-mass approximation and the Kubo formula, is developed to
investigate electronic and optical properties in the
presence/absence of a uniform magnetic field. The unusual electronic
properties cover the stacking-dependent Dirac-cone structures, the
significant energy widths along the stacking direction, the Landau
subbands (LSs) crossing the Fermi level, the $B_0$-dependent LS
energy spectra with crossings and anti-crossings, and the monolayer-
or bilayer-like Landau wavefunctions. There exist the
configuration-created special structures in density of states and
optical spectra. Three kinds of graphites quite differ from one
another in the available inter-LS excitation channels, including the
number, frequency, intensity and structures of absorption peaks. The
dimensional crossover presents the main similarities and
differences between graphites and graphenes; furthermore, the
quantum confinement enriches the magnetic quantization phenomena in
carbon nanotubes and graphene nanoribbons. The
cooperative/competitive relations among the interlayer atomic
interactions, dimensions and magnetic quantization are responsible
for the diversified essential properties. Part of theoretical
predictions are consistent with the experimental measurements.

\vskip 0.6 truecm

$\mathit{PACS}$: 73.20.At, 73.22.-f, 75.70.Ak

\end{abstract}

\par\noindent * Corresponding author. {~ Tel:~ +886-6-275-7575;~ Fax:~+886-6-74-7995.}\\~{{\it E-mail address}: mflin@mail.ncku.edu.tw (M.F. Lin)}


\noindent{Contents}

\pagenumbering{Roman}

\addcontentsline{toc}{chapter}{摘要}

\vskip0.6 truecm

\noindent {1. Introduction} \hfill ............................................................................................................~~~01

\vskip0.5 truecm

\noindent

\noindent {2. Theoretical models}
\hfill...................................................................................................~~~10

\noindent {2.1 The magnetic tight-binding model for layered graphites}
\hfill ........................................~~~10

\noindent {2.1.1 Simple hexagonal graphite graphene}
\hfill....................................................................~~~14

\noindent {2.1.2 Bernal graphite}
\hfill....................................................................................................~~~16

\noindent {2.1.3 Rhombohedral graphite}
\hfill......................................................................................~~~17

\noindent {2.1.4 The gradient approximation for optical properties}
\hfill...............................................~~~20

\vskip0.5 truecm

\noindent {3. Simple hexagonal graphite}
\hfill.......................................................................................~~~21

\noindent {3.1 Electronic structures without external fields }
\hfill.........................................................~~~22


\noindent {3.2 Optical properties without external fields }
\hfill............................................................~~~28

\noindent {3.3 Magnetic quantization}
\hfill.........................................................................................~~~31

\noindent {3.3.1 Landau levels and wave functions }
\hfill.......................................................................~~~31

\noindent {3.3.2 Landau subband energy spectra }
\hfill........................................................................~~~33

\noindent {3.4 Magneto-optical properties}
\hfill.....................................................................................~~~39

\vskip0.5 truecm

\noindent {4. Bernal graphite}
\hfill.......................................................................................................~~~48

\noindent {4.1 Electronic structures without external fields}
\hfill ...........................................................~~~48

\noindent {4.2 Optical properties without external fields}
\hfill.............................................................~~~54

\noindent {4.3 Magnetic quantization}
\hfill..........................................................................................~~~56

\noindent {4.3.1 Landau subbands and wave functions}
\hfill................................................................~~~56

\noindent {4.3.2 Anticrossings of Landau subbands}
\hfill......................................................................~~~60

\noindent {4.4 Magneto-optical properties}
\hfill....................................................................................~~~63

\vskip0.5 truecm

\noindent {5. Rhombohedral graphite}
\hfill..........................................................................................~~~71

\noindent {5.1 Electronic structures without external fields}
\hfill .........................................................~~~71

\noindent {5.2 Anisotropic Dirac cone along a nodal spiral}
\hfill............................................................~~~73

\noindent {5.3 dimensional crossover}
\hfill............................................................................................~~~77

\noindent {5.4 Optical properties without external fields}
\hfill ..............................................................~~~79

\noindent {5.5 Magneto-electronic properties}
\hfill...............................................................................~~~81

\noindent {5.5.1 Tight-binding model}
\hfill..........................................................................................~~~81

\noindent {5.5.2 Onsager quantization}
\hfill..........................................................................................~~~83

\noindent {5.6 Magneto-optical properties}
\hfill.................................................................................~~~88

\vskip0.5 truecm

\noindent {6. Quantum confinement in carbon nanotubes and graphene
nanoribbons}
\hfill................~~~92

\noindent {6.1 Magneto-electronic properties of carbon nanotubes}
\hfill ...............................................~~~93

\noindent {6.2 Magneto-optical spectra of carbon nanotubes}
\hfill.......................................................~~~99

\noindent {6.3 Magneto-electronic properties of graphene nanoribbons}
\hfill..........................................~~~105

\noindent {6.4 Magneto-optical spectra of graphene nanoribbons}
\hfill...................................................~~~112

\noindent {6.5 Comparisons and applications}
\hfill................................................................................~~~120

\vskip0.5 truecm

\noindent {7. Concluding remarks}
\hfill.................................................................................................~~~126

\noindent {Acknowledgments}
\hfill........................................................................................................~~~135

\noindent {References (Ref. 1-Ref. 233)}
\hfill........................................................................................~~~135

\noindent {Figure captions (Fig. 1-Fig. 49)}
\hfill..................................................................................~~~171

\pagebreak
\renewcommand{\baselinestretch}{2}
\newpage

\vskip 0.6 truecm



\setcounter{page}{1}
\pagenumbering{arabic}

\newpage

\section{Introduction}

Carbon atoms can form various condensed-matter systems with unique
geometric structures, mainly owing to four active atomic orbitals.
From three- to zero-dimensional carbon-related systems cover
diamond,\cite{PRSA232;463}
graphite,\cite{PRSLSA181;101,PRSLSA106;749,AP30;139}
graphene,\cite{Science306;666} graphene
nanoribbons,\cite{Science319;1229} carbon
nanotube,\cite{Nature354;56} and carbon
fullerene.\cite{Science280;1253} Most of them have the ${sp^2}$
bonding except for the ${sp^3}$ bonding in diamond. The former might
exhibit similar physical properties, e.g., the $\pi$-electronic
optical excitations.\cite{PRL102;037403,NatCom4;2542}
Graphite is one of the most extensively studied materials
theoretically and experimentally. This layered system is very
suitable for exploring the diverse 3D and 2D phenomena. The
interplane attractive forces originate from the weak Van der Waals
interactions of the ${2p_z}$ orbitals. The honeycomb lattice and the
stacking configuration are responsible for the unique properties of
graphite, e.g., the semi-metallic behavior due to the hexagonal
symmetry and the interlayer atomic interactions. The essential
properties are dramatically changed by the intercalation of various
atoms and molecules. Graphite intercalation compounds could achieve
a conductivity as good as
copper.\cite{Carbon4;125,STAM9;044102,AP30;139,AIP51;1} In general,
there exist three kinds of ordered configurations in the layered
graphites and compounds, namely AA, AB and ABC stackings. Simple
hexagonal, Bernal and rhombohedral graphites exhibit the rich and
diverse electronic and optical properties in the presence/absence of
a uniform magnetic field (${\bf B}$=${B_0\hat z}$). To present a
systematic review of them, the generalized tight-binding model is
developed under the magnetic quantization. This model, combined with
the Kubo formula, is utilized to investigate the essential
properties of layered carbon-related systems. The dimensional
crossover from graphene to graphite and the quantum confinement in
nanotube and nanoribbon systems are discussed thoroughly. A detailed
comparison with the other theoretical studies and the experimental
measurements is also made.

Few- and multi-layer graphenes, with distinct stacking
configurations, are successfully produced by various experimental
methods since the first discovery of monolayer graphene in 2004 by
the mechanical exfoliation.\cite{Science306;666} They possess the
hexagonal symmetry and the nanoscale size, leading to a lot of
remarkable characteristics, e.g., the largest Young’s
modulus,\cite{Science321;385} feature-rich energy
bands,\cite{RevModPhys81;109,RPP76;056503} diverse optical selective
rules,\cite{PCCP18;17597,PRB75;155430,ACSNano4;1465,PRB83;125302,PRL98;197403,PRL111;077402,PRL101;267601,
PRL100;087401} unique magnetic
quantization,\cite{SciRep4;7509,PCCP17;26008,PRB77;085426,PRL96;086805,PRB77;155416,
PRB84;205448,PRB90;205434,RSCAdv4;56552,JPCC119;10623,Carbon94;619}
anomalous quantum Hall
effects,\cite{Nature438;197,Nature438;201,NatPhys2;177,NatPhys7;621,NatPhys7;953}
and multi-mode
plasmons.\cite{Plasmonics10;1409,NanoLett11;3370,ACSNano6;431,NatPho7;394}
Their electronic and optical properties are very sensitive to the
changes in the stacking
configuration,\cite{JPCC119;10623,Carbon94;619,PRB90;205434,PRB77;155416,
PCCP17;26008,RevModPhys81;109} layer
number,\cite{PRB90;205434,PRB77;155416,RevModPhys81;109,PCCP17;26008}
magnetic
field,\cite{RSCAdv4;56552,PRB90;205434,PRB84;205448,PRL96;086805,PRB77;155416,ACSNano4;1465,PRB75;155430,PRB83;125302,PRL98;197403,PRL111;077402,
PRL101;267601,PRL100;087401,PCCP17;26008,PRB77;085426,RPP76;056503}
electric field,\cite{RSCAdv4;56552,RPP76;056503} mechanical
strain,\cite{RevModPhys81;109,RPP76;056503}
doping,\cite{RevModPhys81;109,RPP76;056503} and
sliding.\cite{SciRep4;7509} Five kinds of electronic structures,
linear,\cite{PCCP17;26008}
parabolic,\cite{PRB90;205434,PCCP17;26008,PRB77;085426} partially
flat,\cite{PRB90;205434,PCCP17;26008}
sombrero-shaped,\cite{RSCAdv4;56552,PRB90;205434,PCCP17;26008} and
oscillatory\cite{RSCAdv4;56552,Carbon94;619,SciRep4;7509} energy
bands, are revealed in AB and ABC stacking systems. The last ones
could be created by a perpendicular electric field. However, the AA
stacking only has the first kind. Specifically, the intersection of
the linear valence and conduction bands can form the so-called
Dirac-cone structure. The main features of energy bands are directly
reflected in the other essential properties. The finite-layer
confinement effects are expected to induce important differences
between the layered graphenes and graphites. The close relations
arising from the dimensional crossover deserve a thorough
investigation.


Graphite crystals are made up of a series of stacked graphene plane.
Among three kinds of ordered stacking configurations, the AB-stacked
graphite is predicted to have the lowest ground state energy
according to the first-principles calculations.\cite{CARBON32;289}
Nature graphite presents the dominating AB stacking and the partial
ABC stacking.\cite{PRSLSA106;749,PRSLSA181;101} The AA-stacked
graphite, which possesses the simplest crystal structure, does not
exist naturally. The periodical AA stacking is first observed in the
Li-intercalation graphite compounds\cite{AP30;139} with high free
electron density and super conducting transition temperature at
$~$1.9 K.\cite{SSC69;921} Simple hexagonal graphite is successfully
synthesized by using the dc plasma in hydrogen-methane
mixtures,\cite{JCP129;234709} and AA-stacked graphenes are generated
by the method of Hummers and Offeman and the chemical-vapor
deposition (CVD). \cite{JJAP47;1073,JAP109;093523} Furthermore, the
AA stacking sequence is confirmed from the high-resolution
transmission electron microscopy (HRTEM).\cite{JJAP47;1073}
Specifically, angle-resolved photoemission spectroscopy (ARPES),  a
powerful tool in the direct identification of energy bands, is
utilized to examine two/three pairs of Dirac-cone structures in
bilayer/trilayer AA stacking.\cite{NatMater12;887,NanoLett8;1564}

The AA stacking systems have stirred a lot of theoretical
researches, such as band
structures,\cite{PRB44;13237,PRB83;165429,PRB90;205434,PRB44;13237,PRB77;155416}
magnetic
quantization,\cite{EPJB60;161,JNN12;2557,PRB90;205434,PCCP17;26008,PRB77;155416}
optical properties,\cite{APL97;101905,Carbon54;248,CPC189;60}
Coulomb
excitations,\cite{PRB74;085406,PRB89;165407,AOP339;298,RSCAdv5;53736,JPSJ81;104703}
quantum transport,\cite{PRB82;165404} and phonon
spectra,\cite{PRB88;035428} Simple hexagonal graphite, a 3D layered
system with the same graphitic sheets on the $(x,y)$-plane, is first
proposed by Charlier et al.\cite{CARBON32;289} From the
tight-binding model and the first-principles method, this system
belongs to a band-overlap semimetal, in which the same electron and
hole density originate from the significant interlayer atomic
interactions. There is one pair of low-lying valence and conduction
bands. The critical feature is the vertical Dirac-cone structure
with the sufficiently wide bandwidth of $\sim1$ eV along the
${k_z}$-direction (${\bf k}$ wave vector). Similarly, the AA-stacked
graphenes is predicted to exhibit multiple Dirac cones vertical to
one another.\cite{PCCP17;26008,PRB77;155416,PRB83;165429} Moreover,
Bloch wave functions are only the symmetric or anti-symmetric
superposition of the tight-binding functions on the distinct
sublatticesa and layers.\cite{PCCP17;26008} Apparently, these 3D and
2D vertical Dirac cones will dominate the other low-energy essential
properties, e.g., the special structures in density of states
(DOS),\cite{PCCP17;26008,PRB44;13237,EPJB60;161} the quantized
Landau subbands (LSs) and levels
(LLs),\cite{PCCP17;26008,PRB77;155416,JNN12;2557} the rich optical
spectra,\cite{Carbon54;248,CPC189;60} and the diversified plasmon
modes.\cite{PRB74;085406,PRB89;165407,AOP339;298,RSCAdv5;53736,JPSJ81;104703}
The magnetic quantization is frequently explored by the
effective-mass approximation\cite{PRB77;155416} and the generalized
tight-binding model.\cite{PCCP17;26008,PRB90;205434} It is initiated
from the vertical Dirac points, the extreme points in the
energy-wave-vector space. This creates the specific $B_0$-dependent
energy spectra and the well-behaved charge distributions, thus
leading to the diverse and unique magneto-absorption peaks, e.g.,
the intraband and interband inter-LS excitations, the multi-channel
threshold peaks, and beating-form absorption peaks in AA-stacked
graphite.\cite{JNN12;2557,Carbon54;248,CPC189;60} The predicted band
structures, energy spectra and optical excitations could be verified
by ARPES,\cite{NatMater12;887,NanoLett8;1564} scanning tunneling
spectroscopy (STS)$^{Refs}$ and optical spectroscopy,
\cite{PRB83;125302,PRL98;197403,PRL111;077402,PRL101;267601,PRL100;087401}
respectively.

Bernal graphite is a well-known semimetal,\cite{PRSLSA106;749} with
conduction electron concentration of $\sim{5\times\,10^{18}}/cm^3$.
Furthermore, the AB stacking configuration is frequently observed in
the layered systems, e.g.,
bilayer,\cite{Science306;666,ACSNano6;9790}
trilayer,\cite{PRB77;155426,ACSNano6;9790,SciRep6;33487} and
tetralayer graphenes.\cite{ACSNano6;9790} The AB-stacked graphite
possesses two pairs of low-lying energy bands dominated by the
${2p_z}$ orbitals, owing to a primitive unit cell with two
neighboring layers. The highly anisotropic band structure, the
strong and weak energy dispersions along the ${(k_x,k_y)}$ plane and
$k_z$-axis, respectively, is confirmed by the ARPES
measurements.\cite{PRL100;037601,ASS354;229,PhyB407;827,PRB79;125438,
NatPhy2;595PRL100;037601} The similar examinations are done
for two pairs of parabolic bands in bilayer AB
stacking,\cite{PRL98;206802,Science313;951} and the linear and
parabolic bands in trilayer system\cite{PRL98;206802,PRB88;155439}.
The measured DOS of Bernal graphite presents the splitting $\pi$ and
$\pi^\ast$ peaks at the middle energy,\cite{ASS151;251} reflecting
the highly accumulated states near the saddle points. Furthermore,
it is finite near the Fermi level ($E_F$) because of the
semi-metallic behavior.\cite{PRL102;176804} An
electric-field-induced band gap is observed in bilayer AB
stacking.\cite{PRB77;155426} The magnetically quantized energy
spectra, with many special structures, are also identified using the
STS measurements for the AB-stacked
graphite\cite{NatPhys3;623,PRL94;226403} and
graphenes,\cite{Science324;924,PRB91;115405} especially for the
square-root and linear dependences on the magnetic-field strength
(the monolayer- and bilayer-like behaviors at low energy). From the
optical measurements, Bernal graphite shows a very prominent
$\pi$-electronic absorption peak at frequency $\sim5$
eV,\cite{PR138;A197} as revealed in carbon-related systems with the
${sp^2}$ bonding.\cite{PRL106;046401,PRB81;155413} Concerning the
low-frequency magneto-optical experiments, the measured excitation
spectra due to
LSs\cite{PRB15;4077,PRB74;195404,APL99;011914,PRB80;161410,PRL102;166401,PRB86;155409,JAP117;112803}
or
LLs\cite{PRB75;155430,ACSNano4;1465,PRB83;125302,PRL98;197403,PRL111;077402,PRL101;267601,
PRL100;087401} clearly reveal a lot of pronounced absorption
structures, the selection rule of ${\triangle\,n=\pm\,1}$ ($n$
quantum number), and the strong dependence on the wave vector of
$k_z$ or the layer number ($N$).

The earliest attempt to calculate the band structures of monolayer
graphene and Bernal graphite is done by Wallace using the
tight-binding model with the atomic interactions of ${2p_z}$
orbitals.\cite{PR71;622} The former 2D system has the linear valence
and conduction bands intersecting at $E_{F}$, so that it belongs to
a zero-gap semiconductor with vanishing DOS there. However, the
semi-metallic 3D electronic structure is further comprehended from
the Slonczewski-Weis-McClure Hamiltonian involving the important
intralayer and interlayer atomic
interactions.\cite{PR109;272,PR119;606} The magnetic Hamiltonian
could be solved by the low-energy perturbation approximation, in
which the LS energy spectra exhibit the crossing and anti-crossing
behaviors.\cite{JPSJ40;761,JPSJ17;808} The generalized tight-binding
model, which deals with the magnetic field and all atomic
interactions simultaneously, is developed to explore the main
features of LSs, e.g., two groups of valence and conduction LSs, and
the layer-, $k_z$- and $B_0$-dependent spatial oscillation
modes.\cite{PRB83;121201,APL99;011914} As to the AB-stacked
graphenes, their electronic and magneto-electronic properties
present the bilayer- and monolayer-like behaviors, being associated
with two pair of parabolic bands and a slightly distorted Dirac
cone, respectively.\cite{RevModPhys81;109,PCCP17;26008} The former
is only revealed in the odd-$N$ systems. Optical spectra of
AB-stacked systems are predicted to exhibit the strong dependence of
special structures on the layer number and
dimension.\cite{ACSNano4;1465,PRB75;155430,PRB83;125302,PRL98;197403,PRL111;077402,
PRL101;267601,PRL100;087401,PRB80;161410} The magneto-optical
excitations arising from two groups of LSs in graphite or $N$ groups
of LLs in graphenes could be evaluated from the generalized
tight-binding model.\cite{APL99;011914} On the other hand, those due
to the first group of LLs/LLs is frequently investigated by the
effective-mass approximation in detail.\cite{JPSJ40;761,JPSJ17;808}
The calculated electronic and optical properties are in agreement
with the experimental measurements.\cite{NatPhys3;623,PRL94;226403}

Rhombohedral phase is usually found to be mixed with Bernal phase in
natural graphite. The ABC stacking sequence in bulk graphite is
directly identified from the experimental measurements of
HRTEM,\cite{Carbon50;2347} X-ray
diffraction,\cite{Carbon69;17,PhysPro23;102,Carbon33;1399} and
scanning tunneling microscopy
(STM).\cite{Carbon50;4633,ChemPhys348;233} Specifically, this
stacking configuration could account for the measured 3D quantum
Hall effect with multiple plateau structures,\cite{SSC138;118} since
rhombohedral graphite possesses the well separated LS energy
spectra.\cite{NJP15;053032,SSC197;11} In addition, doped Bernal
graphite is predicted to exhibit only one
plateau.\cite{PRL99;146804} Both ABC- and AB-stacked graphenes can
be produced by mechanical exfoliation of kish
graphite,\cite{JPCM20;323202,NanoLett8;565}
CVD,\cite{NanoLett10;1542,Carbon48;1088,NanoLett9;30,AdvMater21;2328}
chemical and electrochemical reduction of graphite
oxide,\cite{EST45;10454,Carbon45;1558,STAM11;054502} arc
discharge,\cite{Carbon102;494,Carbon47;4939,NanoRes3;661} flame
synthesis,\cite{ChemCom47;3520} and electrostatic manipulation of
STM.\cite{PRB86;085428,APL107;263101} The ARPES measurements have
verified the partially flat, sombrero-shaped and linear bands in
tri-layer ABC stacking.\cite{PRB88;155439} As to the STS spectra, a
pronounced peak at the Fermi level characteristic of the partial
flat band is revealed in tri-layer and penta-layer ABC
stacking.\cite{APL107;263101,ACSNano9;5432,PRB91;035410} Moreover,
infrared reflection spectroscopy and absorption spectroscopy are
utilized to examine the low-frequency optical properties, displaying
a clear evidence of two featured absorption structures due to the
partially flat and sombrero-shaped energy bands.\cite{PRL104;176404}
According to the specific infrared conductivities, infrared
scattering scanning near-field optical microscopy can distinguish
the ABC stacking domains with nano-scaled resolution from other
domains. However, the magneto-optical measurements on ABC-stacked
graphenes are absent up to now.

The ABC-stacked graphite has a rhombohedral unit cell, while the AA-
and AB-stacked systems possess the hexagonal ones. This critical
difference in stacking symmetry is responsible for the diversified
essential properties. Among three kinds of bulk graphites,
rhombohedral graphite is expected to present the smallest band
overlap (the lowest free carrier density), and the weakest energy
dispersion along the
$k_z$-direction.\cite{CJP36;352,Carbon7;425,JMST897;118} There also
exists a robust low-energy electronic structure, a 3D spiral
Dirac-cone structure. This results in the unusual magnetic
quantization,\cite{NJP15;053032,SSC197;11,PRB78;245416} in which the
quantized LSs exhibit the monolayer-like behavior and the
significant $k_z$-dependence. The previous studies show that four
kinds of energy dispersions exist in the ABC-stacked
graphenes.\cite{PCCP17;26008,PRB90;205434} Specially, the partially
flat bands corresponding to the surface states and the
sombrero-shaped bands are absent in bulk system. They can create the
diverse and unique LLs, with the asymmetric energy spectra about
$E_F$, the normal and abnormal $B_0$-dependences, the well-behaved
and distorted probability distributions, and the frequent crossings
and anti-crossings.\cite{PCCP17;26008,PRB90;205434} Apparently,
optical and magneto-optical properties are greatly enriched by the
layer number and
dimension.\cite{PCCP17;15921,JPCM27;125602,APL98;261920,PCCP13;6036}
The quantized LLs of the partially flat bands and the lowest sombrero-shaped band have been verified by the magneto-Raman spectroscopy for a large ABC domain in a graphene multilayer flake.\cite{NanoLett16;3710} Layered graphenes are predicted to have more complicated excitation spectra, compared with 3D system. The former and the latter, respectively, reveal N$^2$ categories of inter-LL transitions and
one category of inter-LS excitations.\cite{PCCP17;15921,PCCP13;6036}

In addition to the stacking configurations, the distinct dimensions
can create the diverse phenomena in carbon-related systems. The
quantum confinement in 1D carbon nanotubes and graphene nanoribbons
play a critical role in the essential properties. The systematic
studies have been made for the former since the successful synthesis
using the arc-discharge evaporation in 1991.\cite{Nature354;56} Each
carbon nanotube could be regarded as a rolled-up graphene sheet in
the cylindrical form. It is identified to be a metal or
semiconductor, depending on the radius and chiral
angle.\cite{PRB46;1804,PRL78;1932,Science292;702} The
geometry-dependent energy spectra, with energy gaps ($E_g$'s), are directly
verified from the STS measurements.\cite{Nature391;59,Nature391;62}
Specifically, the cylindrical symmetry can present the well-known
Aharonov-Bohm effect under an axial magnetic
field.\cite{JPSJ62;2470,PRB67;045405,PRB62;16092,PRB51;7592} This is
confirmed by the experimental measurements on
optical\cite{Science304;1129,AdvMater18;1166} and transport
properties.\cite{Nature397;673,NatMater4;745,PRL93;216803} However,
a closed surface acts as a high barrier in the formation of the
dispersionless LLs, since a perpendicular magnetic field leads to a
vanishing flux through carbon hexagons. It is very difficult to
observe the physical phenomena associated with the highly degenerate
states except for very high field strength.\cite{JPSJ65;505}

The essential properties are greatly enriched by the boundary
conditions in 1D systems. The open and periodical boundaries, which,
respectively, correspond to graphene nanoribbon and carbon nanotube,
induce the important differences between them. A graphene nanoribbon
is a finite-width graphene or an unzipped carbon nanotube. Graphene
nanoribbons could be produced by cutting few-layer
graphenes,\cite{NanoLett9;2083,NanoLett9;2600} unzipping
multi-walled carbon
nanotubes,\cite{Nature458;872,NanoLett9;1527,Carbon48;2596} and
using the direct chemical
syntheses.\cite{NatComm4;2646,Nature466;470,NanoLett8;2773} The cooperative or
competitive relations among the open boundary, the edge structure,
and the magnetic field are responsible for the rich and unique
properties. The 1D parabolic bands, with energy gaps, in armchair
graphene nanoribbons, are confirmed by ARPES.\cite{ACSNano6;6930}
Furthermore, STS has verified the asymmetric DOS peaks and the
finite-size effect on energy
gap.\cite{APL105;123116,SciRep2;983,PRB91;045429,ACSNano7;6123} Optical spectra
are predicted to have the edge-dependent selection
rules.\cite{PRB95;155438,OptExpress19;23350,PRB84;085458} The theoretical
calculations show that only the quasi-LLs (QLLs), with partially
dispersionless relations, could survive in the presence of a
perpendicular magnetic
field.\cite{PRB73;241403,Nanotechnology18;495401} The
magneto-optical selection rule, as revealed in layered graphenes,
sharply contrasts with that in carbon nanotubes with the
well-defined angular momenta along the azimuthal
direction.\cite{JAP103;073709}

In this work, we propose and develop the generalized tight-bindings
model to fully comprehend the electronic and optical properties of
the graphite-related systems. The Hamiltonin is built from
the tight-binding functions on the distinct sublattices and layers,
in which all important atomic interactions, stacking configuration,
layer number and external fields are taken into account
simultaneously. A quite large Hamiltonian matrix, being associated with
the periodical variation of the vector potential, is solved by an
exact diagonalization method. The essential properties can be
evaluated very efficiently. Moreover, the effective-mass
approximation is utilized to provide the qualitative behaviors and
the semi-quantitative results, e.g., the layer-dependent
characteristics. Specifically, the Onsager quantization method is
also introduced to understand the magnetic LS energy spectra in the
ABC-stacked graphite with the unique spiral Dirac cones. Such
approximations are useful in the identification of the critical
atomic interactions  creating the  unusual properties.

The AA-, AB- and ABC-stacked graphites and graphenes, and 1D
graphene nanoribbons and carbon nnaotubes are worthy of a systematic
review of essential properties. Electronic and optical properties,
which mainly come from carbon ${2p_z}$ orbitals, are investigated in
the presence/absence of magnetic field. Electronic structures,
quantized LS and LL state energies, magnetic wave functions, DOS and
optical spectral functions are included in the calculated results.
Band widths, energy dispersion relations, critical points in
energy-wave-vector space, crossings and anti-crossings of
$B_0$-dependent energy spectra, spatial oscillation modes of
localized probability distributions, and various special structures
in DOS are explored in detail. The main features of optical
excitations focus on tht available excitation channels, the form,
number, intensity and frequency of prominent absorption structures,
and the layer/dimension and field dependences. Moreover, the
theoretical predictions are compared with the ARPES, STS and optical
measurements and require more experimental examinations. Chapter 2
covers geometric structures, important atomic interactions, the
generalized tight-binding model, and the Kubo formula, in which the
main issues are the construction of the magnetic Hamiltonians and
the efficient combination two methods. In chapter 3, the stacking-
and layer-enriched essential properties are studied for the
AA-stacked graphite and graphenes, especially for those due to the
vertical multiple Dirac-cone structures. The analytical band
structures and magneto-electronic energy spectra are obtained from
the approximate  expansions about the high symmetry points. They are
available in illustrating the diversified characteristics, e.g., the
determination of the close relations between the absorption spectra
and the important intralayer/interlayer atomic interactions. The
dimensional crossover from monolayer graphene to graphite creates
the critical differences of 2D and 3D phenomena, including the
semiconductor-semimetal transition, the $k_z$-dependent band width,
the LS/LL energy spectra near the Fermi level, the optical gap, and
the low- and middle-frequency absorption structures.

The dramatic transformations of essential properties are clearly
revealed in distinct stacking configurations. As to the AB-stacked
systems, the linear and parabolic energy dispersions, the crossings
and anti-crossings of LS/LL energy spectra, the well-behaved and
perturbed magnetic wave functions, the layer- and
dimension-dependent optical spectra; the rich magneto-absorption
peaks are investigated in Chap. 4. The monolayer- and bilayer-like
behaviors are presented for Bernal graphite and layered AB stacking.
Specifically, ABC-stacked graphenes has the linear, parabolic,
partially flat and Sombrero-shaped energy bands, while rhombohedral
graphite exhibits a 3D spiral Dirac-cone structure, as indicated in
Chap. 5. Such characteristics are expected to create the unique
essential properties. The low-energy approximation and magnetic
quantization are proposed to explain the diversified electronic
properties and optical spectra. In chapter 6, the reduced dimension
in graphene nanoribbons and  carbon nanotubes leads to the rich
essential properties being sensitive to the open/periodical boundary
condition, width/radius, edge/chiral angle, and external fields.
Comparisons among the graphite-related systems and potential
applications are also discussed. Finally, chapter 6 contains
concluding remarks and outlook. The theoretical framework could be
further extended to the other main-stream layered materials.

\section{Theoretical models}


In the presence of a uniform magnetic field,
$\mathbf{B}=B_{0}\hat{z}$, electrons are forced to undergo the
cyclotron motion in the x-y plane. As a result, electronic states
are evolved into highly degenerate states, called Landau subbands
(LSs) in graphites. The 1D LSs are calculated from the subenvelope
functions established on different sublattices in the framework of
the generalized tight-binding model, which simultaneously takes into
account external fields and atomic interactions. The magneto
Hamiltonian is built from the tight-binding functions coupled with a
periodic Peierls phase in an enlarged unit cell; the period depends
on the commensurate relation between the lattice constant and the
Peierls phase. According to the Kubo formula, it could further be
utilized to comprehend the main features of magneto-absorption
spectra, which are closely related to the Landau-level spectrum and
the transition matrix elements. The method provides accurate and
reliable results for a wide-energy range. The three prototypical
configurations of bulk graphites, namely, simple hexagonal, Bernal
and rhombohedral graphites, are chosen for a systematic review. The
magnetic quantization in 3D graphene systems shows interesting
phenomena as a function the stacking configuration and the magnetic
field strength.

\subsection{The magnetic tight-binding model for layered graphites}

The geometric structures of simple hexagonal, Bernal and
rhombohedral bulk graphites are shown in Figs. 1(a)-1(c).
They are, respectively, constructed from 2D graphene layers
periodically stacked along $\widehat{z}$ with AA, AB and ABC
stacking configurations, where the layer-layer distance $I_{z}$ is
set as $3.35$ ${\AA}$. Detailed definitions of the stacking
sequences are made in following sections. The unit cells of
different graphites are marked by the gray shadows, which contain
two sublattices, $A^{l}$ and $B^{l}$, on each layer, where $l$
represents the number of the layer, and the symbols $\alpha$'s,
$\beta$'s and $\gamma$'s indicate the intrelayer and intralayer
hopping integrals. The first Brillouin zone is a hexagonal prism, as
shown in Fig. 1(e), where the highly symmetric points are defined as
$\Gamma$, M, K, A, L and H. The KH lengths for the AA, AB, and ABC
stackings are, respectively, equal to $\pi/I_{z}$, $\pi/2I_{z}$, and
$\pi/3I_{z}$ based on the periods along $\widehat{z}$.

In general, the essential physical properties are mainly determined
by 2$p_{z}$ orbitals of carbon atoms. Built from the subspace
spanned by the tight-binding functions $\varphi_{A^{l}}$ and
$\varphi_{B^{l}}$($l=1,2....$), the wave function is characterized
by their linear combination over all $A$ and $B$ sublattices in a
unit cell:
\begin{eqnarray}
\Psi =\sum_{l} C_{A^{l}}\varphi_{A^{l}}+C_{B^{l}}\varphi_{B^{l}},
\end{eqnarray}
where $C_{A^{l}}$ and $C_{B^{l}}$ are normalization factors. The
tight-binding functions are
\begin{eqnarray}
\begin{array}{l}
\varphi_{A^{l}}=\sum_{{R}_{A^{l}}}
\exp (i\mathbf{k}\cdot \mathbf{R}_{A^{l}})\chi (%
\mathbf{r}-\mathbf{R}_{A^{l}})\\
\varphi_{B^{l}}=\sum_{{R}_{B^{l}}}
\exp (i\mathbf{k}\cdot \mathbf{R}_{B^{l}})\chi (%
\mathbf{r}-\mathbf{R}_{B^{l}}),
\end{array}
\end{eqnarray}
where $\chi (\mathbf{r})$ is the atomic 2$p_{z}$ orbital of an
isolated carbon, and $\mathbf{R}$ is the position vector of an atom.

The effective momentum in the presence of ${\bf B}$ is
$\mathbf{P}-e\mathbf{A}/c$, so that the 3D electronic bands of bulk
graphites are quantized into the so-called 1D LSs. A periodic
Peierls phase $G_{R}\equiv {\frac{2{\pi }}{{\phi }_{0}}}$
$\int_{\mathbf{R}^{\prime }}^{\mathbf{R}}\mathbf{A}(\mathbf{r})\cdot
d\mathbf{r}$ is introduced to the tight-binding functions in Eqs.
(2) and (3), where $\mathbf{A}$ is the vector potential and $\phi
_{0}=2\pi\hbar c/e$ ($4.1356\times10^{-15}$ [T$\cdot$m$^{2}]$) is the flux
quantum. The Hamiltonian element coupled with the Peierls phase
factor is given by
\begin{equation}
H^{B}_{i,j}=H_{i,j}e^{i\Delta G_{i,j}}=H_{i,j}e^{i
{\frac{2{\pi}}{{\phi }_{0}}}\int_{\mathbf{R_{j}}}^
{\mathbf{R_{i}}}\mathbf{A}(\mathbf{r})\cdot d\mathbf{r}}\text{.}
\end{equation}
The phase factor gives rise to an enlargement of the primitive unit
cell (Fig. 1(d)), depending on the commensurate period of the
lattice and the Peierls phase. Using the Landau gauge
$\mathbf{A}=(0, B_{0}x, 0)$, the period of the phase is
$l=3R_{B}b\hat{x}$, in which there are
$2R_{B}$ $A$ and $2R_{B}$ $B$ atoms in an enlarged unit cell
($R_{B}=\frac{\phi_{0}/(3\sqrt{3}b^{\prime 2}/2)}{B_{0}}\simeq \frac{%
79000\text{ T}}{B_{0}}$).
This implies that the wave functions of graphites under a uniform
magnetic field can be characterized by the subenvelope functions
spanned over all bases in the enlarged unit cell, the zero points of
which are used to define the quantum numbers of LSs.
The wavefunction is decomposed into two components in the
magnetically enlarged unit cell as follows:
\begin{equation}
|\Psi _{\mathbf{k}}\rangle =\sum\limits_{m=1}^{2R_{B}-1}
(A_{o}|A_{m\mathbf{k}}\rangle +B_{o}|B_{m\mathbf{k}}\rangle
)+\sum\limits_{m=1}^{2R_{B}}(A_{e}|A_{m\mathbf{k}}\rangle
+B_{e}|B_{m\mathbf{k}}\rangle )\text{,}
\end{equation}%
where $o$ and $e$, respectively, represent the odd-indexed and
even-indexed parts. The subenvelope function $A_{o,e}$ ($B_{o,e}$),
described by an $n$-th order Hermite polynomial multiplied with a
Gaussian function, is even or odd spatially symmetric and represents
the probability amplitude of wavefunction contributed by each carbon
atom. Considering the Peierls substitution for interlayer and
intralayer atomic interactions, we can obtain the explicit form of
the magnetic Hamiltonian matrix of bulk graphites. A procedure for
the band-like Hamiltonian matrix is further introduced to
efficiently solve the eigenvectors and eigenvalues by choosing an
appropriate sequence for the bases. In the following sections, the
Hamiltonian matrices are derived for the simple hexagonal, Bernal
and rhombohedral graphites in the generalized tight-binding model.

\subsubsection{Simple hexagonal graphite}

Simple hexagonal graphite, as shown in Fig. 1(a), has each layer
periodic along $\hat{z}$ with the same x-y projection. The primitive
unit cell includes only two atoms same as that of monolayer
graphene. Four important atomic interactions are used to describe
the electronic properties, i.e., $\alpha_{0}(=2.569$ eV),
$\alpha_{1}(=0.361$ eV), $\alpha_{2}(=0.013)$ eV and
$\alpha_{3}(=-0.032$ eV),\cite{PRB44;13237} respectively coming from
the intralayer hopping between nearest-neighbor atoms, the
interlayer vertical hoppings between nearest- and
next-nearest-neighbor planes, and non-vertical hopping between
nearest-neighbor planes.

The zero-field Hamiltonian matrix in the subspace of tight-binding
basis $\{\varphi_{A},\varphi_{B}\}$ is expressed as
\begin{equation}
H_{AA}=\left\{
\begin{array}{cc}
\alpha_{1}h+\alpha_{2}(h^{2}-2) & f(k_{x},k_{y})(\alpha_{0}+\alpha_{3}h)\\
f^{*}(k_{x},k_{y})(\alpha_{0}+\alpha_{3}h) & \alpha_{1}h+\alpha_{2}(h^{2}-2)  \\
\end{array}%
\right\} \text{,}
\end{equation}%
where $f(k_{x},k_{y})=\sum_{j=1}^{3}\exp(i\mathbf{k}\cdot
\mathbf{r}_{j})=
\exp(ibk_{x})+\exp(ibk_{x}/2)\cos(\sqrt{3}bk_{y}/2)$ represents the
phase summation arising from the three nearest neighbors, and
$h=2\cos(k_{z}I_{z})$. The $k_{z}$-dependent terms are involved in
the matrix elements due to the periodicity along the $z$-direction.
The $\pi$-electronic energy dispersions are obtained from
diagonalizing the Hamiltonian matrix in Eq. (5):
\begin{equation}
E^{c,v}_{\pm}(k_{x},k_{y},k_{z})=\alpha_{1}h+2\alpha_{2}[h^{2}/2-1]\pm
(\alpha_{0}+\alpha_{3}h)|f(k_{x},k_{y})|,
\end{equation}%
and wave functions are
\begin{equation}
\Psi^{c,v}_{\pm,\mathbf{k}}=\frac{1}{\sqrt{2}}\{\Psi_{\mathbf{k}}^{A}\pm
\frac{f^{*}(k_{x},k_{y})}{|f(k_{x},k_{y})|}\Psi_{\mathbf{k}}^{B}\}.
\end{equation}%
The superscripts $c$ and $v$, respectively, represent the conduction
and valence states.

As a result of the vector-potential-induced phase, the number of
bases in the primitive unit cell is increased by $2R_{B}$ times
compared to the zero-field case. Using the Peierls substitution of
Eq. (4) and considering only the neighboring atoms coupled by
$\alpha$'s, one can derive a band-like form for the magnetic
Hamiltonian matrix of the AA-stacked graphite
\begin{eqnarray}
\langle B_{mk}|H|B_{m^{\prime }k}\rangle &=&\langle
A_{mk}|H|A_{m^{\prime }k}\rangle {=[\alpha_{1}h+\alpha
_{2}(h^{2}-2)]{\delta
_{m,m^{\prime}}}}\text{,}\label{7.1}\\
{{\langle A_{mk}|H|B_{m^{\prime }k}\rangle }} &{=}&{{(\alpha
_{0}+\alpha _{3}h)}}[t_{1k}(m)\delta_{m,m^{\prime }}+q\delta
_{m-1,m^{\prime}}] \text{,}\label{7.1}
\end{eqnarray}%
where the eigenvector is expanded in the bases with the specific
sequence \\
$\{{A_{1k},B_{2R_{B}k},
B_{1k},A_{2R_{B}k},........B_{R_{B}k},A_{R_{B}+1k}}\}$. The
independent phase terms are
\begin{equation}
\begin{gathered}
\\t_{1k}(m)=exp\{i[-(k_{x}b/2)-(\sqrt{3}k_{y}b/2)+\pi\Phi(i-1+1/6)]\},
\hfill \\
\\t_{2k}(m)=exp\{i[-(k_{x}b/2)-(\sqrt{3}k_{y}b/2)+\pi\Phi(i-1+3/6)]\},
\hfill \\
\\q=exp\{ik_{x}b\}. \hfill \\.
\end{gathered}
\end{equation}
By diagonalizing the matrix in Eqs. (8) and (9), the
$k_{z}$-dependent energies and wave functions of the valence and
conduction LSs are thus obtained. Such a band-like matrix spanned by
a specific order of the bases is also applicable to other prototypes
bulk graphites. In addition,  when the low-energy approximation,
related to the Dirac points, is made for the Hamiltonian matrix  in
Eq. (5), the LSs spectra are further evaluated from the magnetic
quantization (discussion in 3.3.2).$^{Ref}$ However, the
conservation of 3D carrier density needs to be included in  this
evaluation.

\subsubsection{Bernal graphite}

Bernal graphite is the primary component of the natural graphites.
The primitive unit cell comprises $A^{1}$, $B^{1}$, $A^{2}$ and
$B^{2}$ atoms on two adjacent layers, where $A^{1}$ and $A^{2}$
($B^{1}$ and $B^{2}$) are directly located above or below $A^{2}$
and $A^{1}$ (the centers of hexagons) in adjacent layers, a
configuration namely AB stacking (Fig. 1(b)). The critical atomic
interactions based on Slonczewski-Weiss-McClure (SWM) model cover
$\gamma_{0},.....,\gamma_{5}$, which are interpreted as hopping
integrals between nearest-neighbor and next-nearest-neighbor atoms,
and additionally $\gamma_{6}$, which refers to the difference of the
chemical environments between non-equivalent $A$ and $B$ atoms. The
values are as follows: $\gamma_{0}=3.12$ eV, $\gamma_{1}=0.38$ eV,
$\gamma_{2}=-0.021$ eV, $\gamma_{3}=0.28$ eV, $\gamma_{4}=0.12$ eV,
$\gamma_{5}=-0.003$ eV and $\gamma_{6}=-0.0366$ eV.\cite{CARBON32;289}

The tight-binding Hamiltonian is described by a $4\times4$ matrix,
which, expanded in the basis
$\{{\varphi_{A^{1}},\varphi_{B^{1}},\varphi_{B^{2}},\varphi_{A^{2}}}\}$,
takes the form
\begin{equation}
H_{AB}=\left\{
\begin{array}{cccc}
E_{A} & \gamma_{0}f(k_{x},k_{y})
& \gamma_{1}h & \gamma_{4}h f^{*}(k_{x},k_{y})\\
\gamma_{0}f^{*}(k_{x},k_{y}) & E_{B}
&\gamma_{4}h f^{*}(k_{x},k_{y}) & \gamma_{3}h f(k_{x},k_{y})\\
\gamma_{1}h & \gamma_{4}h f(k_{x},k_{y})
& E_{A} &\gamma_{0}f^{*}(k_{x},k_{y})\\
\gamma_{4}h f(k_{x},k_{y})&\gamma_{3}h f^{*}(k_{x},k_{y})
&\gamma_{0}f(k_{x},k_{y})&E_{B}\\
\end{array}%
\right\} \text{,}
\end{equation}%
where $E_{A}=\gamma_{6}+\gamma_{5}h^{2}/2$ and
$E_{B}=\gamma_{2}h^{2}/2$ indicate the sum of the on-site energy and
the hopping energy of A and B atoms, respectively. Energy bands and
wave functions are easily calculated from diagonalizing the
Hamiltonian matrix.

At $\mathbf{B}=B_{0}\widehat{z}$, the magnetically enlarged unit
cell includes $2\times 4R_{B}$ bases, which constructs the
$k_{z}$-dependent Hamiltonian matrix with non-zero terms only
between neighboring sublattices on same and different layers. An
explicit form of the matrix elements is given by
\begin{eqnarray}
\langle B_{mk}^{1}|H|A_{mk^{\prime }}^{1}\rangle &=&
-\gamma_{0}(t_{1}(m)\delta_{m,m^{\prime}}+q\delta_{m+1,m^{\prime}}),
\label{2.1} \\
\langle B_{mk}^{1}|H|A_{mk^{\prime }}^{2}\rangle &=&
\gamma_{4}h(t_{1}(m)\delta_{m,m^{\prime}}+q\delta_{m+1,m^{\prime}}),
\label{2.2} \\
\langle A_{mk}^{2}|H|A_{mk^{\prime}}^{1}\rangle&=&
\gamma_{1}h\delta_{m,m^{\prime}},
\label{2.3}\\
\langle B_{mk}^{2}|H|B_{mk^{\prime}}^{1}\rangle &=&
\gamma_{3}h(t_{2}(m)\delta_{m,m^{\prime}}+q\delta_{m+1,m^{\prime}}),
\label{2.4} \\
\langle A_{mk}^{2}|H|B_{mk^{\prime}}^{2}\rangle &=&
-\gamma_{0}(t_{3}(m)\delta_{m-1,m^{\prime}}+q\delta_{m,m^{\prime}}),
\label{2.5} \\
\langle A_{mk}^{1}|H|B_{mk^{\prime}}^{2}\rangle &=&
-\gamma_{4}h(t_{3}(m)\delta_{m-1,m^{\prime}}+q\delta_{m,m^{\prime}}),
\label{2.6} \\
\langle A_{mk}^{1}|H|A_{mk^{\prime}}^{1}\rangle &=&\langle
A_{m}^{2}|H|A_{m^{\prime}}^{2}\rangle=E_{A}\delta_{m,m^{\prime}},
\label{2.7} \\
\langle B_{mk}^{1}|H|B_{mk^{\prime}}^{1}\rangle &=& \langle
B_{mk}^{2}|H|B_{mk^{\prime}}^{2}\rangle=E_{B}\delta_{m,m^{\prime}},
\label{2.7} \\
\end{eqnarray}
where $t_{1}(m)$, $t_{2}(m)$, $q$ are shown in Eq. (10), and
$t_{3}(m)$ is expressed as
\begin{equation}
t_{3}(m)=exp\{i[-(k_{x}b/2)-(\sqrt{3}k_{y}b/2)+\pi\Phi(i-1+5/6)]\}\\
+exp\{i[-(k_{x}b/2)+(\sqrt{3}k_{y}b/2)-\pi \Phi(i-1+5/6)]\}.
\end{equation}
It should be noted that the calculations based on the effective-mass
approximation get trouble with an infinite order of the Hamiltonian
matrix induced by the significant interlayer hopping integral of
$\gamma_{3}$\cite{PRB77;155416,PRB84;205448}; this divergence also exists for the ABC-stacked systems.\cite{PRB84;125455} Nevertheless, through a qualitative perturbation analysis
of $\gamma_{3}$ and other interlayer interactions, the minimal
model, which regards $\gamma_{0}$ and $\gamma_{1}$ as the
unperturbed terms, well describes the low-energy dispersions in the
vicinity of the vertical edges in the first Brillouin zone. The
generalized Peierls tight-binding model, which retains all important
atomic interactions and magnetic field, however, can provide
comprehensive descriptions for graphites more than the limitation of
the accuracy at low energies.

\subsubsection{Rhombohedral graphite}

For the ABC-stacked graphite, called rhombohedral graphite, the unit
cell is chosen along the $z$-direction (Fig. 1(c)). There are six
atoms in a unit cell. The interlayer atomic interactions, based on
SWM model, take into account the nearest-neighbor intralayer
interaction $\beta _{0}=-2.73$ eV and five interlayer interactions
$\beta _{1}=0.32$ eV, $\beta _{2}=-0.0093$ eV, $\beta _{3}=0.29$ eV,
$\beta_{4}=0.15$ eV and $\beta _{5}=0.0105$ eV, in which the former
two refer to vertical atoms and the latter three are
non-vertical.\cite{NJP15;053032} The Hamiltonian matrix can be
expressed as a combination of nine $2\times2$ submatrices for
simplicity
\begin{equation}
H_{ABC}=\left\{
\begin{array}{ccc}
H_{1} & H_{2} & H_{2}^{\star}\\
H_{2}^{\star} & H_{1} & H_{2}\\
H_{2} & H_{2}^{\star} & H_{1}\\
\end{array}%
\right\} \text{,}
\end{equation}%
where $H_{1}$ and $H_{2}$ take the forms
\begin{equation}
H_{1}=\left\{
\begin{array}{cc}
0 & \beta_{0}f(k_{x},k_{y})\\
\beta_{0}f^{*}(k_{x},k_{y}) & 0\\
\end{array}%
\right\} \text{;}
\end{equation}%

\begin{equation}
H_{2}=\left\{
\begin{array}{cc}
(\beta_{4}\exp(ik_{z}I_{z})+\beta_{5}\exp(-i2k_{z}I_{z}))f^{*}(k_{x},k_{y})
& \beta_{1}\exp(ik_{z}I_{z}) +\beta_{2}\exp(i2k_{z}I_{z}) \\
(\beta_{3}\exp(ik_{z}I_{z})+\beta_{5}\exp(-i2k_{z}I_{z}))f(k_{x},k_{y})
&(\beta_{4}\exp(ik_{z}I_{z})+\beta_{5}\exp(-i2k_{z}I_{z}))f^{*}(k_{x},k_{y})\\
\end{array}%
\right\} \text{.}
\end{equation}%

It is also noted that the hexagonal unit cell used here is not the
primitive unit cell of rhombohedral graphite. The primitive unit
cell should be a rhombohedral form that consists of 2 atoms and
inclined to the $z$-axis by an angle $\theta =\tan
^{-1}(\frac{b/2}{I_{z}})$ (Fig. 1(c)). That is to say, the bases of
the primitive unit cell are reduced from 6 to 2, as the rhombohedral
unit cell is selected instead of the hexagonal
one.\cite{JMST897;118} The Hamiltonian has an analytic solution near
the zone edges H-K-H by using a continuum
approximation.\cite{NJP15;053032,Carbon7;425,CJP36;352} This
reflects the fact that the energy dispersions in the hexagonal cell
can be zone-folded to the primitive rhombohedral one, and that the
inversion symmetry is characterized, similarly to that of simple
hexagonal graphite. As a result, the physical properties of
rhombohedral graphite might present certain features similar to
those of monolayer graphene or simple hexagonal graphite, and their
difference is only the degeneracy of energy states. A comparison
between rhombohedral and hexagonal unit cells is made in detail in
Chapter 5.

At $\mathbf{B}=B_{0}\widehat{z}$, the magnetically enlarged
rectangle cell is chosen as the enlargement of the hexagonal unit
cell along the $z$-axis for the convenience of calculations. Such a
rectangular cell includes $3\times 4R_{B}$ atoms and the Hamiltonian
matrix elements are given by
\begin{eqnarray}
\langle B_{m\mathbf{k}}^{1}|H|A_{m^{\prime }\mathbf{k}}^{1}\rangle
&=&\beta _{0}[t_{1}(m)\delta_{m,m^{\prime }}+q\delta
_{m,m^{\prime }-1}], \\
\langle B_{m\mathbf{k}}^{2}|H|A_{m\mathbf{k}}^{2}\rangle
&=&\beta_{0}
(t_{3}(m)\delta_{m,m^{\prime }-1}+q\delta_{m,m^{\prime }}), \\
\langle B_{m\mathbf{k}}^{3}|H|A_{m\mathbf{k}}^{3}\rangle
&=&\beta_{0} (t_{3}(m)\delta_{m,m^{\prime }}+q\delta_{m,m^{\prime
}+1}),
\label{3.1} \\
\langle A_{m\mathbf{k}}^{1}|H|B_{m\mathbf{k}}^{2}\rangle &=&\langle
A_{m\mathbf{k}}^{2}|H|B_{m\mathbf{k}}^{3}\rangle=\langle
B_{m\mathbf{k}}^{1}|H|A_{m\mathbf{k}}^{3}\rangle=(\beta
_{1}e^{i3k_{z}I_{z}}+ \beta_{2}e^{-6ik_{z}I_{z}})\delta_{m,m^{\prime
}},
\label{3.2} \\
\langle A_{i\mathbf{k}}^{3}|H|B_{j\mathbf{k}}^{2}\rangle &=&(\beta
_{3}e^{ik_{z}I_{z}}+\beta_{5}e^{-6ik_{z}I_{z}})(
t_{1}(m)\delta_{m,m^{\prime }-1}+q\delta_{m,m^{\prime }}),
\label{3.3} \\
\langle A_{m\mathbf{k}}^{2}|H|B_{m\mathbf{k}}^{1}\rangle &=& (\beta
_{3}e^{i3k_{z}I_{z}}+\beta
_{5}e^{-6ik_{z}I_{z}})(t_{2}(m)\delta_{m,m^{\prime
}}+q\delta_{m,m^{\prime }+1}),
\label{3.3} \\
\langle A_{i\mathbf{k}}^{1}|H|B_{j\mathbf{k}}^{3}\rangle &=&(\beta
_{3}e^{ik_{z}I_{z}}+\beta_{5}e^{-6ik_{z}I_{z}})(
t_{3}(m)\delta_{m,m^{\prime }-1}+q\delta_{m,m^{\prime }}),
\label{3.3} \\
\langle B_{m\mathbf{k}}^{1}|H|B_{m\mathbf{k}}^{2}\rangle &=& (\beta
_{4}e^{i3k_{z}I_{z}}+\beta _{5}e^{-6ik_{z}I_{z}})
(t_{2}(m)\delta_{i,j}+q\delta_{m,m^{\prime }+1})\delta_{m,m^{\prime
}-1},
\label{3.4} \\
\langle A_{m\mathbf{k}}^{1}|H|A_{m\mathbf{k}}^{2}\rangle &=& \langle
B_{m\mathbf{k}}^{2}|H|B_{m\mathbf{k}}^{3}\rangle = (\beta
_{4}e^{-i3k_{z}I_{z}}+\beta _{5}e^{6ik_{z}I_{z}})
(t_{3}(m)\delta_{m,m^{\prime }-1}+q\delta_{m,m^{\prime }}),
\label{3.5} \\
\langle A_{m\mathbf{k}}^{2}|H|A_{m\mathbf{k}}^{3}\rangle &=&(\beta
_{4}e^{-i3k_{z}I_{z}}+\beta _{5}e^{6ik_{z}I_{z}})
t_{2}(m)\delta_{m,m^{\prime }}+q\delta_{m,m^{\prime }+1}),
\label{3.6} \\
\langle B_{j\mathbf{k}}^{3}|H|B_{j\mathbf{k}}^{1}\rangle &=&(\beta
_{4}e^{-i3k_{z}I_{z}}+\beta _{5}e^{6ik_{z}I_{z}})
t_{2}(m)\delta_{m,m^{\prime }}+q\delta_{m,m^{\prime }+1}),
\label{3.7} \\
\langle A_{j\mathbf{k}}^{3}|H|A_{i\mathbf{k}}^{1}\rangle &=&(\beta
_{4}e^{-i3k_{z}I_{z}}+\beta _{5}e^{6ik_{z}I_{z}})(
t_{1}(m)\delta_{m,m^{\prime }}+q\delta_{m,m^{\prime }+1}),
\label{3.8}. \\
\end{eqnarray}

The independent phase terms are shown in Eqs. (10) and (21). The
generalized tight-binding model, accompanied with an with exact
diagonalization method, can further be applied to study other
physical properties, such as the optical absorption
spectra\cite{ACSNano4;1465,APL97;101905,Carbon54;248,
CPC189;60,APL99;011914,PCCP17;15921,JPCM27;125602} and plasma
excitations.\cite{AOP339;298,PRB89;165407,PRB74;085406} Different
kinds of external fields, for example, a modulated magnetic
field,\cite{PRB83;195405} a periodic electric
potential\cite{JVSTB28;386} and even a composite
field,\cite{OptEx22;7473} could also be involved in the calculations
simultaneously. Furthermore, this model can also be applicable to
other layered materials with a precisely chosen layer sequence, such
as graphene, MOS$_{2}$ and silicene, germanene, tinene, and
phosphorene.\cite{PRB94;205427,NJP16;125002,RSCAdv5;51912,
SciRep7;40600,PRB94;045410}
The electronic structures and characteristics of wave functions
could be well depicted and the results are accurate and reliable
within a wide energy range.

\subsubsection{The gradient approximation for optical properties}

When graphite is subjected to an electromagnetic field, the optical
spectral function $A(\omega )$ is used to describe its optical
response. At zero temperature, $A(\omega )$ is expressed as follows
according to the Kubo formula,
\begin{eqnarray}
A(\omega
)&\propto&\sum_{n^{v},n^{c}}^{}\int_{1stBZ}\frac{d\textbf{k}}{{(2\pi)^{2}
}} \left\vert \left\langle \Psi ^{c}_{\textbf{k}}(n^{c})\left\vert
\frac{ \widehat{\mathbf{E}}\cdot \mathbf{P}}{m_{e}}\right\vert \Psi
^{v}_{\textbf{k}}
(n^{v})\right\rangle \right\vert^{2} \\ \nonumber%
&\times& Im\left\{\frac{{
f[E^{c}_{\textbf{k}}(n^{c})]-f[E^{v}_{\textbf{k}}(n^{v})]}}
{E^{c}_{\textbf{k}} (n^{c})-E^{v}_{\textbf{k}}(n^{v}){-\omega
}{-\imath \gamma }} \right\},
\end{eqnarray}
where $\widehat{\mathbf{E}}$ is the direction of electric
polarization, $\textbf{P}$ the momentum operator,
$f[E_{\textbf{k}}(n)]$ the Fermi-Dirac distribution, $m_{e}$ the
electron mass and $\gamma$ the phenomenological broadening
parameter. $\widehat{\mathbf{E}}$ lies on the ${(x,y)}$ plane is
chosen for a model study. $n^{c,v}$ is the energy band index
measured from the Fermi level at zero field, or it represents the
quantum number of each LS. The integration for all wave vectors is
done within  a hexahedron (a rectangular parallelepiped) at zero
(non-zero) magnetic field. The initial and final state satisfy the
condition of $\Delta \mathbf{k}=0$, responsible for the zero
momentum of photons. This implies that only the vertical transitions
are available in the valence and conduction bands. Using the
gradient approximation,\cite{Carbon54;248,ACSNano4;1465} the
velocity matrix element element is evaluated from
\begin{eqnarray}
M^{c,v}_{\textbf{k}}(n^{c},n^{v})\sim\frac{\partial}{\partial k_{x}}\left\langle
\Psi ^{c}_{\textbf{k}}(n^{c})\left\vert H \right\vert \Psi ^{v}_{\textbf{k}}
(n^{v})\right\rangle \quad \text{for} \quad     \widehat{E}\|\widehat{x}.
\end{eqnarray}
Substituting the Hamiltonian matrix of graphite into Eq. (38), and
integrating all the available transitions over the first Brillouin
zone and the quantum numbers, the spectral absorption function
$A(\omega)$, is obtained. In addition, the absorption spectra are
almost independent of the polarization direction, when
$\widehat{\mathbf{E}}$ is on the x-y plane.

The velocity matrix significantly depends on the relation between
the initial- and final-state wave functions, a main factor in
determining the transition intensity and the optical selection rule.
In the absence of external fields, what should be especially noticed
is the optical transitions centered about the highly symmetric
$\mathbf{k}$ points, e.g., $\Gamma$, M, K...., where the joint
density of states (JDOS) and $M^{c,v}_{\textbf{k}}(n^{c},n^{v})$
have relatively large values.
Under a magnetic field, the Bloch function at a fixed $k_{z}$ is a
linear combination of the products of the subenvelope function and
the tight-binding function on each sublattice site in the enlarged
unit cell. That is,
\begin{equation}
\begin{gathered}
\left\vert\Psi_{\textbf{k}})\right\rangle
=\sum_{m=1}^{2R_{B}}A_{m}\left\vert \ A_{m\textbf{k}}\right\rangle
+B_{m}\left\vert \ B_{m\textbf{k}}\right\rangle,
\end{gathered}
\end{equation}
where $A_{m}$ and $B_{m}$ are the subenvelope functions, and $m$
indicates the $m$-th atom. In consequence,
$M^{c,v}_{\textbf{k}}(n^{c},n^{v})$ is simplified as the product of
three matrices: the operator $\frac{\partial H}{\partial k}$ and the
subenvelope functions of the initial and final states. Moreover,
$M^{c,v}_{\textbf{k}}(n^{c},n^{v})$ can be deduced as a simple inner
product of the subenvelope functions, due to the fact that the
Peierls phase slowly changes in the enlarged unit cell so that this
derivative term $\frac{\partial H}{\partial k}$ can be taken out of
the summation in Eq. (39). Considering both interlayer and
intralayer atomic interactions, one find that while all the hopping
integrals, $\alpha$'s, $\beta$'s or $\gamma$'s, make contributions
to the absorption spectrum, the relatively stronger in-plane atomic
interaction, $\alpha_{0}$, $\beta_{0}$ or $\gamma_{0}$, plays the
most important role in the optical transitions. When the occupied
LSs are excited to the unoccupied ones, the available excitation
channels satisfy the general selection rule, $\Delta
n=n^{c}_{A^{l}(B^{l})}-n^{v}_{B^{l}(A^{l})}=\pm1$, where
$n_{A(B)}^{l}$ is the quantum mode for the $n^{c,v}_{A^{l}(B^{l})}$
sublattices on the $l$-th layer. The detailed calculation results
are discussed in the following chapters.


\section{Simple hexagonal graphite}
The AA-stacked graphite possesses the highest stacking symmetry
among the layered graphites. The hexagonal symmetry, the AA stacking
configuration and the significant interlayer atomic interactions are
responsible for the unusual essential properties. The non-titled
Dirac-cone structure is formed along the $k_z$-direction, in which
its width is more than 1 eV. The 3D Dirac cone covers free electrons
and holes with the same density, leading to the semi-metallic
behavior with an obvious plateau structure in the low-energy DOS. It
is further quantized into the 1D parabolic LSs without any crossings
or anti-crossings. Each well-behaved LS contributes two asymmetric
square-root-form peaks in DOS. A lot of LSs, which can cross the
Fermi level, belong to the valence or conduction ones. Specifically,
this creates the intraband and the interband inter-LS
magneto-optical excitation channels. The quantized energies have a
simple dependence on (${B_0,n^{c,v},k_z}$), so that the
magneto-absorption spectra present the beating features. Such
phenomena are never predicted or observed in the other
condensed-matter systems. On the other hand, the zero-field
absorption spectrum is largely suppressed and almost featureless at
low frequency because of many forbidden vertical transitions. The
AA-stacked graphite and graphenes quite different from each other in
electronic and optical properties. The experimental verifications on
energy bands, DOSs and absorption spectra of simple hexagonal
graphite could be utilized to determine the critical intralayer and
interlayer atomic interactions.

\subsection{Electronic structures without external
fields}

The 2D $\pi$-electronic structure of a monolayer graphene is
reviewed first. Given the interlayer atomic interactions
$\alpha_{1}=0$, $\alpha_{2}=0$ and $\alpha_{3}=0$ in Eq. (5), one
can obtain the band structure of monolayer graphene, i.e.,
$E^{c,v}(k)=\pm\alpha_{0}|f(k_{x},k_{y})|=\pm\alpha_{0}
\{1+4\cos(3bk_{x}/2\cos(\sqrt{3}bk_{y}/2+4\cos^{2}
(\sqrt{3}bk_{y}/2\}^{1/2}$. The band structure is simplified as the projection of the energy dispersion of the simple hexagonal graphite on the $k_{z}=0$ plane (the red hexagon in Fig. 2(a)).
Both conduction and valence bands are symmetric about the Fermi
level ($E_F=0$) along K$\rightarrow\Gamma\rightarrow$
M$\rightarrow$K. In the low-energy region, the energy dispersion is
described by $E^{c,v}=\pm3\alpha_{0}bk/2$, which characterizes an
isotropic Dirac cone centered at the K point (the Fermi level).
There are special band structures at highly symmetric points in the
1st BZ, e.g., the local maximum $E^{c}=3\alpha_{0}$ and the local
minimum $E^{v}=-3\alpha_{0}$ at the $\Gamma$ point, and the saddle
points $E^{c,v}=\pm\alpha_{0}$ at the M point. Such critical points
in the energy-wave-vector space would induce Van Hove singularities
in DOS. The band width is evaluated as $6\alpha_{0}$, which is
determined by the difference between the two local extreme values at
the $\Gamma$ point. Monolayer graphene is a zero-gap semiconductor
with a vanishing DOS at $E_F$ (Fig. 3(b)); that is, free carriers
are absent at zero temperature.

The interlayer atomic interactions can dramatically change
electronic structures. According to Eq. (5), the energy dispersions
of simple hexagonal graphite without magnetic field are shown by the
black curves in Fig. 2(a). There exists one pair of valence and
conduction bands, in which the former is no longer symmetric to the
latter about the Fermi level, $E_{F}=0.016$ eV. Energy bands are
highly anisotropic and strongly dependent on $k_{z}$. At a fixed
$k_{z}$, the $(k_{x},k_{y})$-dependent energy dispersions resemble
those of a monolayer graphene. Moreover, the critical points are
very sensitive to the change of $k_z$, e.g., those at the corners (K
and H; Fig. 1(e)), the middle points  between two corners (M and L);
the centers of the  ${k_x-k_y}$ plane ($\Gamma$ and A). The energy
spacing of a monolayer-like band structure grows when $k_{z}$ moves
from K to H. That is, the Dirac-cone structures could survive and
remain similar in the increase/decrease of $k_z$. This will be
directly reflected in the magnetic quantization. The middle points,
which correspond to the saddle points with high DOS, are expected to
present the strong absorption spectra. Overall, the $\pi$-electronic
width is evaluated as the energy difference between the maximum
energy at the $\Gamma$ point and the minimum energy at the A point:
that is,
$[2(\alpha_{1}+\alpha_{2})+3(\alpha_{0}+2\alpha_{3})]-[-2(\alpha_{1}-
\alpha_{2})-3(\alpha_{0}-2\alpha_{3})=4\alpha_{1}+6\alpha_{0}$.

In the low-energy approximation around the corners along the K-H
direction, Eq. (5), used to describe the Dirac-type energy
dispersions, can be expressed as
\begin{equation}
E^{c,v}_{\pm}(k_{x},k_{y},k_{z})=E_{D}\pm v_{F}|\mathbf{k}|
\end{equation}%
where $E_{D}=\alpha_{1}h+2\alpha_{2}[h^{2}/2-1], v_{F} =
3b(\alpha_{0}+\alpha_{3}h)/2$ (the Fermi velocity) and
$|\mathbf{k}|=\sqrt{k_{x}^{2}+k_{y}^{2}}$. The first term $E_{D}$
indicates the Dirac-point energy. The second term represents the
conical energy dispersion of which the slope particularly shows a
slight discrepancy on $k_{z}$.

A closer examination is necessary to explore the dependence of the
Dirac cone on $k_{z}$. Given by $E_{D}$ in Eq. (41), the
localization of the Dirac point is described as a correspondence to
the energy dispersion along the K-H line (indicated by the arrow in
Fig. 2). In the vicinity of the zone corners, the conduction and
valence Dirac cones overlap each other. At the K point, the state
energy of the Dirac point, $2(\alpha_{1}+\alpha_{2})$, is higher
than $E_{F}$. This indicates that the valence states between $E_{F}$
and the Dirac point are regarded as free holes in the low-lying
valence bands. As the states gradually move away from K toward H,
the carrier density of free holes decreases because the Dirac point
gets lower. It is not until the Dirac point approaches the Fermi
level that there are no free carriers. With a further increase of
$k_{z}$ ($E_{D}<E_{F}$), the free carrier change into electrons,
being determined by the Fermi level in the conduction Dirac cone.
Its density reaches a maximum value at the H point. In short, it
means that the interlayer atomic interactions induce free-hole
(free-electron) pockets in the low-energy valence (conduction) bands
near the K (H) point. Two kinds of free carriers have the same
density. Furthermore, the Dirac points of cone structures are
located at the corners of the 1st BZ during the variation of $k_z$
(Fig. 1(e)). These are expected to play an important role in the
essential physical properties, e.g., optical properties,
magneto-electronic and magneto-optical properties, electronic
excitations, and transport properties.

In the case of 2D multi-layer AA-stacked graphene, there are $N$
pairs of valence and conduction Dirac cones, mainly owing to the
highest stacking symmetry. For example, there are two and three
pairs in bilayer and trilayer graphenes, respectively (Figs.
2(b)$\&$ 2(c)), in which the overlap of valence and conduction cones
indicates the semimetallic behavior. The Dirac-cone structures,
which are initiated from the K point, are almost symmetric about the
Fermi level. The Dirac-point energies between ${-2\alpha_1}$ and
${2\alpha_1}$ are described by\cite{JAP110;013725}
\begin{eqnarray}
E_{D}=2\cos[j\pi/(N+1)]\alpha_{1},
\end{eqnarray}
in the low-energy approximation (ignoring $\alpha_2$ and
$\alpha_3$), where $j=1,2,......N$. When $N$ is an odd number, the
Dirac point of the middle cone structure touches with the Fermi
level. With an increase of layer number, the multi cone structures
are gradually evolved into a 3D one with a significant
$k_z$-dependent band width. However, it might have certain important
differences between the AA-stacked few-layer graphenes and graphite
in the essential properties as a result of the confinement effect
along the $z$-direction, e.g., the optical threshold frequency, DOS,
and the features of magneto-absorption peaks.

On the experimental side, ARPES can directly identify the
wave-vector-dependent energy bands. Using the high-resolution ARPES
measurements, the dimension-created unusual electronic structures
have been verified for the carbon-related systems with the hexagonal
symmetry, including graphene nanoribbons, number- and
stacking-dependent graphenes, and AB-stacked graphite. The confirmed
characteristics cover the confinement-induced energy gap and 1D
parabolic bands in finite-width
nanoribbons,\cite{ACSNano6;6930,PRB73;045124} the Dirac-cone
structure in monolayer
graphene,\cite{PRL98;206802,PRL110;146802,NatPhys3;36} two/three
pairs of linear bands in bilayer/trilayer AA stacking,
\cite{NatMater12;887,NanoLett8;1564} two pairs of parabolic bands in
bilayer AB stacking,\cite{PRL98;206802,Science313;951} the partially
flat, sombrero-shaped and linear bands in tri-layer ABC
stacking,\cite{PRB88;155439} and the bilayer- and monolayer-like
energy dispersions in Bernal graphite at the K and H points,
respectively.\cite{PRL100;037601,ASS354;229,PhyB407;827,PRB79;125438,NatPhy2;595} The 3D band structure of AA-stacked graphite is worthy of the detailed ARPES examinations, especially for the Dirac-cone structures and the saddle points along the K-H and M-L lines, respectively. Such measurements can determine the intralayer and interlayer hopping integrals and the significant effects due to them.

The primary characteristics of electronic structures directly
reflect on DOS. The low-energy special structures in DOS are
dominated by the stacking configuration or the interlayer atomic
interactions. In the range of ${|E|\le\,0.3}$ eV and ${B_0=0}$,
simple hexagonal graphite presents a plateau structure centered
about $E=0$ (the Fermi level), as shown in Fig. 3(a).
This originates from the superposition of all $k_z$-dependent
Dirac-cone structures with various disks in the $(k_x,k_y)$ plane. A
finite DOS at ${E=0}$ clearly illustrates the semi-metallic
behavior. DOS grows quickly in the increase of $E$. There exist two
very cusp structures at the middle energies of
$[-2(\alpha_1-\alpha_2)+(\alpha_0-2\alpha_3)]\leq E
\leq[2(\alpha_1+\alpha_2)+(\alpha_0+2\alpha_3)]$ and $
[-2(\alpha_1-\alpha_2)-(\alpha_0-2\alpha_3)]\leq
E\leq[2(\alpha_1+\alpha_2)-(\alpha_0+2\alpha_3)]$ (Eq. (6)), mainly
owing to the saddle points along the M-L line (Fig.
1(e)).\cite{NJP12;083060} On the other hand, monolayer graphene
exhibits a V-shape DOS near ${E=0}$, as shown in Fig. 3(b)). DOS
vanishes at the Fermi level, leading to  the semiconducting
behavior. For N-even systems, the low-energy DOS corresponds to  a
plateau structure, e.g., that of bilayer graphene (Fig. 3(c)).
However, it is a superposition of the plateau and V-shape structures
for N-odd systems, such as DOS of trilayer graphene (Fig. 3(d)).
Apparently, the AA-stacked graphenes of  N${\ge\,2}$  belong to
semimetals. At middle energy, the symmetric peaks of the logarithmic
form mainly come from the saddle point (the M point in Figs.
2(a)-2(c)), in which their number is proportional to that of layer
(Figs. 3(b)-3(d)).

\subsection{Optical properties without external
fields} The main features of absorption spectra are determined by
the velocity matrix element, carrier distribution and DOS. At low
frequency, the first one is just the Fermi velocity in the AA
stacking systems, mainly owing to the similar Dirac-cone with the
isotropic linear dispersions.\cite{NJP12;083060,APL103;041907} That is, the AA-stacked
graphite and multi-layer graphenes have the identical excitation
strength for each available channel. The former, as indicated in
Fig. 4(a), presents a largely reduced low-frequency absorption
spectrum and a shoulder structure at ${\omega\sim4\alpha_1}$. For
any given $k_z$, the vertical transitions are forbidden when half of
excitation frequency is smaller than the energy difference
($E_{th}$) between the Fermi-momentum state ($\bf k_F$) and the
Dirac point. $E_{th}$ is about $2\alpha_1$ for various $k_z$’s, so
that absorption spectrum is very weak at $\omega\,<2\alpha_1$. With
the increase of frequency, it exhibits a shoulder structure and
grows quickly, since the deeper or higher electronic states make
more contributions. On the other hand, absorption spectrum of
monolayer graphene is linearly proportional to excitation frequency
(Fig. 4(b)), directly reflecting the linear energy dependence of DOS
(Fig. 3(b)). As for the middle-frequency absorption spectrum,
graphite and graphene, respectively, exhibit the very prominent
plateau and symmetric peak at
${2\alpha_0-4\alpha_0\le\,\omega\le\,2\alpha_0+4\alpha_0}$ and
${\omega\,=2\alpha_0}$. The former originates from the saddle points
along the M-L line (Eq. (6)), with a rather high DOS. Such
structures are the so-called $\pi$-electronic absorption peaks,
frequently observed in the carbon-related systems with the sp$^2$
bondings (discussed later).

The AA-stacked layered graphenes present the unusual low-frequency
absorption spectra during the variation of layer number, as clearly
indicated in Fig. 4(b)-4(d).
The critical factor is the well-behaved
$N$ pairs of Dirac cones almost symmetric about the Fermi
level.\cite{APL103;041907} The wave functions of these cone
structures are  the symmetric or anti-symmetric linear superposition
of the layer-dependent tight-binding functions (e.g., Eq. (7)),
leading to the available excitation channels only arising from the
same Dirac cone. That is, the inter-Dirac-cone vertical transitions
are absent. The $N$-odd systems have the zero threshold frequency,
since the Dirac point of the middle cone structure touches with the
Fermi level, e.g., the trilayer system (Fig. 4(d)). However, the
optical gaps, which they are characterized the energy spacing of the
highest occupied and the lowest unoccupied Dirac points (Eq. (42)),
are finite in the N-even systems. They decline with the increasing
layer number, and the highest threshold frequency is ${2\alpha_1}$
for the bilayer AA stacking (Fig. 4(c)).\cite{PRB74;085406} As to
the other intra-Dirac-cone excitations, their threshold spectra
exhibit the shoulder structures with absorption frequency determined
by the Fermi-momentum state (or about double that of energy
difference between the Dirac point and the Fermi level). In
addition, absorption spectra might reveal two sub-shoulders because
of the slightly asymmetric Dirac-cone structures due to the
interlayer atomic interactions, e.g., those of N=4 and
5\.cite{APL103;041907} Specifically, the change of layer number
results in the crossing behavior. The AA-satcked graphenes and
graphite possess the almost identical low-frequency optical
properties when $N$ grows to 30 (detailed discussions in
Ref.\cite{APL103;041907}) The dimension-induced important
differences could be observed under the obvious confinement effect.

The above-mentioned features of vertical excitation spectra could be
verified by optical spectroscopies, such as the absorption,\cite{PRL102;037403,NatCom4;2542}
transmission,\cite{PRB83;125302,PRL102;037403,PRL100;087401,
PRL101;267601,PRL98;197403} reflection,\cite{PRB15;4077,PRB74;195404,PRL102;037403,PRL111;077402} Raman scattering\cite{NanoLett14;4548,PRL107;036807,NanoLett16;3710} and Rayleigh scattering spectroscopies.\cite{Science306;1540}
Experimental measurements have confirmed the rich and diverse
optical properties in the carbon-related systems, such as, Bernal
graphite,\cite{PRB80;161410,PRL102;166401,PRB86;155409,JAP117;112803} graphite intercalation compounds,\cite{JMR2;858,JPDAP48;485304}
layered graphenes,\cite{PRB83;125302,PRL98;197403,
PRL111;077402,PRL101;267601,PRL100;087401} graphene nanoribbons,\cite{NatComm4;2646,NatChem6;126} carbon
nanotubes,\cite{Science304;1129,AdvMater18;1166} and carbon fullerenes.\cite{PRB49;16746,PRB49;7012} Such systems
possess the ${\sim5-6}$ eV $\pi$ peak arising from the
${2p_z}$-orbital bondings; that is, all the ${sp^2}$-bonding systems
can create this prominent peak. The AB- and ABC-stacked graphenes
quite differ from each other in the absorption frequencies, spectral
structures and electric-field-induced excitation spectra.\cite{PRL106;046401,PRB81;155413,PRB85;245410,PRL104;176404,NatPhys7;944}
Moreover, carbon nanotubes exhibit the strong dependence of
asymmetric absorption peaks on radius and chirality.\cite{Science304;1129} The important features in AA-satcked graphenes and graphite  are worthy
of  systematic experimental investigations, especially for the
dependence of optical gap, shoulder structure, $\pi$ peak, and
spectral intensity on the layer number.

\subsection{Magnetic quantization}

\subsubsection{Landau levels and wave functions}

In the presence of $\mathbf{B}=B_{0}\hat{z}$, electrons are flocked
on the $x-y$ plane to form the transverse cyclotron motions, while
the motion along the field direction remains intact. The 3D
electronic states in simple hexagonal graphite are evolved into one
group of so-called LSs, which are dispersed along the
$\widehat{k_{z}}$ direction, but highly degenerate on the
$k_{x}-k_{y}$ plane. This implies that the $k_{z}$-dependent LSs are
directly quantized from the corresponding $k_{z}$-dependent Dirac
cones along K-H in the absence of a magnetic field. The study on the
magnetic quantization of a Dirac cone in monolayer graphene is the
first step to realize the magneto-electronic properties in
graphites.

The electronic states of a Dirac cone are magnetically quantized
into one group of valence and conduction LLs. Each LL,
dispersionless along $k_{x}$ and $k_{y}$, is fourfold degenerate
without the consideration of spin degeneracy. The occupied valence
and unoccupied conduction LLs are symmetric about $E_{F}=0$, as
shown in Fig. 5 (a).
The quantum numbers, characterized by the zero-point ones of the
subenvelpe functions, are indicated by $n^{c}$ and $n^{v}$ for the
conduction and valence LLs, respectively. Considering the sequence
of LLs, one can find that the $n^{c,v}=0$ LLs are located at
$E_{F}=0$, and the $n^{c,v}=1,2,3...$ LLs are counted away from the
Fermi level (Fig. 4(a)). At ${(k_x=0,k_y=0)}$, the corresponding
wave functions of the four-fold degenerate states are localized
around four different centers: $1/6$, $2/6$, $4/6$ and $5/6$
positions of the enlarged unit cell. The main features of LLs can be
realized by discussing one of the four-fold degenerate states, e.g.,
the $1/6$-localized Landau states (Fig. 5(b)). The quantum number is
determined by the normal mode in $B_o$ sublattice. For the cases of
$n^{c,v}\geq1$ LL, the subenvelope functions of $A_{o}$ and $B_{o}$
sublattices are presented in the (${n^{c,v}-1}$)-th and an
$(n^{c,v})$-th order Hermite polynomials, respectively. They have
the following relationship between conduction and valence states:
$A_{o}^{c}=A_{o}^{v}$ $\&$ $B_{o}^{c}=-B_{o}^{v}$ for the same
atoms, and $A_{o}^{v}(n^{v})\propto B_{o}^{c}(n^{c}=n^{v}-1)$ $\&$
$B_{o}^{v}(n^{v})\propto A_{o}^{c}(n^{c}=n^{v}+1)$ for the different
atoms. It can be deduced that as to the inter-LL optical
transitions, the simple linear relationships account for the
specific selection rule $\Delta n=n^{c}-n^{v}=\pm 1$, according to
the spectral function in Eq. (38).

For simple hexagonal graphite, the formation of the LSs corresponds
to the magnetic quantization of the Dirac cones that are distributed
along the K-H line as described by Eq. (41). The LS energy
dispersions strongly depend on $k_{z}$, and the relationship between
Landau states and wave functions at a fixed $k_{z}$ resembles that
of a monolayer graphene. These purely arise from the highly
symmetric AA stacking with the same ${(x,y)}$-plane projection. At
the K point, the conduction and valence LLs are symmetric about
$E^{c,v}(n^{c,v}=0,k_{z}=0)=E_{D}(k_{z}=0)\simeq 0.283\alpha_{0}$, as
indicated in Fig. 6(a). The similar LL spectrum is revealed at the H
point, while it is centered about $E_{D}(k_{z}=\pi/I_{z})$$\simeq-0.279\alpha_{0}$ (Fig. 6(b)).
With the same quantum number, the AA-stacked graphite and monolayer
graphene have the same relationship of two subenvelope functions
with respect to the amplitude, spatial symmetry, phase and zero
points, as shown in Figs. 6(c) and 5(b). Moreover, the linear
relationship between two subenvelope functions remains the same,
clearly illustrating that the specific optical selection rule of
$\Delta n=\pm1$ is also applicable to the inter-LS transitions in
simple hexagonal graphite. In short, 3D simple hexagonal graphite
consisting of the same projection graphenes layers exhibits the
essential 2D quantum phenomena, mainly owing to the Dirac-type
energy dispersions.


\subsubsection{Landau subband energy spectra}

The LS spectrum in the 1st BZ ($0\leq k_{z}\leq \pi/I_{z}$) is
essential for understanding the magneto-electronic properties of
bulk graphites. In the K-H direction, the LSs at a fixed $B_{0}$
exhibit a parabolic dispersion with two band-edge states at the two
edges of the 1st BZ, i.e., K ($k_{z}=0$) and H ($k_{z}=\pi/I_{z}$),
as shown for $B_{0}=40$ T in Fig. 7 (a).
There are no crossings and anticrossings of LSs, directly reflecting
the monotonous dependence of energy bands on wave vectors (the black
curve in Fig. 2 (a)). In particular, the $k_{z}$-dependent
dispersion of the $n^{c,v}=0$ LS is consistent with that of the
Dirac points along the K-H direction in the absence of external
fields, i.e., ${E^{c,v}(n^{c,v}=0,k_{z})=E_{D}(k_{z})}$. A slice of
the LS spectrum with respect to a specific $k_{z}$ can be regarded
as a combination of massless-Dirac LLs with the zeroth LL given by
$E_{D}(k_{z})$. Furthermore, according to Eq. (41), the energy width
of a LS corresponds to the energy difference between the Dirac
points at the zone edges, K and H: that is,
$E_{D}(k_z=0)-E_{D}(k_z=\pi/I_{z})=(2\alpha_{1}+2\alpha_{2})-(2\alpha_{1}\cos(\pi)+2\alpha_{2}\cos(2\pi))=4\alpha_{1}\simeq1.444$
eV. It should be noticed that simple hexagonal graphite retains the
semi-metallic characteristics in the presence of a magnetic field,
implying that free carrier pockets near the K-H edge might cause the
optical transitions between two valence or conduction LSs (intraband
excitations).

On the other hand, with the variation of $B_{0}$, the
field-dependent energy spectrum displays a form similar to that of
monolayer graphene,
${E^{c,v}(n^{c,v},k_z)-E_D(k_z)}\propto\sqrt{n^{c,v}B_{0}}$, while
the proportional constant ($\propto$ the Fermi velocity) is weakly
dependent on $k_{z}$, as shown in Fig. 7 (b). In the low-energy
approximation, the analytic solution of LS energies is derived by
introducing the quantization condition to the Dirac cone of
graphites as in Eq. (41) to have
\begin{equation}
\begin{gathered}
E^{c,v}(n^{c,v},k_z)\approx E_{D}(k_z)\pm\hbar
v_{F}\sqrt{2eB_{0}n^{c,v}/\hbar}.
\end{gathered}
\end{equation}
By the detailed calculations, the four atomic interactions,
$\alpha_{0},\alpha_{1},\alpha_{2}$ and $\alpha_{3}$, can be
expressed in terms of the low-lying LS energies at the K and H
points as follows:
\begin{equation}
\begin{gathered}
\alpha_{0}=\frac{l_{B}}{3\sqrt{2}b}\{E^{c}(n^c=1,k_z=0)-E^{c,v}(n^{c,v}=0,k_z=0)\\
+E^{c}(n^c=1,k_z=\pi/I_{z})-E^{c,v}(n^{c,v}=0,k_z=\pi/I_{z})\},\\
\alpha_{1}=\frac{1}{4}\{E^{c,v}(n^{c,v}=0,k_z=0)-E^{c,v}(n^{c,v}=0,k_z=\pi/I_{z})\},\\
\alpha_{2}=\frac{1}{4}\{E^{c,v}(n^{c,v}=0,k_z=0)+E^{c,v}(n^{c,v}=0,k_z=\pi/I_{z})\};\\
\alpha_{3}=\frac{l_{B}}{6\sqrt{2}b}\{E^{c,v}(n^c=1,k_z=0)-E^{c,v}(n^{c,v}=0,k_z=0)\\
-E^{c}(n^c=1,k_z=\pi/I_{z})+E^{c,v}(n^{c,v}=0,k_z=\pi/I_{z})\}.
\end{gathered}
\end{equation}
${l_B=\sqrt{\hbar c/eB_{0}}}$ is the magneto length related to the
effective localization range of LS. Equation (44) means that the
atomic interactions can be determined by the STS and magneto-optical
measurements on the LS energies. Based on the band structure, 3D
graphite is expected to display the massless Dirac-like
magneto-optical properties. However, as a result of the strongly
dispersed LSs across the Fermi level, the greatly enhanced free
carrier pockets near the edges of the first Brillouin zone is
responsible for the spectral features that are considerably differ
from the essential quantum phenomena in 2D graphenes.

The magnetically quantized DOS has a lot of special structures,
depending on 1D LSs or 0D LLs. Simple hexagonal graphite exhibits
many peaks in the square-root form arising from the quantized LSs
with the 1D parabolic dispersions, as shown in Fig. 8(a).
Each LS contributes two asymmetric peaks corresponding to the
band-edge states at the K and H points. For example, at $B_0=40$ T,
the $n^{c,v}=0$ ($n^{c}=1$) LS has two peaks at $E=0.272\alpha_{0}$ and
$-0.272\alpha_{0}$ ($E=0.35\alpha_{0}$ and $-0.195\alpha_{0}$), as indicated by the red (blue)
arrows. On the other hand, few-layer graphenes present a plenty of
delta-function-like symmetric peaks due to the dispersionless LLs,
e.g., monolayer, bilayer and trilayer systems in Figs. 8(b)-8(c),
respectively. The initial peak of the zeroth mode (the red arrows)
corresponds to the Dirac point (Figs. 2(b)-2(d)).

STS is an efficient method in examining energy spectra of
condensed-matter systems. The tunneling differential conductance
(dI/dV) is approximately proportional to DOS and directly presents
the main features in DOS. The STS measurements have been
successfully utilized to identify the diverse electronic properties
in graphene-related systems with the  $sp^2$ bondings, such as,
few-layer graphenes,\cite{PRL106;126802,NatPhys6;109,PRB91;155428,
PRB91;035410,PRB77;155426,PRB87;165102,APL107;263101, ACSNano9;5432}
Bernal graphite,\cite{ASS151;251,PRL102;176804} graphene
nanoribbons,\cite{PNAS110;11256,PRB77;075422,PRB77;205421} and
carbon nanotubes.\cite{Nature391;59,Nature391;62} Specifically, two
low-lying DOS characteristics, a linear $E$-dependence vanishing at
the Dirac point and a ${\sqrt B_0}$-form LL energy spacing, are
confirmed for monolayer graphene.\cite{PRL106;126802,NatPhys6;109,Science324;924,PRB91;115405} A sufficient-wide plateau
and a lot of square-root LS peaks in AA-stacked graphite require
further experimental verifications. The STS measurements on them are
useful in the identifications of the intralayer and interlayer
atomic interactions.

\subsection{Magneto-optical properties}

The AA-stacked graphite exhibits the unique magneto-optical
properties, since the 1D LSs have the sufficiently wide band widths
and the specific energy dispersions. The intraband and the interband
inter-LS vertical excitations appear in the low-frequency absorption
spectra, as clearly indicated in Figs. 9(a)-9(c).\cite{Carbon54;248}
The former originate from the valence and conduction LSs across the
Fermi level. Only the occupied $n^v$ ($n^c$) LS to the unoccupied
${n^v-1}$ (${n^c+1}$) one is the effective excitation channel; that
is, the optical excitations between the well-behaved Landau
wavefunctions need to satisfy the selection rule of
${\Delta\,n=\pm\,1}$. The intraband absorption peaks are denoted as
${\omega^{vv}_{nn-1}}$ and ${\omega^{cc}_{nn+1}}$ (Fig. 9(a)). They
are closely related to the $k_z$-dependent Fermi-momentum state of
each LS ($k_F^{n^{c,v}}$ in Figs. 10(a) and 10(c)).
For example, the ${\omega_{1615}^{vv}}$ peak comes from all the vertical excitations in the range of ${k_F^{16^v}\le\,k_z\le\,k_F^{15^v}}$ (Fig. 10(a)).
${\omega^{vv}_{n+1n}}$ is close to ${\omega^{cc}_{nn+1}}$, so their
absorption peaks are merged together, e.g., those for ${n^v\le\,15}$
and ${n^c\le\,14}$ at ${B_0=40}$ T (Fig. 9(a)). Such two-channel
peaks are observable for ${n^{c,v}\le\,6}$. As a result of the
smaller frequency differences, the other peaks become a broad and
prominent structure, i.e., they behave as a multi-channel threshold
peak. This composite structure is absent in the layered graphenes
and other graphites.

The interband absorption peaks come to exist in the frequency range
of ${\omega\,>} 0.039\alpha_{0}$, as clearly shown in Fig. 9(b). They
originate from the ${(n+1)^{v}\rightarrow\,n^c}$ and
${n^{v}\rightarrow\,(n+1)^c}$ vertical excitations, respectively,
corresponding to the allowed ranges in
${k_F^{(n+1)^v}\le\,k_z\le\,k_F^{n^c}}$ and
${k_F^{n^v}\le\,k_z\le\,k_F^{(n+1)^c}}$. ${k_F^{n^c}-k_F^{(n+1)^v}}$
is almost identical to ${k_F^{(n+1)^c}-k_F^{n^v}}$, and
corresponding excitation frequencies behave similarly (Fig. 10(b)).
Two kinds of interband channels can create nearly the same
absorption spectrum. The effective $k_z$-ranges are sufficient wide
except for very small quantum numbers, so that the distinct
curvature variations of the ${(n+1)^v}$ and ${n^c}$ LSs result in
two specific absorption frequencies due to the Fermi-momentum states
${k_F^{(n+1)^v}}$ and ${k_F^{n^c}}$. Furthermore, such ranges cover
the ${k_z=\pi\,/2I_z}$ state with the lowest DOS. These are
responsible for the existence of many double-peak structures in the
cusp form. Such double peaks have the non-uniform intensity, and
their widths grow with the increasing frequency because of the
enlarged range between two associated Fermi-momentum states.

The interband magneto-absorption spectra present the unique beating
phenomena, as clearly indicated in Figs. 11(a)-11(b).
The beating oscillations, which include several groups of diversified absorption
peaks, are very sensitive to the change of field strength. With the
increase of absorption frequency, the widened double-peak structures
might overlap each other or one another. The first group at lower
frequency is composed of the isolated double peaks. The second group
arises from a combination of two neighboring double peaks, and their
composite peak intensity is twice that of the original peaks.
Concerning the third group, three neighboring peaks are merged to a
single structure and its intensity is enhanced to almost three times
the pristine one. As a result, the spectral intensity is
proportional to the number of the combined double-peak structures.
The unusual association of absorption peaks directly reflects the
specific $k_z$-dependence of each LS parabolic dispersion (Eq. (43);
details in \cite{CPC189;60}). It should be noticed that this is the
first time to predict the beating phenomenon in optical properties.

The magneto-absorption peaks of few-layer graphenes and graphite,
with the exception of the optical selection rule
${\triangle\,n=\pm\,1}$, reveal very distinct features. For the
former, the dispersionless LLs create the delta-function-like
symmetric structures with a uniform intensity, as shown in Figs.
12(a).
The multi-channel threshold peak is absent. Only one
two-channel peak, belonging to the intraband absorption channel, is
present in bilayer and tri-layer AA stackings (red and blue curves
in Fig. 12(a)), in which it does not have a complete dispersion
relation with the $B_0$-field strength because of the variation of
the highest occupied LL (Fig. 14(b) and (d)).\cite{APL97;101905} All
the AA-stacked systems exhibit a plenty of interband absorption
peaks, but the main differences lie in the peak structures.
Monolayer system shows the isolated symmetric peaks, while bilayer
AA stacking displays the pair-peak structures. Furthermore, the
N-odd systems correspond to the superposition of the monolayer- and
bilayer-like absorption peaks. Some initial ${n^v\rightarrow\,n^c}$
excitations are forbidden in N=2 ${\&}$ 3 systems, reflecting the
Fermi-Dirac distribution of multi-Dirac ones. In addition, the
well-behaved beating oscillations are not presented in layered
graphenes.

The $B_0$-dependent absorption frequencies provide the important
information for the experimental verifications and in understanding
the effects due to  dimensions and stacking configurations. All peak
frequencies of AA-stacked graphite, as shown in Figs. 13(a) and
13(b), grows with an increasing field strength.
They present the
complete dispersion relations with $B_0$, in which the
field-strength dependence is roughly proportional to ${\sqrt B_0}$
except for the multi-channel peak (solid circles in Fig. 13 (a)).
The observable intraband excitations cover the multi-channel peak
and five two-channel peaks. The multi-channel threshold frequency
does not exhibit a ${\sqrt B_0}$-dependence, since the initial
intraband excitation channels dramatically change with field
strength. Concerning the interband excitations, there exist two
splitting absorption frequencies at sufficiently high magnetic
field. The critical field strength is reduced in the
higher-frequency absorption peaks. It is relatively easy to observe
the double-peak structures for large $\omega$ and ${B_0}$. On the
other side, the layered graphenes exhibit the unique intraband and
interband absorption frequencies, as clearly indicated in Figs.
14(a)-14(e). Monolayer graphene has a regular ${\sqrt
B_0}$-dependence at low magneto-absorption frequency (${\omega\,<1}$
eV in Fig. 14(a)). As to bilayer and trilayer AA stackings, they
show the discontinuous $B_0$-dependences in the two-channel
intraband peaks (Figs. 14(b) and 14(d)).
This mainly stems from the
fact that the highest occupied LL becomes the smaller-$n^{c,v}$ one
in the increase of $B_0$. Furthermore, their pair-peak interband
excitations cannot survive when both $n^v$ and ${(n+1)^c}$ are
occupied or unoccupied, i.e., more interband absorption peaks are
absent at low field strength (Figs. 14(c) and 14(e)). In addition,
the critical differences among three kinds of graphites will be
discussed in Chap. 6.

As for magneto-optical measurements, the infrared transmission
spectra have identified the ${\sqrt B_0}$-dependent absorption
frequencies of the interband LL transitions in mono- and multi- graphene.\cite{PRL98;197403,JAP117;112803,PRL101;267601,PRL100;087401} Furthermore, the magneto-Raman spectroscopy is utilized to observe the low-frequency LL excitation spectra for the AB-stacked graphenes up to 5
layers.\cite{NanoLett14;4548} The unique magneto-excitation spectra of simple
hexagonal graphite deserve thorough experimental examinations, such
as, the multi-channel threshold peak, the intraband two-channel
peaks, the interband double-peak structures, and the magneto-optical
beating phenomenon. Similar measurements could be done for
AA-stacked graphenes to verify the dimension-induced differences in
the channel, structure, number, frequency and intensity of
magneto-absorption peaks. Such comparisons are useful in
illustrating the diversified magnetic quantization of the multiple
Dirac-cone structures in the AA stacking systems.

\section{Bernal graphite}

Bernal graphite, with band profiles of monolayer and bilayer graphenes, is a critical bulk material for a detailed inspection of the massless and massive Dirac fermions. Theoretical and experimental researches show that the essential properties of graphite can be described by the quisiparticles at the high symmetry points of the Brillouin zone: massless Dirac fermions at the H point and massless ones at the K point. In particular, with the dimensional crossover from 3D to 2D, the many exciting properties of fewlayer graphenes originate from the interlayer couplings in bulk graphite. The optical excitation channels are only allowed between the respective monolayer-like subbands or between the bilayer-like subbands, regardless of external fields. The anticrossings of LLs/LSs and the electron-hole induced twin-peak structures are revealed in both 2D graphene and 3D graphite, while they are are more obvious in graphene with the increase of the layer number.

are crucially massive Dirac fermions dependent on understanding the interlayer coupling that originates in bulk graphite.

However, based on the interlayer atomic interactions of the dimensional crossover, the measured profiles of the B0-dependent peaks, e.g., threshold channels and peak intensity, spacing and frequency, can be used to distinguish the stacking layer, configuration and dimensionality.

\subsection{Electronic structures without external
fields}

The band structure of the Bernal graphite in the absence of external
fields are shown in Figs. 15 (a) and (b).
With a slight overlap of conduction and valence subbands, Bernal
graphite is classified as a semimetal due to the low-density free
carriers. The in-plane energy dispersions considerably depends on
the value of the momentum $k_{z}$, which contain the characteristics
of 2D monolayer and AB-stacked bilayer graphenes at certain special
$k_{z}$'s. In Eq. (11), $h=2\cos(k_{z}I_{z})$ indicates the factor
of the effective interlayer interactions in Bernal graphite. In the
HLA plane ($k_{z}=\pi/2I_{z}$ and $h=2\cos(\pi/2)=0$), the
Hamiltonian matrix can be reduced to a $2\times2$ matrix of
monolayer graphene, because the elements coupling by the
nearest-layer interactions are equal to zero and the on-site energy
$\gamma_{6}$ can be negligible. It is shown that the occupied valence
bands $E_{v}$ are symmetric to the unoccupied conduction bands
$E_{c}$ about $E_{F}=0$ (Fig. 15 (b)). The low-energy band structure displays a massless-Dirac-like linear dispersion with the Dirac point located near the H point, while the energy states are double degenerate.

The energy dispersions in the M$\Gamma$K plane show another graphene
properties. Substituting the condition $k_{z}=0$ and $h=2\cos(0)=2$
into the Hamiltonian matrix in Eq. (11), one gets a $4\times4$ bilayer-like Hamiltonian matrix, while the effective interlayer interactions are twice as large as those of bilayer graphene. The in-plane energy subbands are
asymmetric about the Fermi level due to the influence of the
interlayer atomic interactions, $\gamma_{2}$,...$\gamma_{6}$. In the
vicinity of the K point, the low-energy dispersions are
characterized by massive-Dirac quasi-particles. The coordinate of
the band-edge states are consistent with those of AB-stacked bilayer
graphene, i.e., at the M and K points. However, the effective
interlayer atomic interaction 2$\gamma_{1}$ gives rise to the double
band-edge state energies $\sim2\gamma_{1}$ at the K point as
compared to the bilayer graphene.

Along KH, the strongly anisotropic energy dispersions on $k_{z}$ are
mainly caused by the interlayer interactions. The cosine and the
flat dispersions along KH (Fig. 15(a)) are responsible for the two
types of atom chains; one is a straight chain of sublattices coupled
by $\gamma_{1}$ along $\widehat{z}$, and the other is a zigzag chain
of sublattices coupled by $\gamma_{4}$ in the $yz$-plane. In the
minimum model, the former and the latter are, respectively, described by
$E^{c,v}\simeq\gamma_{1}h=2\gamma_{1}\cos(k_{z}I_{z})$ and
$E^{c,v}\simeq0$.\cite{JPSJ40;761} When the state grows from K($k_{z}=0$) to
H($k_{z}=\pi/2I_{z}$), the two dispersions gradually get closer and become degenerate at the H point. Also, the in-plane dispersions are bilayer-like, while their behavior transforms into monolayer-like at the H point. That is to say, Beranl graphite exhibits both the massless and massive Dirac
fermions in the vicinity of the H and K points, respectively.
ARPES has been used to measure the 3D energy dispersions all around
the 1st BZ from the hole pocket at the H point to the electron
pocket at the K point. \cite{PRL100;037601,ASS354;229,PhyB407;827,PRB79;125438,NatPhy2;595}.
Both the massless and massive Dirac fermions are verified in terms
of the linear and parabolic dispersions, respectively. Furthermore,
the measured small hole pocket at the H point is in agreement with
the theoretical model and the quantum oscillation measurements.\cite{PRL93;166402} Remarkably, the Dirac quasi-particles are
responsible for special structures in the DOS and dominate the
optical excitations.

On the other hand, N-layer AB-stacked graphenes could exhibit
massless and massive Dirac fermions; the band structure resembles
bilayer case or a hybridization of monolayer and bilayer cases,
depending on whether the layer number is odd or even. The trilayer
graphene displays a hybridization of band structure by a monolayer
and a bilayer graphenes, while the even-layer graphene consist of
only pairs of bilayer-like parabolic subbands, as shown in Fig. 16. Near the K point, the intersection of low-energy subbands
indicates that AB-stacked graphenes are gapless 2D semimetals (the
insets of Figs. 16 (a) and (b)). With an increment of the graphene layer, the band structure in cases of even (odd) N consists of N (N-1) pairs of
bilayer-like parabolic bands, while it owns a particular pair of
monolayer-like linear bands near the Fermi level if N is odd.

The main characteristics of electronic structures, dominated by the
stacking configuration or the interlayer atomic interactions, are
directly reflected in the DOS. In Beranl graphite, the DOS mainly
originates from the bilayer-like and monolayer-like in-plane
dispersions, respectively, corresponding to the K- and H-point band-edge along the $k_{z}$ dispersions (Fig. 17).
The DOS VHSs marked by black, red and blue colors correspond to the band-edge and saddle-point states in Figs. 15 and 16.
A finite DOS at $E=0$ clearly indicates its semi-metallic
properties of Bernal graphite. Besides, the low-energy intensity smoothly grows with the frequencies, which can be regarded as a superposition of the linear and pearabolic dispersions. The former and the latter are, respectively, verified in DOS by the roughly linear and quadratic $B_{0}$-dependent tunneling energies.\cite{PRB87;165102} However, a shoulder spreads out near $\pm2\gamma_{1}$, attributed to the band-edge states of bilayer-like parabolic subbands. Such structure is reflected by a VHS at $\sim\pm\gamma_{1}$ in bilayer and trilayer graphenes.\cite{PRB77;155426} With the increasing energies,
the DOS exhibits two prominent asymmetric peaks at the middle energies of $E^{c,v}\simeq\pm\gamma_{0}$. This is a superposition of all saddle points distributed along M$\rightarrow$L during the band structure transformation from bilayer-like to monolayer-like. In contrast, the trilayer graphene display three prominent peaks: one comes from band-edge state of monolayer-like subband and two from those of bilayer-like ones. Some of the main features in DOS are verified by STS\cite{PRB77;155426,PRB87;165102} and the measured VHSs could lead to special structures in absorption spectra.

\subsection{Optical properties without external
fields}

The absorption is determined by the relationship between the
electronic structures (or DOS) and the optical excitation
transitions. In Beranl graphite, it demonstrates that $A(\omega)$ is identical for all polarization directions, $\widehat{E}$, on the graphene plane, indicating the isotropy of the frequency distribution of the absorption intensity over all frequencies. The optical responses due to
massless and massive Dirac fermions are, respectively, reflected by
the optical excitations channels in the vicinity of the H and K
points, as shown in Fig. 18 (a).
At low energies, one weak shoulder is revealed at $E\simeq2\gamma_{1}$ as a result of the excitations between the two low-energy parabolic
bands. Moreover, at middle energies, a single sharp peak is
accompanied by two shoulder on its both sides. They are responsible for the multi saddle-point channels of all the bilayer-like and monolayer-like band structures as $k_{z}$ moves from M to L.


Some of the features of the optical spectrum are consistent with the
experimental results. In the study by Obraztsov et al.\cite{APL98;091903}, they study the optical spectra of
polarized beam in cases of the different polarizations on the
graphene plane and the stacking direction, which are, respectively,
indicated by p-polarized and s-polarized. All the cases show a
similar behavior between the spectral intensity and polarization of
the laser beam, while the photoresponse for p-polarized excitation
beam is relatively strong than for the s-polarized one because of
the relatively strong energy dispersions on the in-plane direction.
The results might reflect the relatively strong energy dispersions
for the in-plane direction than for the out-plane direction.

The optical response of the Dirac quasi-particles is also a dominant
contributor for 2D AB-stacked graphenes. It leads to two kinds of
special structures: discontinuities at low frequencies and
logarithmic divergences at middle frequencies. The former and the
latter, respectively, come from the vertical transitions around the
K point and those around the M point, as shown by Fig. 18 (b). For the
bilayer case, the absorption spectrum exhibits a single shoulder at
$\omega\simeq \gamma_{1}$ and four peaks at $\omega\simeq
2\gamma_{0}$. Infrared spectroscopy have shown a clear picture for the
low-energy excitations around the K point.\cite{PRB78;235408,PRB79;115441} Besides, the spectral intensity grows with the higher frequency, until in the middle-frequency spectrum, four logarithmic saddle-point peaks spread around $\omega\simeq 2\gamma_{0}$. On the other hand, the excitation
channels of the trilayer graphene are only allowed between the
respective monolayer-like subbands or between the bilayer-like
subbands. Infrared spectroscopy has verified the absorption
spectrum, which is a combination of a monolayer and a bilayer
graphehes. The aforementioned results indicate that optical
spectroscopies can be used to verify the AB stacking domains on the
surface or the bulk domains of graphite.\cite{ACSNano9;6765}

\subsection{Magnetic quantization}

\subsubsection{Landau subbands and wave functions}

The $k_{z}$-dispersed LSs are depicted from the zone boundary point
K to H for $B_{0}=40$ T (Fig. 19 (a)).
According to the zero-field band structure, the monolayer-like and
bilayer-like signatures are deduced to coexist in Bernal bulk
graphite. A series of subenvelope functions distributed among the
four constituent sublattices are illustrated in Figs. 19 (b) and (c)
for the Landau states at K and H points. The LSs can be classified
into two groups (blue and red) according to the characteristics of
the energy dispersions and the subenvelope functions. In the
vicinity of the K point, the two groups are attributed to the
magnetic quantization of the respective parabolic subband (blue and
red in Fig. 15). In the 1st BZ, the onset LS energies are consistent
with the cosine $E^{c,v}=2\gamma_{1}\cos(k_{z}I_{z})$ and flat
$E^{c,v}=\sim0$ dispersions along KH. As $k_{z}$ changes from K to
H, the two groups merge to a series of double degenerate
monolayer-like Landau states reflecting the zero interlayer atomic
interactions. However, the splitting of the lowest $n^{c,v}=0$ LSs
directly reflect the non-equivalent on-site energies of A and B
sublatices. In general, away from the H point, the lift of
degeneracy can be mainly attributed to the interlayer atomic
interactions $\gamma_{1}$, $\gamma_{3}$ and
$\gamma_{4}$.\cite{PRB83;121201}
The energy spacings of LSs are determined by the curvatures of the
parabolic subbands.
Near the K point, electron-hole asymmetry of LSs is presented under the influence of the interlayer atomic interactions, while it becomes symmetric for the monolayer-like LSs at the H point.
Moreover, the K and H points correspond to band edges of LSs, where the Dirac quasi-particles are the dominant contributor in the magneto-optical
properties. The effective mass model (only considering $\gamma_{0}$
and $\gamma_{1}$) can obtain qualitatively consistent calculations
for the first few LSs. However, it misses the feature of the
electron-hole asymmetry, which may be hardly observable in
STS but has been validated to be significant in magneto-reflectence/absorption,\cite{PRB84;153405,PRB80;161410,PRB85;245410} and magneto-Raman measurements.\cite{PRB84;235138,PRB85;121403}

At the K point, the first group of LSs appears at $E^{c,v}\simeq0$,
and the second group begin at $E^{c,v}\simeq2\gamma_{1}$, where the
subenvelope functions are associated in the way similar to the case
of bilayer graphene. Accordingly, the numbers of the zero points of
$B_{1}$ and $A_{1}$ are employed to define quantum numbers of the
first- and second-group LSs, $n_{1}$ and $n_{2}$, respectively. That
is, the relationship of the two groups is
$A_{1}:A_{2}:B_{1}:B_{2}=n-1:n-2:n:n-1$ for $n_{1}=n\geq2$ and
$n:n-1:n+1:n$ for $n_{2}=n\geq1$. As $k_{z}$ moves to H, with the
increasing energy, the first group ascends while the second group
descends according to the cosine dispersion $2\gamma_{1}\cos(k_{z}I_{z})$.
Furthermore, the H-point subenvelope functions behave monolayer-like relationship, as depicted by the wave functions in Fig. 19 (c).


The profile of monolayer-like (bilayer-like) LSs in the H (K) point
can be clearly seen from their energy evolution with the variation
of the field strength, as shown in Fig. 20.
Notably, the former is linearly dependent on $B_{0}$ and the latter
exhibits a square-root $B_{0}$ dependence, as a consequence of
massless and massive Dirac quasi-particles. It is shown in the
K-point energy evolution that as compared to bilayer graphene, the
energy spacings are significantly reduced and the onset energy of
the second group of LSs is increased to twice the value of
$\gamma_{1}$. This can be simply explained by the minimum model in
which the in-plane dispersion at the K point is coupled with an vertical
effective hopping energy of 2$\gamma_{1}$ between two neighboring
layers. On the other hand, the monolayer-like energy evolution at H
point comes from the fact that the state energies are correlated
with only the nearest-neighbor in-plane hopping $\gamma_{0}$.
However, the splitting of the lowest $n^{c,v}=0$ LSs is revealed only in the case of the non-equivalent on-site energies for A and B sublatices.

The DOS also reveals the prominent peaks with both the
monolayer-like and bilayer-like signatures, which respectively
originate from the vicinity of local extreme values of LSs at the H
and K points, in Fig. 21 (a).
This implies that the essential properties in Bernal graphite can be
regarded as a combination of monolayer and bilayer graphenes, as
well as display the linear and square-root dependence on $B_{0}$. In
STS measurements,\cite{NatPhys3;623} the bilayer-like spectral features are relatively dominant and the valance DOS peaks are stronger than the conduction ones because the corresponding band curvature is relatively small. Furthermore, the energy dependences of the monolayer-like and bilayer-like LLs are also observed in the tunneling spectra of decoupled monolayer, bilayer and trilayer graphenes.\cite{PRB91;115405} It should be noticed that the reduced peak spacings, Fermi velocity, effective mass and onset energy of the second group $2\gamma_{1}$ are important features to distinguish the bulk graphite from a bilayer graphene. While the graphene properties have been verified with STS in bulk graphites, there are still unsolved issuers for the tiny peaks resulting from the bulk properties of LSs between the K and H points.

\subsubsection{Anticrossings of Landau subbands}

The evolution of the LSs from K to H is responsible for the magnetic
quantization for the transformation of the subbands from parabolic
dispersion to linear dispersion, as shown in Fig. 22.
As a result of the anticrossing of LSs, there are some tiny peaks
appearing in couples between monolayer-like and bilayer-like LSs, as
indicated by green arrows in Fig. 21(a). Such tiny peaks come
from the band extrema of the reversed LSs around $k_{z}=0.8$, as shown in
Fig. 22 (a). In the paper,\cite{JPSJ40;761} Nakao had explained the
anticrossing phenomenon. They applied the perturbation method to
calculate the lift of degeneracy and led to the conclusion that
these level anticrossings are due to the trigonal warping effect of
$\gamma_{3}$ in the LS spectrum. Also, the event is deduced to
appear at the crossover of two LSs that satisfy the condition,
$n_{1}-n_{2}=3I+1$, where I is an integer, that is; the two LSs have
the same quantum mode for a certain kind of subenvelope function.
The opening energy of the anticrossing LSs is more obvious in the
stronger field. This phenomenon is also predicted in the LL spectra
of AB- and ABC-stacked few-layer graphenes.\cite{PCCP17;26008,PRB90;205434,PRB83;165443} In addition,
$\gamma_{2}$ and $\gamma_{5}$ also induce band edges for the first
few valence subbands near the Fermi level.\cite{PRB83;121201}
These VHSs between K-H might be a cause for the unresolved DOS peaks
in previous works.\cite{PRL94;226403,NatPhys3;623}


In the anticrossing region, the mixture of the LLs caused by the
trigonal parameter $\gamma_{3}$ can be realized by the evolution of
the subenvelope functions of the LSs. The weight of the amplitude
with respect to each sublattice has a significant hybridization
alone $\widehat{k_{z}}$ for the $n^{c}_{1}=5$ LS that couples with
$n^{c}_{2}=1$ LS, as shown in Figs. 22 (f)-(i). The behavior is distinct from the unhybridized LS,
e.g., $n^{c}_{1}=2$ LS in Figs. 22 (b)-(e). For
$n^{c}_{1}$=2 LS, as the state moves from K to H, its bilayer-like
subenvelope functions gradually transform into monolayer-like ones,
i.e., carrier distribution of two layers transfers into one of the
two layers. The quantum mode of the dominating sublattice $B^{1}$ transforms from two to one. On the other hand, the evidence of
the state hybridization of $n^{c}_{1}=5$ and $n^{c}_{2}=1$ LSs can be shown by the perturbed behavior of the subenvelope functions around the anticrossing center, $k_{z}\sim0.8(\pi/2I_{z})$. A such state is a multi-mode state, composed of the
main mode $n^{c}_{1}=5$ and the side mode $n^{c}_{2}=1$ LS, as
depicted in the dashed rectangle. Nevertheless, at the
H point, the wave function of the LS converts to the monolayer-like
$n^{c}=1$ LS as a result of the vanished perturbation. It should be noted
that the transition channels of the hybridized LSs might be too weak
to observe in optical spectroscopy measurements, but might be
observable in STS measurements.

\subsection{Magneto-optical properties}

Bernal graphite, with band profiles of monolayer and bilayer
graphenes, is a critical bulk material for a detailed inspection of
the massless and massive Dirac fermions. The recent surge in
interest in 2D graphenes is based on the properties of bulk
graphite. The monolayer-like and bilayer-like
absorption spectra are predicted to coexist in the bulk spectrum as
a result of the excitation channels between intragroup LSs near the
H point and between two intragroup and intergroup LSs near the K
point. Whether the optical transitions actually take
place is subject to the relationship of the initial- and final-state
subenvelope functions. According to $A(\omega)$, in Eqs. 38 and 39,
it is deduced to that they must have the quantum-mode difference by
one with regard to a same sublattice. The spectral profiles, such as
peak intensity, frequency and numbers, can thus be described by the
responses of the optical channels of the 1D LSs. Especially, near
the band edges of the 1D LSs at the K and H points, charge carriers
chiefly accumulate and predominate the optical excitations.


The 1D-LS channels result in the square-root divergent peaks in the
absorption spectra where two series of peaks, marked by red and black dots, correspond to the massive and massless Dirac fermions, respectively, as shown in Figs. 23.
The K-point associated peaks can mainly be classified into four
groups of peaks, resulting from two inter- and intra-group LS excitations.
However, the intergroup ones are obscured due to the broadening peak
width and relatively weak coupling of wavefunctions between the
initial state and final state, especially for the region at higher
frequencies. Except that the threshold peak comes from only one single channel, most of the K-point associated peaks come from pair channels, so called twin peaks. The splitting energy between the
pair is induced by the electron-hole asymmetry; it decreases with
increased frequencies or decreased field strengths.
In contrast, owing to the subband symmetry at the H point, the
degenerate channels give rise to only single-peak spectrum with
relatively symmetric divergence form. All the such peaks can be
precisely described by a monolayer graphene with only $\gamma_{0}$.
Accordingly, the absorption spectrum of Bernal graphite displays both bilayer-like twin-peak structure and monolayer-like single-peak structure.


Spectral intensity is mainly determined by two factors: the
$k_{z}$-dependent band-edge curvatures and the velocity matrix. It
turns out that the H-point and K-point excitations have almost the
same contribution to the magneto-absorption spectra, in spite of two
times more excitation numbers for the former, according to the DOS
shown in Fig. 2-4. The optical transition rate depends on the
expectation value of the velocity matrix, which can be divided into
several components of wavefunction products each with their own
respective integral hoppings $\gamma'$s. However, because
$\gamma_{0}$ is at least one order larger than the others, the
optical transition rate is simplified as an inner product of the
same-layer A and B subenvelope functions of the initial and final
states. Based on the $k_{z}$-evolution of the subenvelope functions
described in Chapter 4.3.1, the strong-strong combination at H is a
evidence of the comparable intensity of the Dirac quasi-particles in
the H- and K-related spectra.


There are some inconspicuous absorption peaks that come from the band-edge states of the anticrossing LSs, which satisfy the selection rules of modulo 3, instead of the principle selection rule that characterizes the
prominent K-point and H-point peaks, in Fig. 23. One can ascribes the specific selection rules to the hybridization of the anticrossing LSs, which is determined by interlayer atomic interactions $\gamma_{3}$ (Chapter 4.3.2). Therefore, the inconspicuous peaks have relatively weak intensities than the K-point and H-point peaks, not only because of the smaller DOS but also because of the smaller velocity matrix. These extra peaks are also
studied in graphene systems,\cite{PCCP17;15921,arXiv1603.02797} especially for the severely symmetry breaking structures, such as AAB-\cite{PCCP18;17597} and sliding bilayer graphenes.\cite{SciRep4;7509} This indicates that obtaining a comprehensive description of graphene and graphite systems requires the full atomic interactions of SWM model for its accurate calculations of the velocity matrix and characterization of the wave functions.

In graphenes, the spectral peaks are shaper and more distinguishable
than those in graphite due to the stronger Landau quantization
effect in 2D materials, as shown in Fig. 24.
The excitation channels in bilayer graphene resemble the K-related
ones in Bernal graphite, while the absorption peaks are
delta-function-like, reflecting the 0D dispersionless LLs as in the
2D systems. This is in contrast to the square-root-like divergent
form of the absorption peaks in 3D bulk systems as a result of the
1D dispersive LSs along out-of plane direction. Also, the absorption
spectrum presents twin-peak structures due to the splitting of dual
channels, with an smaller splitting energy of $\sim$10 meV and
half-reduced onset energy $\gamma_{1}$, as compared to the Bernal
graphite. In the case of trilayer AB-stacked graphene, the
magneto-absorption spectrum is regarded as a combination of a
monolayer-like and a bilayer-like spectra. The former exhibits a
predominantly uniform intensity with single-peak structures similar
to the H-point characteristics in graphite. However, the latter
displays half-intensity twin peaks corresponding to the splitting of
the dual channels in the electron-hole asymmetric LL spectrum. In
addition, the absence of intergroup excitations between bilayer-like
and monolayer-like LSs is due to the anti-symmetric phase relation
in the velocity matrix.


The optical channels at K and H points show two kinds of field
evolutions of the absorption frequencies, as shown in Fig. 25.
The former type behaves as a linear dependence of $B_{0}$ whereas
the latter is square-root like, as depicted by the black and red solid
curves in Fig. 25(a). The absorption frequencies related to H point
are identical to those of monolayer graphene and can well described
by a $\sqrt{nB_{0}}$ relationship. (Fig. 14(a)). However, compared
to the bilayer spectrum (Fig. 25(b)), the splitting in double peaks
is a bit enhanced for the K-point channels because of the
amplification of interlayer atomic interactions in graphite. In
short, both AB-stacked graphene and graphite exhibit massless and
massive Dirac-fermion properties in the optical absorption spectra
with or without magnetic fields. The induced spectral features, such
as the frequency dependence, peak width, divergent form and onset energy
$\gamma_{1}$, are the signatures that can be used to distinguish
between Bernal graphite and graphene. It should be noticed that the
twin peaks are sensitive to the magnetic field strength. However,
the observability in optical spectroscopy depends on how the
magnetic field compete with the experiment resolution and ambient
temperature. These main features are very useful information in
identifying the stacking configurations and dimensionality of
systems from experimental measurements.




Magneto-optical spectroscopy provides useful insight into the LSs in
graphite materials. The optical response of the Dirac
quasi-particles is a dominant contributor in the magneto-optical properties. Infrared
magneto-transmission studies mainly focus on detailed information
about the differences between graphites and multilayer epitaxial
graphenes.\cite{PRL100;136403,PRB80;161410,PRL102;166401,PRB83;073401,PRB85;245410,
PRB86;155409,JAP117;112803,PRL111;096802}
Reflect the bilayer-like parabolic dispersion, A series of absorption peaks of energy scaled as linear-$B_{0}$ is present in the spectra and identified as contributions from massive Dirac fermions in the vicinity of the K point.\cite{PRL100;136403} By fitting the linear relationship and selection rule $\Delta n=\pm1$, the effective interlayer interaction $2\gamma_{1}$ at the K point is obtained in the framework of the minimal model. The deduced value is actually about double of that in bilayer graphene. However, reflecting the
inherent complexity of the SWM model, there is an evidence of the
splittings of channels at the K point, attributed to electron-hole
asymmetry. This is also observed in magneto reflection\cite{PRB84;153405}, magneto-absorption\cite{PRB80;161410,PRB85;245410} and
magneto-Raman experiments.\cite{PRB84;235138,PRB85;121403} The experimental results indicate that the full SWM model including additional interlayer atomic interactions well describes the electron-hole LS asymmetry.



The monolayer-like spectra are verified by a series of inter-LS
transitions with a characteristic magnetic field frequency
dependence $\omega\propto\sqrt{nB_{0}}$ at the H point.\cite{PRB85;245410,PRB86;155409,PRB83;073401,JAP117;112803,
PRL100;136403} The measured
dependence can be used to directly obtained the value of
$\gamma_{0}$ in graphite. The full SWM model provide a basic
interpretation in such a case to clarify the optical response of the
graphene layers and get a quantitative agreement between optical
experiments and theory. However, there is also an evidence of
splitting of the degenerate channels the H
point.\cite{PRB85;245410,PRB86;155409,PRB83;073401,JAP117;112803} It
is confirmed that the observed splitting is not associated to the
electron-hole asymmetry of the Dirac cone. The splitting of the
degenerate channels might be attributed to the on-site energy
difference between A and B sublattices, spin-orbital coupling,
anticrossing of LSs or parallel magnetic flux. Nevertheless, these
results require a more elaborated model and clearer experimental
evidences.
While the trigonal warping affects the anticrossings of the LSs in the
low-energy region, a new series of absorption peaks obeying new
selection rules might possibly be observed with the increased
hybridization of the coupled LSs.

Magneto-optical spectroscopies have also been used to study the
massless and massive Dirac fermions in 2D AB-stacked few-layer
graphenes. The electron-hole asymmetry is also reported on bilayer
graphene with cyclotron resonance\cite{PRL100;087403}, ARPES\cite{Science313;951} and infrared spectroscopy,\cite{PRL102;037403,PRB79;115441} which is mainly under the influence of $\gamma_{2}$ and the in-equivalent environments of the two sublattices. The infrared transmission spectrum of ultrathin graphene (3-5 layers) indicates that two series of $B_{0}$- and $\sqrt{B_{0}}$-dependent frequencies are observed for the low-lying inter-LL excitations of the massless and massive quasiparticles, respectively.\cite{PRL97;266405}
Magneto-Raman spectroscopy has also been used to probe the those
Dirac-like optical excitations in few-layer graphenes.\cite{PRL107;036807,NanoLett14;4548} However, further experiments of the higher excitation channels is needed for identifying the other energy deoendence of the higher LLs away from the Fermi level. The optical experiments can be used to determine the interlayer atomic interactions that dominate the electron-hole asymmetry and LL and LS dispersions. The linear and square-root energy relationships of Dirac-like magneto-channels can be found in few-layer graphenes and
graphite, while the differences between the interpreted values of
$\gamma$'s can distinguish the stacking layer, configuration and
dimensionality.



\section{Rhombohedral graphite}

\subsection{Electronic structures without external
fields}

In the hexagonal unit cell with $p_{3}$ symmetry,\cite{PCCP13;6036}
the energy dispersions of the rhombohedral graphite are depicted
along different symmetric directions, as shown in Figure 26.
The band structure consisting of three pairs of occupied valence and
unoccupied conduction subbands is highly anisotropic and asymmetric
about the Fermi level. The in-plane energy bands show linear or
parabolic dispersions, whereas they weakly depend on $k_{z}$. For
the KM$\Gamma$K plane at $k_{z}=0$, the K, M, and $\Gamma$ points
are the local maximum (minimum), saddle, and the maximum points,
respectively. They would induce large DOS and greatly affect optical
excitations. Near the K point, the first pair of subbands, crossing
across the Fermi level, exhibit linear dispersion without degeneracy.
The second pair are double-degenerate parabolic subbands; however,
the broken degeneracy along $\Gamma$K leads to three non-degenerate
parabolic subbands at middle energies $\sim \pm2B_{0}$. When the
plane is shifted from $k_{z}=0$ to $k_{z}=\pi/3I_{z}$, similar
in-plane dispersions are also revealed in the HLAH plane. However,
the pair of linear bands is located at the HA line instead of the HL
line. In particular, the energy subband along KH becomes a
three-fold degeneracy with a very weak $k_{z}$-dependent dispersion
(inset in Fig. 26). Moreover, small free-carrier pockets are formed
near K-H region, because the intersection of valence and conduction
linear subbands crosses $E_{F}=0.007$ with a very weak dispersion.

Remarkably, one pair of linear subbands always show up at the
$(k_{x},k_{y})$ plane, regardless of the value of $k_{z}$. This
means that the existence of Dirac cones in Rhombohedral graphite
with the Dirac points spirally distributing with $k_{z}$ about the
high symmetry line along K-H. Besides, the parabolic subbands are
attributed to zone folding,\cite{NJP15;053032} because the
Hamiltonian (Eqs. (22)-(24)) is built in the triple hexagonal unit
cell instead of in the primitive unit cell, which is a rhombohedron
with space group symmetry $R\overline{3}m$.

In the next section, a analytic solution for the 3D Dirac cones is
calculated in the primitive unit cell along the highly symmetric
points by using the effective-mass model with only $\gamma_{0}$ and
$\gamma_{1}$.\cite{Carbon7;425,PRB78;245416,PRB90;085312} The
trajectory of the Dirac-cone movement is a function of $k_{z}$.
Furthermore, the distortion and anisotopy of the Dirac structures
have also been studied separately under the influence of the other
interlayer atomic interactions.\cite{NJP15;053032}

\subsection{Anisotropic Dirac cone along a nodal spiral}

The primitive unit cell of a rhombohedral graphite is defined in
Fig. 27 (a), which is 1/3 of the volume of the hexagonal unit cell in
Fig. 26.
The three primitive unit vectors $a_{1}$, $a_{2}$, and $a_{3}$ are
related to the c-axis:
3$I_{z}\widehat{\mathbf{z}}=\sum_{i=1}^{3}\mathbf{a}_{i}$, where
$a_{1}=a_{2}=a_{3}$, and the angles between two primitive vectors
are the same. A rhombohedron with six identical faces is referred to
as the primitive cell in ABC-stacked graphite. The Dirac-type
dispersion is obtained under a continuum approximation for the
low-energy band structure in the vicinity of the H–K–H hexagonal
edges specified in Fig. 27 (b). With the the BZ edge served as a
reference line, the energy dispersion is described as a function of
the 3D wave-vector measured from the BZ edges.

Based on the two sublattices, a full tight-binding Hamiltonian is
represented by a $2\times2$ matrix
\begin{equation}
H_{ABC}=\left\{
\begin{array}{cc}
H_{1} & H_{2}\\
H_{2}^{\ast} & H_{1}\\
\end{array}%
\right\} \text{,}
\end{equation}%
where $H_{1}$ and $H_{2}$ take the form
\begin{eqnarray}
\begin{array}{l}
H_{1}=2v_{4}\hbar k \cos(\phi+k_{z})+2v_{5}\hbar k
\cos(\phi-2k_{z})\text{,}\\
H_{2}=-v_{0}\hbar k \exp(-i\phi)+\beta_{1}\exp(-ik_{z})+\beta_{2}\exp(-i2k_{z})\\
+v_{3}\hbar k \exp[i(\phi-k_{z})]+v_{5}^{\prime}\hbar
k\exp[i(\phi+2k_{z}) \text{.}
\end{array}
\end{eqnarray}
The perpendicular wave-vector components $k_{z}$ is scaled by $1/d$
and the variables $k$, $\phi$ and $v_{m}^{\prime}$ are defined as
follows: $k=\sqrt{k_{x}^{2}+k_{y}^{2}}$, $\phi=\arctan(k_{y}/k_{x}-
7\pi/6)$ and $v^{(\prime)}_{m}=3b|\beta^{(\prime)}_{m}|/2\hbar (m =
0, 3, 4, 5)$. This chiral Hamiltonian characterizes the inversion
symmetry, and its eigenvalues are calculated as
\begin{equation}
E=H_{1}\pm |H_{2}|\text{.}
\end{equation}%
In Eq. (45), the off-diagonal elements can be written as
$H_{2}=f(k_{x}, k_{y}, k_{z})+\beta_{1}\exp(ikz)$, where
$|f|\simeq[v_{0}^{2}+v_{3}^{2}-2v_{0}v_{3}\cos(2\phi-k_{z})]^{1/2}\hbar
k$ as $\beta_{2}$ and $\beta_{5}^{\prime}$ are neglected. This
indicates the same chirality in rhombohedral graphite and in, while
$\beta_{1}$ induces an offset energy from the hexagonal edge for the
Dirac point. On the other hand, the identical diagonal elements lead
to gapless Dirac cones, and their linearity in $k$ implies the
possibility of cone tilting. Moreover, the zone-folded parabolic
bands are absent in the primitive rhombohedral representation.

By ignoring $\beta_{2}$, $v_{4}^{\prime}$ and $v_{5}$, the
coordinate $(k_{D},\phi_{D})$ and the energy $E_{D}$ of the Dirac
point are, respectively, expressed as follows in the case of
$H_{2}=0$:
\begin{eqnarray}
k_{D}=\beta_{1}(v_{0}\hbar)^{-1}[1+(v_{3}/v_{0})\cos(3k_{z})]\\
\phi_{D}=-k_{z}+(v_{3}/v_{0})\sin(3k_{z})\text{,}
\end{eqnarray}
and
\begin{eqnarray}
E_{D}(k_{D}(k_{z}),\phi_{D}(k_{z}))=2\beta_{1}(v_{5}/v_{0}+v_{3}v_{4}/v_{0}^{2})\cos(3k_{z})\text{,}
\end{eqnarray}
up to first-order perturbation $\mathcal{O}(v_{3}/v_{0})$. In
particular, the Dirac point displays a spiral dispersion as a
function of $k_{z}$. In terms of polar coordinates $(q,\theta)$ and
the coordinate transformation
$q^{2}=k^{2}+k_{D}^{2}-2k_{D}k\cos(\phi-\phi_{D})$, the Dirac-type
energy dispersion in Eq. (47) can be simply expressed as
\begin{equation}
E(q,\theta,k_{z})=E_{D}\pm\epsilon(q,\theta,k_{z})\text{,}
\end{equation}
where
\begin{equation}
\epsilon(q,\theta,k_{z})=[v_{0}-v_{3}\cos(2\theta-k_{z})\pm2v_{4}
\cos(\theta+k_{z})]\hbar q
\end{equation}
describes the dispersion of the anisotropic Dirac cone, and the +
and - signs refer to the upper and lower half Dirac cones,
respectively.
By the minimal model with only $\beta_{0}$ and $\beta_{1}$, the
Dirac cone is identical to that of monolayer graphene with a Fermi
velocity $v_{0}$.\cite{PRB78;245416} The spiral of the Dirac point
is described as a function of $v_{0}$ in the case of
$v_{3}=v_{4}=0$, which lies on a cylindrical surface of radius
$\beta_{1}(v_{0}\hbar)^{-1}$ with the spiral angle $\phi_{D}$ in
sync with $-k_{z}$.
However, taking into account the interlayer hoppings $v_{3}$ and
$v_{4}$, we find that the spiral becomes non-cylindrical and
exhibits $k_{z}$-dependent anisotropy. The former and the latter,
respectively, keeps and reverses the sign under a phase shift
$\theta\rightarrow \theta+\pi$; they, respectively, cause a rotation
and tilt of the Dirac cones as a function of $k_{z}$. The previous
works\cite{NJP15;053032} have demonstrated the anisotropic tilt of
the Dirac cones that vary in orientation and shape with $k_{z}$
along the Dirac-point spiral. The Dirac-point spiral across $E_{F}$
indicates the semimetallic properties, while the band overlap
$\simeq10$ meV according to Eq. (50) is one (two) order of magnitude
smaller than that in Bernal graphite (simple hexagonal) graphite.
Near $k_{z}=0(\pm\pi/3)$, there is an electron (hole) pocket, and at
$k_{z}=\pm\pi/6$, the free-carrier pockets shrink to the Dirac
point, i.e, the location of the Fermi level
$E_{F}=E_{D}(±\pi/6)=0$. This phenomenon is also demonstrated in
the hexagonal unit cell (Fig. 26). Similar analysis can be
performed for the H-K-H edge extension, around which the spiral
angle synchronizes with $k_{z}$ and the chirality is reversed.

\subsection{dimensional crossover}

The 3D characters of the electronic properties are significant for
the dimensional crossover from the 2D few-layer graphene to the 3D
bulk graphite.\cite{PRB93;075437} ABC-stacked configuration gives
rise to different lattice symmetries for the 2D and 3D systems. With
the periodic stacking of graphene sheets, the bulk graphite has a
biparticle lattice symmetry belonging to the space group
$R\overline{3}m$. Its primitive unit cell is a rhombohedron
containing two atoms, designated as A and B in Fig. 27 (a). In
contrast, the N-layer ABC-stacked graphene has two atoms on each
layer, i.e., total of 2$N$ atoms. Therefore, it is clear to figure
out the distinction of the energy bands between both systems. The
low-energy electronic properties in the bulk graphite is described
by the 3D anisotropic Dirac cones tilted relative to
$\widehat{k_{z}}$.
On the other hand, the few-layer case is characterized by one pair
of partially flat subbands at $E_{F}=0$, which is mainly contributed
by the surface-localized states.\cite{PCCP17;26008,PRB84;165404}
Such appearances of subbands, irrelevant to the bulk subbands,
indicate the dimensional crossover from graphite to few-layer
graphene.\cite{PRB93;075437} Nevertheless, the weakly
$k_{z}$-dependent dispersion across $E_{F}=0$ displays a
semi-metallic behavior for the bulk graphite, making it a candidate
system for the observation of 3D QHE.\cite{SSC138;118,PRB78;245416}
Besides, there are sombrero-shaped subbands near $\beta_{1}$, which
can be used as an interpretation of the dimensional crossover of the
3D case to the 2D limit.



The DOS is very useful for understanding the optical properties. Its
main characteristics are responsible for the Dirac cones that spiral
down as $k_{z}$ varies. In the low-energy region, the DOS intensity
increases nonlinearly with the increasing $\omega$, as shown in Fig. 28 (a). At $\omega=0$, a sharp valley is formed due to the fluctuation of Dirac-point energies within $E_{F}-$5 meV $\thicksim E_{F}-$5 meV. Furthermore, a cave structure consisting of a local maximum and a local minimum
is formed near $\omega\simeq$ (-) 0.09 $\beta_{0}$ for conduction
(valence) DOS.\cite{PCCP13;6036} The nonlinear dependence of
$D(\omega)$ is attributed to the deformation of the isoenergy
surface of the anisotropic Dirac cones, as indicated by the insert
of Fig. 28 (a). However, when $\omega$ exceeds the nonmonotonous
structure, $D(\omega)$ becomes linearly dependent on $\omega$ as a
consequence of the restoration of Dirac cones. In the middle-energy
region, the smoothly enhanced DOS is contributed by the parabolic
subbands, while the prominent peak near $E^{c,v}\simeq\pm\beta_{0}$ is a bit broader than that of monolayer graphene as a combination of a series of in-plane saddle-point states in the 3D $\mathbf{k}$ space.

For ABC-stacked graphene, the dimensional crossover of electronic
properties is revealed by 2D divergent structures in the
DOS.\cite{JPSJ76;024701} The DOS is nearly symmetric about the Fermi
level, as shown in Fig. 28 (b). At low energies, an evidence for the partial flat bands near the $E_{F}$ is revealed by a symmetric broadening peak in the STS measurements.\cite{ACSNano9;5432,PRB91;035410} Those away from $E_{F}$ originate from the sombrero-shaped and
parabolic subbands.
Consequently, the low-energy features in ABC-stacked graphenes are
dominated by the surface-localized states. Beside, the saddle-point
states at middle energies induce three symmetric peaks. These are
contrast to the valley and cusp DOS in the 3D bulk $\mathbf{k}$
space. The differences between ABC-stacked graphenes and graphites
are mainly caused by the different stacking symmetries and the
reduction of dimension, which would be reflected in the absorption
spectra $A(\omega)$.

\subsection{Optical properties without external fields}

Absorption spectrum of rhombohedral graphite reflects the Dirac-cone
energy dispersions, as shown in Fig. 29 (a). In general, the
low-frequency intensity increases approximately linearly with the
increasing $\omega$, as a result of the excitations within the Dirac
cones that spiral around K (K$^{\prime}$) corner, where the Dirac
points are fluctuated within a narrow range $E_{F}-5$ meV $\thicksim
E_{F}$+5 meV.\cite{SynMet162;800} However, the interlayer atomic
interactions distort the Dirac cones and slightly break the linear
dependence of the absorption intensity on the
frequency.\cite{PCCP13;6036} The small valley at $\sim0.2$
$\beta_{0}$ is associated with the transition between the caves in
the DOS (Fig. 28 (a)). As the energy dispersion transforms from
linear to parabolic ($\omega\gtrsim1.0$ $\beta_{0}$), the spectra
deviate from the linear dependence.
In the middle-energy region, the enhanced $k_{z}$ dependence of the
parabolic subbands induces a wider distribution of the spectral
structure. There are two separated peaks at $\omega\sim 1.95$
$\beta_{0}$ and $\omega\sim 2.0$ $\beta_{0}$ (blue circles); they
solely come from the excitations from the BZ edges M and L,
respectively. The second peak is higher than the first one because
of the higher JDOS and transition probability.

In a N-layer ABC-stacked graphene, the optical absorption spectrum
is richer than that of graphite in both the low- and
middle-frequency regions, as a result of the more N$\times$N kinds
of excitation channels.\cite{arXiv1603.02797,NatPhys7;944} For
example of the trilayer case, the vertical transitions near the K
point among different low-lying subbands give rise to feature-rich
structures at low energies, including asymmetric peaks and
shoulders, as shown in Fig. 29 (b). The middle-frequency channels
also lead to several obvious peaks associated with the saddle-point
states near the M point. It should be noted that the threshold peak,
due to the vertical transitions between the surface-localized and
sombrero-shaped subbands, is prominent at $\omega\sim\beta_{1}$ as a
dominance of the surface-localized states in the DOS. Optical
transmission spectroscopy has been used to verify the optical
excitations related to the surface-localized and sombrero-shaped
subbands.\cite{NatPhys7;944,PRL104;176404}. These experimental
evidences identify the dimensional crossover from ABC-stacked 3D
graphite to 2D graphene.

\subsection{Magneto-electronic properties}

\subsubsection{Tight-binding model}

The LS spectra in rhombohedral graphite are calculated by a
diagonalization scheme designed for the Peierls tight-binding
Hamiltonian in the representation of the triple hexagonal unit
cell.\cite{JPSJ81;024701} The main characteristics of the spectra
exhibit a definitely discernible 3D semi-metallic behavior, as
compared to that of Bernal graphite and simple hexagonal graphite
(small by one or two orders of magnitude), as shown in Fig. 30(a)
for $B_{0}=40$ T.
Due to the triple-size enlarged unit cell, the degeneracy of LSs is
deduced to be $3\times4$ for a single LS at a specific $k_{z}$ point
in the hexagonal 1st BZ. With the ABC-sacking sequence, its bulk
limit has the specific group symmetry $R\overline{3}m$, containing
two atoms in a primitive unit cell in comparison to 2N atoms for a
N-layer 2D case. This results in very distinct magneto-electronic
properties between both systems. Accordingly, the LSs (LLs) for the
cases of bulk graphite and the N-layer graphene are, respectively,
classified as one-group and N-group\cite{PCCP17;26008,PRB90;205434,JPSJ81;024701}.

The energy dispersions of LSs actually display a $k_{z}$ dependence
consistent with the behavior of the zero-field band structure, in
which the Dirac-point spiral dispersion in Eq. (50) corresponds to
the $n^{c,v}=0$ LSs.
The behavior of the subenvelope functions provides an evidence that
also suggests the magnetic quantization of Dirac
cones.\cite{SSC197;11} Their relationship between A and B sublattices
is independent on $k_{z}$, and the same as that obtained from a
comparative diagonalization for monolayer graphene ( e.g., at
$k_{z}=\pi/6$ $I_{z}$ in Figs. 30 (b)-(g)). The subenvelope function of
the $n^{c,v}$ LS consists of quantum modes $n^{c,v}$ and $n^{c,v}-1$
on sublattices B and A, respectively, where $n^{c,v}-1>0$. In
particular, the $n^{c,v}=0$ LSs are characterized by the same pseudo
spin polarizations on A sublattice. The reversed pseudo spin
polarization on B is held by part of degenerate states. Within the
first-order minimal model, the Dirac-point spirals can be
topologically stable by the chiral symmetry, meaning that the
interlayer atomic interactions are not obvious in the
diagonalization results.\cite{JETP93;59}


Remarkably, in the minimal model, as a consequence of the magnetic
quantization on the Dirac cones, the LS spectrum is dispersionless as function of $k_{z}$, as
shown by the red curve in Fig. 30 (c). However, the effect of full
interlayer interactions reflects the dimensional crossover for the
Dirac cone and LS spectra in the rhombohedral graphite. While the
properties of Dirac cones are preserved during the variation of
$B_{0}$, these discernible 3D characteristics, deviated from
$\sqrt{n^{c,v}B_{0}}$ dependence, are presented by $\beta_{3}$ and
$\beta_{4}$ in terms of the tilt and distortion of the Dirac cones.
In the next section, Onsager quantization method is used to give
analytic energy solutions, and leads to the identification of the
effects of the critical atomic interactions on the LSs along a nodal
spiral.

\subsubsection{Onsager quantization}

The Onsager quantization rule is used to obtain the quantized
energies for the Landau states of an isoenergetic surface along the
Dirac point spiral.\cite{NJP15;053032} According to the energy
dispersion in Eq. (51), the area $S(\epsilon,k_{z})$ enclosed by the
contour of energy $\epsilon$ is calculated from
\begin{eqnarray}
S(\epsilon,k_{z})=\int_{0}^{2\pi}d\theta \int_{0}^{Q(\epsilon)}
\epsilon(q,\theta,k_{z}) q\cdot dq \text{.}
\end{eqnarray}
Using Eq. (51) and trigonometric substitutions, the integration is
approximated as
\begin{eqnarray}
S(\epsilon,k_{z})\simeq\pi\epsilon^{2}/v_{0}^{2}\hbar^{2}[1-(v_{3}/v_{0})^{2}]^{-3/2}\{1+6(v_{4}/v_{0})^{2}[1+2(v_{3}/v_{0})\cos(3k_{z})]\}\text{.}
\end{eqnarray}
The Onsager quantization condition is given by
\begin{eqnarray}
S(\epsilon,k_{z})=2\pi e B_{0} \hbar n^{c,v} \text{,}
\end{eqnarray}
where the zero phase shift results from the same electron chirality
and Berry phase as in monolayer graphene, regardless of the
anisotropy of the Dirac cones. By neglecting interlayer atomic
interactions, the quantized Landau energies are obtained for an
isolated Dirac cone. i.e.,
\begin{eqnarray}
E^{(0)c,v}(n^{c,v})=\pm \hbar
v_{0}\sqrt{2eB_{0}n^{c,v}/\hbar}\text{,}
\end{eqnarray}
which turns into
\begin{eqnarray}
E^{(0)c,v}(n^{c,v})F(k_{z})
\end{eqnarray}
for the tilted Dirac cone in rhombohedral graphite, where
$F(k_{z})=[1-(v_{3}/v_{0})^{2}]^{3/4}\times\{1-3(v_{4}/v_{0})^{2}[1+2(v_{3}/v_{0})\cos(3k_{z})]\}$
is used to describe the dispersion factor as a consequence of the
variation of the enclosed area. The energies of LSs is then obtained
by superimposing the dispersions, Eq. (57), on the Dirac point
$E_{D}$, Eq. (50), i.e.,
\begin{eqnarray}
E^{c,v}(n^{c,v},k_{z})=E_{D}+E^{(0)c,v}(n^{c,v})F(k_{z})
\end{eqnarray}
According to the lowest LS in Eq. (58),
$E^{c,v}(n^{c,v}=0,k_{z})=E_{D}\propto \cos(3k_{z})$, the Fermi
level is determined at the point $k_{z}=\pi/6$, as the field
strength allows the formation of LS bulk gaps, i. e., $B_{0}\geq
[E_{D}(k_{z}=0)-E_{D}(k_{z}=\pi/3)]^{2}/v_{0}^{2}e\hbar\simeq 0.11$
T. Therefore, the renormalized Fermi velocity is given by
\begin{eqnarray}
v_{F}(k_{z}=\pi/6)=v_{0}[1-(v_{3}/v_{0})^{2}]^{3/4}[1-3(v_{4}/v_{0})^{2}]^{1/2}.
\end{eqnarray}

At $B_{0}=40$ T, the energy dispersions of LSs are plotted from
$k_{z}=0\sim k_{z}=\pi/3I_{z}$, as shown in Fig. 30. Also plotted
are the calculations from the numerical diagonalization and the
minimal model.
In the minimal model, the LS dispersion and the Fermi velocity are
obtained by keeping only $v_{0}$ and $v_{1}$ in Eqs. (58) and (59).
The calculated LSs are dispersionless in the 3D momentum space and
the Fermi velocity is $v_{F}=v_{0}$, as a result of the identical
isotropic Dirac cones along a dispersionless Dirac-point spiral.
However, the full tight-binding model brings about characteristics
beyond the minimal model. Within a Dirac cone, the quantized Landau
energies are symmetric about its Dirac point, based on Eq. (56),
while the spiral localization of the Dirac points gives rise to the
electron-hole asymmetry, as shown in Figs. 3-4 (a) and (b). This is
interpreted as a consequence of due to $v_{3}$ and $v_{4}$, causing
a narrow excitation energy range for a single channel,
$n^{v}\rightarrow n^{c}$, which might be observed in optical
experiments.
Low-lying LSs are weakly dispersive in contrast to both AA- and
AB-stacked graphites. In particular, only the $n^{c,v}=0$ LS moves
across the Fermi level. The LS spacings are also reduced due to the
renormalization of the Fermi velocity. Moreover, the bulk gap can be
closed for higher LSs ($n^{c,v}\gtrsim25$), because of the enhanced
$k_{z}$ dispersion. The LSs calculated from Onsager quantization are
consistent with those from the numerical diagonalization for
$n=0,1,2,3,4$. The agreement of the magnitude holds up to
$n^{c,v}=18\simeq\pm730$ meV. However, the inconsistency coming from
the continuum approximation is more apparent for higher $n^{c,v}$.

The magnetically quantized DOS of rhombohedral graphite is plotted
in Fig. 32 (a). In the framework of the minimal model, the DOS
peaks of rhombohedral graphite are identically reduced to 2D
delta-function-like peaks of monolayer graphene, because the Dirac
cones are isotropic and circularly distributed at $E_{F}=0$. In
particular, all the peaks are equal-intensity and symmetric about
the Dirac points at $E_{F}=0$.
Under the influence of $\beta_{3}$ and $\beta_{4}$, the isotropic
Dirac cones become tilted, anisotropic and spiral near the edge of
BZ. The equal-intensity peaks transform into nonequal-intensity
double peaks whose widths are determined by the $k_{z}$
dispersions of the 1D LSs. Each peak has two square-root divergent
forms corresponding to the band-edge energies at the K and H points, (green and blue dots).
Dirac-point spiral causes the particle-hole asymmetry, which
furthermore destroys the anti-symmetric dispersions of LSs in the
BZ, leading to different DOS intensities for valence and conduction
LSs.

On the contrary, the DOS of ABC-stacked trilayer graphene exhibits
three groups of symmetric delta-function-like peaks, which are not
regularly sequenced according to the dispersionless LLs of three
different subbans, as shown in Fig. 32 (b). Different
characteristics of the three groups of LLs are clearly shown. The
first peak at the Fermi level is composed of three surface-localized
LLs; therefore, its intensity is approximately estimated to be three
times than other peaks. Furthermore, peaks are densely formed for
$\omega\gtrsim \beta_{1}$, which approaches to the crossover of the
onset energies for the second and third groups. While the low-lying
peaks in ABC-stacked fewlayer graphenes have been confirmed by STS,
the essential differences between ABC-stacked graphites and
graphenes need to be further verified.

\subsection{Magneto-optical properties}

The spiral Dirac cones in rhombohedral graphite contribute to a
one-dimensional magneto-optical structure, different from monolayer
graphene (minimal model) as a result of the tilted anisotropic Dirac
cone and the renormalization of the Fermi velocity, as shown in Fig.
33 (a) for $B_{0}=40$ T.
Across the Dirac points, the interband optical transitions between
the LSs of $n^{c}$ and $n^{v}$ obey a specific selection rule,
$n^{c}-n^{v}=\pm1$, at a fixed $k_{z}$ for the tiled Dirac cones
along a nodal spiral.\cite{JPCM27;125602} Since $E_{D}(k_{z})$
breaks the anti-symmetric dispersions of $E^{c,v}(n^{c,v},k_{z})$
along the 1st BZ, the Fermi level is determined at
$E_{D}(k_{z}=\pi/6)$, exactly across the middle of the zero-mode
LSs; therefore, the range of $0\leq k_{z}< \pi/6$ is unoccupied and
that of $\pi/6\leq k_{z}\leq \pi/3$ is occupied. The spectral
intensity exhibits a variation during $K\rightarrow H$. For a single
channel $n^{v} \rightarrow n^{c}(=n^{v}\pm 1)$, the peak frequency
is calculated from analytic LS solutions in Eq. (58)
\begin{eqnarray} \omega^{vc}_{nn\pm 1}=(\sqrt{2B_{0}}\hbar
v_{F}/l_{B})(\sqrt{n}+\sqrt{n\pm 1})F(k_{z}=\pi/6)\text{,}
\end{eqnarray} accompanied with a frequency
distribution $(\sqrt{2B_{0}}\hbar v_{F}/l_{B})(\sqrt{n}+\sqrt{n\pm 1})
(F(k_{z}=0)-F(k_{z}=\pi/3))$.

Near the BZ edges, K and H, the vertical transitions have only tiny
energy difference and consequently merge to form single peaks.
Notably, within the minimal model, the spectra converts to
uniform-intensity single-peak structure because the LSs are even
irrelevant to $k_{z}$ and identical to those of monolayer graphene
as described by Eq. (58) with $E_{D}=0$ and $F(k_{z})=1$. However,
functions $E_{D}(k_{z})$ and $F(k_{z})$ vary as $k_{z}$ in terms of
other interlayer atomic interactions than $\beta_{1}$. Consequently,
under the perturbative $k_{z}$-dependent interlayer hoppings
$\beta_{3}$, $\beta_{4}$, etc., the deviation from the massless
Dirac-like Landau energies indicates the distortion of the isotropic
Dirac cones.


With the dimensional crossover from 3D to 2D, ABC-stacked graphene
exhibits magneto-optical properties sharply contrast to the bulk
graphite because of the different lattice
symmetries.\cite{PCCP17;15921,arXiv1603.02797} The interband
transitions among N groups of conduction and valence LLs contribute
to N$\times$N groups of absorption peaks, each of which displays
different $B_{0}$-dependence regarding the frequencies, intensities
and numbers, as shown in Fig. 33
(b).\cite{PCCP17;15921,arXiv1603.02797} In general, the intragroup
peaks are relatively stronger than the intergroup ones. The
absorption frequencies and intensities increase with the magnetic
field strength, and also obey the particular selection rule $\Delta
n=\pm 1$. However, the inter-LL excitations for the sombrero-shaped bands give rise to converted frequencies. This abnormal phenomenon is enhanced with the increase of the ABC-stacked graphene layers.
Recently, the inter-LL excitation for the partially flat bands and the lowest sombrero-shaped band have been verified by the magneto-Raman spectroscopy for a large ABC domain in graphene up to 15 layers.\cite{NanoLett16;3710}

However, their abnormal $B_{0}$-dependent properties are
exhibited by perturbed LLs in small anticrossing-LL regions, as
shown in the dashed green ellipse in Fig. 34 (b).
The corresponding
peak intensities and frequencies are discontinuous as a function of
$B_{0}$ and extra optical selection rules are induced for such LLs
with hybridized quantum modes. The larger the layer number N is, the
more complex the absorption spectrum will be. Furthermore, the
spectrum can't converge to 3D spiral-Dirac absorption spectrum due
to the absence of $\overline{P}3m$ symmetry in the 2D limit.

The identification of the interband transitions is available in
magneto-absorption, reflection and transmission measurements. The
energy width of double peaks, with a separation of $\sim10$ meV, can
be resolved; besides, it is feasible to probe the Fermi velocity
through the measurements of the cyclotron resonance by the
far-infrared spectroscopy. Moreover, near the Fermi level,
rhombohedral graphite has bulk gaps between the low-lying LSs in a
very weak magnetic field (estimated in Eq. (58)). The achievement of
3D QHE is attainable within the region of bulk gaps; the expected
QHE plateaus are different than the experiments reported in
ABC-stacked trilayer graphenes.\cite{NatPhys7;953} Furthermore, this is in
contrast to the cases of AA- and AB-bulk graphite which owns
unattainable field strengths that are required to open the bulk gaps
to observe a series of 3D QHE plateaus.

\section{Quantum confinement in carbon nanotubes and graphene nanoribbons
}

The periodical boundary condition in a cylindrical nanotube surface
can quantize the electronic states with the angular momenta (${J^{c,v}}$'s)
corresponding to the well-behaved standing waves. This leads to a
specific optical selection rule, when the electric polarization is
parallel to the
surface.\cite{Carbon42;3169,PRB62;13153,PRB51;7592,JPSJ62;2470,
PRB67;045405,PRB62;16092}
The cooperation with magnetic field greatly enriches the fundamental
properties. Magneto-electronic and optical spectra are very
sensitive to the changes in the magnitude and direction of magnetic
field and the nanotube geometry (radius and chiral angle). The ${\bf
B}$-field can create the metal-semiconductor transition, drastically
change the 1D energy dispersions, obviously destroy the state
degeneracy, and induce the coupling of different angular momenta. As
a result, there are more ${\bf B}$-dependent absorption peaks in the
square-root asymmetric form. Specifically, the magnetic
quantization, with high state degeneracy, is absent except for very
large radii and strong fields.

On the other hand, a finite-size graphene nanoribbon does not have a
transverse wave vector, owing to the open boundary condition. The
nanoribbon width, edge structures, and external field are
responsible for the unusual properties. It is predicted to present
the edge-dependent optical selection rules in terms of the subband
indices (${J^{c,v}}$'s).\cite{PRB95;155438,OptExpress19;23350,PRB84;085458} By the
detailed analyses, they principally come from the peculiar spatial
distributions of the edge-dominated standing waves. A perpendicular
magnetic field could result in QLLs, while the magnetic length is
longer than the nanoribbon width. Each QLL is composed of the
partially flat and parabolic dispersions. It dramatically changes
the main features of electronic and optical spectra. The
magneto-optical selection rule of QLLs is similar to that of
monolayer graphene. Graphene nanoribbons are quite different from
carbon nanotubes in quantum numbers, wave functions, energy gaps,
state degeneracies, selection rules, and magnetic-field effects,
clearly illustrating the critical roles of the boundary conditions.

\subsection{Magneto-electronic properties of carbon nanotubes
}
 Each carbon nanotube can be constructed from by rolling up a
graphene from the origin to the lattice vector ${{\bf R_x}=P{\bf
a_1}+Q{\bf a_2}}$, where ${\bf a_1}$ and ${\bf a_2}$ are primitive
lattice vectors of a 2D sheet (Fig. 35 (a)).
A ($P$,$Q$) nanotube has a chiral angle of ${\theta\,=tan^{-1}[\sqrt
3\,Q/(2P+Q)]}$ and radius of ${r=b\sqrt {3(P^2+PQ+Q^2)}/2\pi}$.
($P$,$P$) and ($P$,0), respectively, correspond to nonchiral
armchair and zigzag systems (${\theta\,=30^\circ}$ and $0^\circ$).
The number of carbon atoms in a primitive unit cell is
$N_{c}=4\sqrt{(P^{2}+PQ+Q^{2})(R^{2}+RS+S^{2})/3}$, where $(R,S)$
correspond to the primitive lattice vector along the nanotube axis.
%


The misorientation of 2$p_z$ orbitals (the curvature effect) on the
cylindrical surface leads to the change in the hopping integral,
i.e., $\gamma _{i}$=$V_{pp\pi }$ $\cos
$($\Phi_{i}$)+$4(V_{pp\pi}-V_{pp\sigma}$) [${r}/{b}$ $\sin
^{2}$(${\Phi_{i}}/{2}$)]$^{2}$, where $i$(=1, 2, 3) corresponds to
the three nearest neighbors, and $\Phi_{i}$
($\Phi_{1}=-b\cos(\pi/6-\theta)/r$,
$\Phi_{2}=b\cos(\pi/6+\theta)/r$; $\Phi_{3}=-b\cos(\pi/2-\theta)/r$)
represents the arc angle between two nearest-neighbor
atoms. The Slater-Koster parameters $V_{pp\pi}$$(=-2.66$ eV) and $%
V_{pp\sigma}$$(=6.38$ eV), respectively, represent the $\pi$ and
$\sigma$ bondings between two $2p$ orbitals in a graphene
plane.\cite{PRB57;15037} The Hamiltonian matrix will be built for
any magnet-field directions. $\alpha$ is the angle between magnetic
field and axis. As for a zero or parallel magnetic field, a
${2\times\,2}$ Hamiltonian, accompanied with ${J^{c,v}=1, 2, 3 ...;
N_c/2}$, is sufficient in calculating the essential
properties.\cite{PRB52;8423} On the other hand, the different J's
would couple one another as the magnetic field deviates from the
nanotube axis. The perpendicular component
$B_{\perp}=B_{0}=\sin\alpha$ leads to the J coupling and the total
carbon atoms in a primitive cell are included in the tight-binding
calculations (details in \cite{JPSJ81;064719,PRB67;045405}).

Each carbon nanotube has a lot of 1D energy subbands at zero field,
being defined by angular momenta of ${J^{c,v}}$'s, as shown in Fig. 36. For
a ${(P,P)}$ armchair nanotube, a pair of linear valence and
conduction bands, with ${J^{c,v}=P}$, intersects at ${E_F=0}$, e.g., those
of ${J^{c,v}=50}$ in (50,50) nanotube (gray-dotted curves in Fig. 36(a)).
They create a finite DOS there (Fig. 38(a)) and thus behave as a 1D
metal. That the metallic behavior is not affected by the curvature
effect could be identified from the periodical boundary condition
and the specific changes of the nearest-neighboring hopping
integrals.\cite{PRL78;1932} That is, the Fermi-momentum states in
armchair carbon nanotubes are sampled from the Dirac points of a
graphitic sheet. In addition, the misorientation of ${2p_z}$
orbitals only leads to a slight redshift in the Fermi momentum
($k_F$), compared with 2/3 (in unit of ${\pi\,/I_x}$; $I_x$ the
periodical distance along the nanotube axis). The higher/deeper
energy subbands are doubly degenerate, since they correspond to
${J^{c,v}}$ and ${N_c/2}-{J^{c,v}}$ simultaneously. They present
parabolic dispersions near the band-edge state of ${k_x=k_F}$.

The radius and chiral angle can determine energy gap and state
degeneracy. ${(P,Q)}$ carbon nanotubes, respectively, belong to
narrow- and moderate-gap semiconductors, being characterized by
${(2P+Q=3I}$ $\&$ ${P\neq\,Q)}$ and ${2P+Q\neq\,3I}$. The former and
the latter, with energy gaps inversely proportional to radius and
square of radius, arise from the periodical boundary condition and
the curvature effect, respectively.\cite{PRB52;8423,PRL78;1932} For
example, the narrow-gap ${(90,0)}$ and ${(80,20)}$ nanotubes and the
moderate-gap ${(91,0)}$ nanotube, respectively, have
${E_g\sim\,0.0005}$ $\alpha_0$  (Figs. 36(b) and 36(d)) and 0.03
$\alpha_0$ (Fig. 36(c)). In general, the low-lying valence
and conduction bands in semiconducting nanotubes possess the
parabolic dispersions with double degeneracy.

State degeneracy, band gap and energy dispersions strongly depend on
the direction and strength of a uniform magnetic field. When ${\bf
B}$ is parallel to the nanotube axis, the angular momentum becomes
${{J^{c,v}}+\phi\,/\phi_0}$ (magnetic flux ${\phi\,=\pi\,r^2B_0}$).
The gapless linear bands of armchair nanotubes are changed into the
separate parabolic bands (black-dotted curves in Fig. 36(a)), i.e.,
the metal-semiconductor transition occurs in the presence of $\phi$.
Furthermore, the magnetic flux destroys the double degeneracy in the
higher/deeper energy subbands; that is, two subbands characterized
by the angular momenta of ${J^{c,v}}$ and ${N_c/2-J^{c,v}}$ are not
identical to each other because of the magnetic splitting effect.
Such effects are relatively prominent, when the direction of
magnetic field approaches the nanotube axis. On the other hand, a
non-parallel magnetic field will result in the coupling of different
${J^{c,v}}$'s, being stronger at large $r$, ${B_0}$ and $\alpha$.
Specifically, a perpendicular magnetic field can break the state
degeneracy, in which the spitting subbands are symmetric about
${k_x=2/3}$ (green-dotted curves). An armchair nanotube keeps the
metallic bahavior at ${\alpha\,=\pi\,/2}$. Whether the intersecting
linear energy bands become the gapless parabolic ones depend on
radius and field strength. For example, (50,50) and (100,100)
nanotubes, respectively, exhibit the linear and parabolic
dispersions (green- and orange-dotted curves). Moreover, the ${\bf
B}$-induced effects on band structures are clearly revealed in
non-armchair carbon nanotubes (Figs. 36(b)-36(d)).

There exist the diverse relations between energy gap and magnetic
field. Under a parallel magnetic field, energy gaps present a linear
and non-monotonous relation with a period of ${\phi_0}$, as clearly
indicated in Fig. 37.
$E_g$ of an armchair nanotube grows in the increase of $\phi$,
reaches a maximum value at ${\phi\,=\phi_0\,/2}$, and then recovers
to zero at ${\phi_0}$ (black curve in Fig. 37(a)). The metal-semiconductor are also revealed in  the narrow- and middle-gap carbon nanotubes (black curves in Figs. 37(b) and 37(c)) near small $\phi$ and ${\phi_0\,/3}$ (or
$\phi_0$ $\&$ ${2\phi_0\,/3}$). In addition, they might be absent in part of narrow-gap systems (inset in Fig. 37(d)).
Obviously, the periodical Aharonov-Bohm effect
is presented in energy gaps of any carbon nanotube. It would be very
difficult to observe this effect in the presence of a dominating
perpendicular magnetic field, e.g., ${E_g}$'s at large ${\alpha}$'s.
There are no simple relations between energy gaps and $\phi$ under
the non-parallel fields. As for  the middle-gap nanotubes (Fig.
47(c)), the larger is $\alpha$, the higher is the $B_0$-field
strength of the semiconductor-metal transition. Specifically, at
${\alpha\,=90^\circ}$, both armchair  and narrow-gap nanotube keep
the gapless feature (green curves in Figs. 37(b) $\&$ 37(d)), i.e.,
the metal-semiconductor transition does not happen during the
variation of field strength. The predicted strong dependence of
energy gap  on the strength and direction of magnetic field and the
nanotube geometry could be further verified from the STS
measurements.

The special structures in DOS are greatly enriched by the nanotube
geometry and the magnetic-field direction, as shown in Fig. 38.
The 1D band-edge states in parabolic and linear dispersions,
respectively, create the square-root asymmetric peaks and plateaus.
All VHSs belong to the former except for that near ${E=0}$ in
metallic nanotubes at zero field (gray solid curve in Fig. 38(a)).
For an armchair nanotube, a parallel magnetic field makes the
plateau change into a pair of asymmetric peaks with an energy gap
(black curve), and the number of the intensity-reduced asymmetric
peaks at other energies becomes double. The energy spacing between
two neighboring peaks declines as ${\alpha}$ grows (red, blue and
green curves). Furthermore, they are merged together under a
perpendicular field (green (orange) curve), in which there is a
plateau structure (a symmetric peak) near ${E=0}$ because of a pair
of gapless linear (parabolic) bands (Fig. 36(a)). Compared with
metallic armchair nanotubes, the narrow-gap systems, as shown in
Figs. 38(b) and 38(d), present the almost same structures except for
a pair of very close asymmetric peaks  near the Fermi level at zero
field (insets). However, there are more peaks structures in the
moderate-gap nanotubes (Fig. 38(c)). On the experimental side, the
zero-field DOS characteristics, including the asymmetric peaks, the
energy spacings, the metallic
behaviors,\cite{Nature391;59,Nature391;62} and the middle and narrow
gaps,\cite{Science292;702} have been verified by the STS
measurements. The rich peak structures and the metal-semiconductor
transitions due to a uniform magnetic field require further
experimental verifications on DOS.

\subsection{ Magneto-optical spectra of carbon nanotubes}

The standing waves on a cylindrical surface plays a critical role in
determining the available optical excitation channels. At zero
field, they present the well-behaved spatial distributions, with the
forms of sine/cosine functions closely related to the angular
momenta, as revealed in Figs. 39(a)-39(c).
For an armchair nanotube, the linear energy bands possess a uniform
distribution along the azimuthal direction (gray curves in Fig.
39(a)), and the higher-energy bands correspond to the normal
oscillations of one, and two wavelengths (Figs. 39(b) and 39(c)).
Such features are independent of $k_x$’s and keep the same in the
presence of a uniform magnetic field, leading to the identical
optical selection rule at zero and parallel magnetic fields. In
addition, these two factors only create the rigid shifts in the
azimuthal distributions. As to the non-parallel magnetic fields, the
normal standing waves are changed into the distorted ones,
especially for a perpendicular one (green curves in Figs.
39(a)-39(c)). However, the latter could be regarded as a
superposition of the $J$-decoupled normal modes, since a
perpendicular field creates the coupling of distinct angular
momenta. It is relatively easy to observe the coupling effect  in
larger nanotubes. For example, an armchair (100,100) nanotube
exhibits the highly distorted standing waves with more zero points
(orange curves in Figs. 39(e) and 39(f)), corresponding to the
gapless parabolic energy bands near ${E_F}$ and the oscillatory ones
(Fig. 36(a)).

The electric polarization is assumed to lie on the nanotube surface
(parallel to the nanotube axis), as considered in layered graphites
(Eq. (38)). All the carbon nanotubes only exhibit the asymmetric
absorption peaks in the square-root form, as shown in Fig. 40.
The available excitation channels arise from the occupied valence
subbands and the unoccupied conduction ones, with the same angular
momentum. The selection rule of ${\Delta\,J=J^v-J^c=0}$ is determined by the
$J$-decoupled standing waves. It could be further applied to any
magnetic fields even with the coupling of $J$'s by using the
superposition of distinct components. At zero field (gray curve in
Fig. 40(a)), an armchair nanotube does not have the threshold
absorption spectrum, since the vanishing velocity matrix elements
(Eq. (39)) prevent the interband excitations due to a pair of
linearly intersecting energy bands. The featureless optical spectrum
is also revealed in monolayer graphene (Fig. 4(b)). The absorption
peaks are closely related to the band-edge states of the parabolic valence
and conduction bands; that is, absorption frequency is their energy
spacing. The weaker is the energy dispersion, the stronger is the
asymmetric peak.

The number, frequency and intensity of absorption peaks are very
sensitive to the magnetic field. The threshold absorption peak of an
armchair system is generated by a parallel magnetic field (triangle
related to the black curve in Fig. 40(a)). The number of other
absorption peaks becomes double (two rectangles) and their
intensities are getting weak, directly reflecting the splitting of
the $J$-dependent degeneracy (Fig. 36(a)). All the absorption peaks
agree with the ${\Delta\,J=0}$ rule. This rule is also suitable for
most of peak structures even when a magnetic field gradually
deviates from the nanotube axis. The first peak declines quickly and
two neighboring peaks approach to each other (red and blue curves),
in which the former disappears and the latter change into a single
peak under a perpendicular magnetic field (green curve). It should
be noticed that the absence of threshold peak at
${\alpha\,=\pi\,/2}$ is independent of the radius of armchair
nanotube and field strength. On the other hand, there exists an
extra low-frequency absorption peak (circle in red, blue and green
curves), mainly owing to the vertical excitations of the first
(second) valence band and the second (first) conduction band (Fig.
36(a)). This peak does not satisfy the selection rule of
${\Delta\,J=0}$, while it could be identified  from the coupling of
the neighboring angular momenta.

The main features of absorption peaks strongly depend on the
geometric structures of carbon nanotubes. The narrow-gap systems
exhibit a zero-field threshold peak at very low frequency (insets in
Figs. 40(b) and 40(d)). For a non-perpendicular $B_0$-field, they
might have the merged double-peak structures (arrows in Fig. 40(b)),
reflecting the small energy spacings between two neighboring
parabolic subbands with high DOSs (Fig. 37(b)). In addition, an
extra  prominent  magneto-absorption peak of ${\Delta\,J\neq\,0}$ is
revealed in all carbon nanotubes  at ${\alpha\neq\,0^\circ}$
(circles in Fig. 37) because of the coupling of distinct angular
momenta. As for the moderate-gap nanotubes, there are more prominent
absorption peaks (Fig. 40(c)) arising from the rich  low--lying
parabolic subbands (Figs. 36(c) and 37(c)).

The dependence of absorption frequencies on the magnetic-field
direction is important in understanding the characteristics of
prominent peaks, mainly owing to the composite effects arising from
the splitting and coupling of angular momenta. For example, within
the frequency range of ${\omega\le\,0.25}$ $\gamma_0$, all the
magneto-absorption peak frequencies monotonously grow or decline as
the magnetic field gradually deviates from the nanotube axis, as
clearly shown in Fig. 41. This is related to the band-edge state
energies of the split subbands. The transition intensities will
present the drastic change during the variation of ${\alpha}$, so
that absorption peaks might disappear or  come to exist. The
threshold peak (triangle) becomes absent in the metallic
(narrow-gap) carbon nanotubes  under a  large ${\alpha}$, e.g.,
${\alpha\,=86^\circ}$ for (50,50) nanotube in Fig. 41(a).
However, it could survive in the middle-gap systems for any angles
(Fig. 41(b)). Its disappearance and existence, respectively,
correspond to the intersecting linear bands and the gapped parabolic
bands (green curves in Figs. 36(a) and 36(c)). An extra peak
(circle), reflecting the significant coupling effect, is revealed at
a enough large $\alpha$ (${\sim\,3.6^{\circ}-5.4^{\circ}}$). Most of
magneto-absorption peaks  (rectangles) present the splitting
behavior. Whether they are merged together or become absent depends on $\alpha$. Up to now, the splitting peaks due to a parallel magnetic
field have been confirmed by the optical measurements.\cite{Science304;1129} The
predicted strong dependence on the magnetic-field direction and the
nanotube geometry is worthy of further examinations.

\subsection{Magneto-electronic properties of graphene nanoribbons}

Two typical kinds of achiral graphene nanoribbons, with the hexagons
normally arranged along the edge structure, are chosen for a model
study. Zigzag and armchair graphene nanoribbons, respectively,
correspond to armchair and zigzag carbon nanotubes. Their widths
could be characterized by the numbers of zigzag and dimer lines
($N_y$) along the transverse $y$-direction, respectively (Figs.
42(a) and 42(b)).
The low-energy band structures are evaluated from the ${2N_y}$
tight-binding functions of ${2p_z}$ orbitals. Such functions are
combined with the Peierls phases to explore the magneto-electronic
and optical properties.
For a zigzag ribbon, the Peierls-tight-binding Hamiltonian matrix
can be expressed as:
\begin{equation}
H_{ij}=\ \ \left\{
\begin{array}{cc}
2\gamma _{0}\cos (k_{x}\frac{\sqrt{3}b}{2}+\Delta{G}_{1}) &  \\
\gamma _{0} &  \\
0 &
\end{array}%
\begin{array}{c}
\text{for} \\
\text{for} \\
\text{others,}%
\end{array}%
\begin{array}{ccc}
\text{j=i+1,} & \text{j} & \text{is even,} \\
\text{j=i+1,} & \text{j} & \text{is odd,} \\
&  &
\end{array}%
\right.
\end{equation}%
where the Peierls phase difference $\Delta{G}_{1}=-\pi\phi(j/2-[N_{y}+1]/2)$.

As to an armchair ribbon, the Hamiltonian matrix is given by
\begin{equation}
H_{ij}=\ \ \left\{
\begin{array}{cc}
\gamma _{0} \exp i(-k_{x}b+\Delta{G}_{1}) &  \\
\gamma _{0} \exp i(k_{x}b/2+\Delta{G}_{2})&  \\
\gamma _{0} \exp i(-k_{x}b/2+\Delta{G}_{3})&  \\
0 &
\end{array}%
\begin{array}{c}
\text{for} \\
\text{for} \\
\text{for} \\
\text{others,}%
\end{array}%
\begin{array}{ccc}
\text{j=i+1,} & \text{j} & \text{is even,} \\
\text{j=i+3,} & \text{j} & \text{is even,} \\
\text{j=i+1,} & \text{j} & \text{is odd,} \\
&  &
\end{array}%
\right.
\end{equation}%
where $\Delta{G}_{1}=\pi\phi(j/2-[N_{y}+1]/2)$, $\Delta{G}_{2}=-\pi\phi/2(j/2-1-[N_{y}]/2)$ and $\Delta{G}_{3}=\pi\phi/2[(j+1/2)-1-[N_{y}]/2]$.

The dimension of Hamiltonian keeps the same even in the presence of a perpendicular magnetic field, while it is largely enhanced for layered
graphenes\cite{PCCP17;26008} or graphites (Chap. 2). The strong competition between the finite-size confinement and the magnetic quantization
will greatly diversify the essential properties.

The finite-size effect directly induces a plenty of 1D energy
subbands, as shown in Fig. 43. Each subband does not have a
transverse quantum number under the open boundary condition;
furthermore, the ${J^{c,v}}$ index only represents the arrangement
ordering measured from the Fermi level. This is one of the most
important differences between graphene nanoribbons and nanotubes.
Electronic structures strongly depend on the edge structures. A
${N_y=100}$ zigzag nanoribbon exhibits the band-edge states near
${k_x=2/3}$ (Fig. 43(a)), being similar to those in a ${(50,50)}$
armchair nanotube (Fig. 36(a)). All the low-lying energy subbands
possess the parabolic dispersions except for the first pair with the
partially flat ones at larger $k_x$'s. The latter belong to the
edge-localized states (discussed later). On the other hand, the
band-edge states are situated at ${k_x=0}$ for an armchair
nanoribbon e.g., ${N_y=180}$ in Fig. 43(e). Each energy subband has
no double  degeneracy except for the spin degree of freedom. The
energy gap, which is determined by the parabolic valence and
conduction subbands of ${J^{c,v}=1}$, declines in the increase of
ribbon width. The width-dependent energy gaps have been confirmed
by the STS measurements.\cite{APL105;123116,SciRep2;983,PRB91;045429,ACSNano7;6123}

The energy dispersions are dramatically changed by the magnetic
quantization. The energy spacings of a zigzag nanoribbon are getting
nonuniform during the variation of $B_0$, as shown for ${N_y=100}$
in Fig. 43(b) at ${B_0=40}$. When field strength is enough high, the
lower-energy 1D parabolic subbands will evolve into the composite
subbands (QLLs) with the parabolic and dispersionless relations
simultaneously, e.g., ${J^{c,v}=2}$ and 3.
A sufficiently large nanoribbon width, being comparable to or longer
than the magnetic length (${l_B}$), accounts for the creation of
QLLs. With a further increase of field strength, the dispersionless
$k_x$-ranges become wider, and the higher-energy subbands might be
transformed into the QLLs, such as magneto-electronic structures in
Figs. 40(c) and 40(d) at ${B_0=60}$ and 80 Ts, respectively.
Specifically, the initial energies of QLLs will approach to the LL
ones in monolayer graphene (red lines; the 2nd term in Eq. (43)).
These clearly indicate that electronic states are gradually
quantized into Landau modes from the lower-energy subbands with the
increasing $B_0$, since they have the smaller kinetic energies. It
is relatively easy to observe the dispersionless Landau states under
the enhanced field strength and the extension of nanoribbon width.
The formation centers of QLLs, respectively, correspond to
${k_x=2/3}$ and 0 for zigzag and armchair nanoribbons (Fig. 40(f) at
40 T). It should be noticed that two neighboring subbands of the
latter will gradually approach to each other  in the increase of
$B_0$, covering the gradual couplings of (${J^{c}=1,J^v=1}$),
(${J^{c}=1,J^v=1}$) and (${J^{v},J^{v}+1}$). As a result,
magneto-electronic structures of armchair nanoribbons might behave
as the splitting QLLs except for the pair of valence and conduction
QLLs. There exist the extra band-edge states and energy gap is
vanishing under a strong magnetic field.

DOSs of the 1D energy subbands present a lot of asymmetric peaks in
the square-root form, as clearly indicated in Fig. 44. The peak
height, which is inversely proportional to the square-root of
subband curvature, grows as state energy increases. It is remarkable
that the peak spacings are almost uniform in zigzag system (Fig.
44(a)), but non-uniform in armchair one (Fig. 44(e)).
Specifically, the former has a pair of merged peaks near ${E=0}$ or
an obvious symmetric peak there (inset in Fig. 44(a)). The low-lying
asymmetric peaks become the delta-function-like symmetric ones, when
the QLLs could be created by a  $B_0$-field (Figs. 44(b), 44(c),
44(d) and 44(f)). Furthermore, their heights decline quickly in the
increase of QLL energy, reflecting the diminish of the
dispersionless $k_x$-ranges (Fig. 43). Under the increasing field
strength, all the peak energies are enhanced for zigzag systems,
while a simple dependence is absent for armchair ones.$^{61}$ As a
result, QLLs of the former will recover to LLs of monolayer graphene
if the magnetic field plays a  dominating role (Figs. 43(b)-43(d)).
The latter could exhibit the complex structures covering single and
double peaks.

\subsection{Magneto-optical spectra of graphene nanoribbons}
When electrons are confined in the finite-width nanoribbons, their
spatial distributions reveal as the regular standing waves (Fig.
45).
The oscillatory patterns, with the specific number of nodes, are
very sensitive to the edge structures, sublattices, state energies,
and wave vectors. They are quite different from those in carbon
nanotubes (Fig. 38), and so do the optical properties. For zigzag
nanoribbons, the wave functions could be decomposed into the
subenvelope functions on the A and B sublattices at the odd and even
zigzag lines (Figs. 45(a)-45(c)). The spatial distributions of wave
functions present the alternative change between the symmetric and
anti-symmetric forms as the subband index increases (more detailed
relations in Ref. 41). On the other hand, those of armchair systems
lie at the ${3m}$-, ${(3m + 1)}$-, and ${(3m + 2)}$-$th$ dimer lines
of the A and B sublattices (Fig. 45(d)-45(f)). There are two kinds
of unique relations. For a specific subband, the subenvelope
functions on two sublattices possess the same phase or the phase
difference of $\pi$. Furthermore, the valence and conduction
subbands, with the identical index, exhibit the similar phase
relation in the subenvelope functions of the A ($B$) sublattice. It
is very significant that the special relations in wave functions are
edge-dependent and thus dominates the distinct selection rules.
However, the main features of the angular-momentum-dominated
standing waves in carbon nanotubes (Fig. 39) are independent of the
chiral angle.

A dominating magnetic field can thoroughly alter the characteristics
of wave functions, regardless of the edge structures. In general,
the standing waves are changed into the well-behaved LL wave
functions, as clearly indicated in Figs. 46(a)-46(e) for a zigzag
system.
The ${k_x=2/3}$ states present the symmetric or anti-symmetric
distributions about the nanoribbon center. They are identical to
those of monolayer graphene (Fig. 5(b)) except that the first pair
of QLLs possesses the edge-localized distributions (Fig. 46(c) and
46(d)). With the different wave vectors, the spatial distributions,
being revealed in Figs. 46(f)-46(j), are getting asymmetric.
Furthermore, the number of zero points might be lost as the wave
vectors are far away from ${2/3}$. Apparently, the competition
between magnetic quantization and finite-size confinement will lead
to the coexistent behavior, in which LL modes and standing waves,
respectively, correspond to low- and high-lying electronic states.
This unusual feature could be observed by changing the $B_0$-field
strength or nanoribbon width.

Graphene nanoribbons possess the edge-dependent optical selection
rules, as shown in Fig. 47.
The optical vertical excitations arise from the interband
transitions of the $J^v$-th valence band and the $J^c$-th conduction
band. For zigzag nanoribbons, a lot of asymmetric absorption peaks
in the square-root form are characterized by the selection rule of
${\Delta\,J=2I+1}$ (Fig. 47(a)). The strong absorption peaks might
come from the multi-channel excitations simultaneously, especially
for the higher-frequency ones. This rule could be directly derived
from the non-vanishing velocity matrix elements (Eq. (39)) by using
the special relations in the subenvelope functions (detailed
calculations in Ref. 41). On the other side, the available
excitation channels in armchair systems agree with the $\Delta J=0$
rule, so the number of absorption peaks is reduced. The valence and
conduction bands, with the identical $J$, could also be revealed as
the prominent absorption peaks of any carbon nanotubes (Fig. 40),
while the angular momentum is conserved  during the vertical
transitions.

The unusual transformation between the edge- and QLL-dominated
absorption peaks could be presented in the variation of
magnetic-field strength (or nanoribbon width), as clearly shown in
Figs. 48(a)-48(c).
The QLL wave functions have the well-behaved spatial symmetry, so
that the effective optical transitions associated with the symmetric
absorption peaks are governed by the QLL-dependent selection rule.
At lower frequency, the magneto-absorption peaks of
${\Delta\,J\neq\pm\,1}$ are absent or become very weak; furthermore,
the ${\Delta\,J=\pm\,1}$ rule is equivalent to the
${\Delta\,n=\pm\,1}$ one in monolayer graphene (red curve in Fig.
48(a)). This is independent of edge structures; that is, the number,
frequency and intensity of prominent absorption peaks are identical
for zigzag and armchair nanoribbons with the almost same width
(Figs. 48(a) and 48(d)). Suck peaks have the symmetric form and
stronger intensities  at high field strengths (Figs. 48(c) and
48(d)). In addition, some extra lower absorption peaks are revealed
by the ${J^{c,v}=1}$ QLLs of the former (e.g., blue circles in Fig.
48(a)) because of the impure LL wavefunctions (Figs. 46(c) and
46(d)), and the splitting QLLs of the latter could create the
double-peak structure (blue circles in Fig. 48(d)). Concerning the
higher-frequency asymmetric peaks (the right-hand side of the
gray-dashed vertical line), there are more prominent structures,
especially for the complex absorption peaks in armchair nanoribbons
(Fig. 48(d)). They originate from the band-edge states of the
parabolic valence and conduction bands, being characterized by the
strong competition of the edge- and QLL-dependent selection rules.

The complicated relations among the lateral confinement, magnetic
quantization and dimension deserve closer investigations. They could
be understood from the initial six prominent magneto-absorption
peaks, as clearly revealed in Fig. 49 for their $B_0$-dependent
frequencies.
Zigzag nanoribbons exhibit the monotonic magnetic-field dependence
(open circles in Fig. 49(a)). In general, peak frequencies grow with
the increasing $B_0$ in the absence of a specific relation. They are
very different from those of monolayer graphene (red dots) except
for the sufficiently high field strength. At low $B_0$, the lateral
confinement (black circles) dominates the frequency, intensity, form
of absorption peaks, e.g., the more higher frequencies compared with
the 2D results. It will be seriously suppressed the magnetic
quantization, when the field strength is over a critical one (red
circles). The number of prominent peaks keeps the same during the
variation of $B_0$, since the ${\Delta\,J=odd}$-induced peaks (Fig.
47(a)) cover the QLL-created ones. However, absorption peaks in
armchair systems might come to exist or disappear frequently as
$B_0$ gradually increases (Fig. 49(b)), being closely related to the
QLL splitting and the high competition of two selection rules.
Obviously, a simple $B_0$-dependence of magneto-absorption
frequencies is absent. The zero-field absorption peaks of
${\Delta\,J=0}$ (black circles; Fig. 47(b)) are not consistent with
the ${\Delta\,J=0}$ rule due to the QLLs, so that their intensities
decay rapidly and vanish after the critical $B_0$s. On the other
hand, the new peaks initiated by the strong magnetic quantization
might appear in the double-peak structures (two close red-circled
curves) or the merged ones. They could behave as the inter-LL peaks
in 2D monolayer (red dots) only at higher field strengths. Up to
now, there are no optical and magneto-optical measurements on
graphene nanoribbons. The theoretical predictions of the edge- and
QLL-dominated selection rules could be verified by optical
spectroscopies, as done for graphite, layered graphenes, and carbon
nanotubes.

\subsection{Comparisons and applications}

The distinct stacking symmetries in three kinds of layered graphites
have created the diverse and novel physical phenomena. The critical
differences cover electronic and optical properties in the
absence/presence of magnetic quantization. The AA-, AB- and
ABC-stacked graphites, respectively, have one, two and one pairs of
$\pi$-electronic valence and conduction bands, in which the
${k_z}$-dependent band widths are ${\sim\,1}$ eV, ${\sim\,0.1-0.2}$
eV and ${\sim\,0.01}$ eV, and the band-overlap widths near the Fermi
level behave similarly. The carrier density of free electrons and
holes is highest in simple hexagonal graphite, while it is lowest in
rhombohedral graphite. Band structures of three systems could be
regarded as the 3D vertical Dirac cone, the composite of monolayer-
and AB-bilayer-like ones, and the spiral cone structure. The 3D
energy dispersions determine the DOS characteristics, such as, the
semi-metallic behavior and van Hove singularity-induced distinct
structures. The band-dominated optical spectra exhibit the
dimension- and stacking-dependent characteristics at low and middle
frequencies, including the frequency, number, form and spectral
width of the special absorption structures.  The main features of
zero-field band structures are directly reflected in the magnetic
LSs. In general, the band-edge states of the 1D parabolic LSs are
shown as many asymmetric peaks in DOS. AA system exhibits a lot of
valence and conduction LSs intersecting with the Fermi level under
the specific energy spacings, while ABC system only presents one
crossing LS of ${n^{c,v}=0}$. Both of them possess the
monolayer-like subenvelope functions and $\sqrt{B_0}$-dependent
energy spectra. However,  AB system has two groups of LSs with the
normal and perturbed modes, leading to the coexistence of crossing
and anticrossing behaviors. Only the initial two LSs of the first
group cross the Fermi level. The $B_0$-dependence of LS energies is
sensitive to the dimension-induced $k_z$, such as the square-root
and linear dependences at ${k_z=0}$ and $\pi$ (K and H points),
respectively. Furthermore, the magnetic subenvelope functions might
dramatically change during the variation of $k_z$. Specifically,
even without any crossings and anticrossings, the rich and unique
magneto-optical spectra are revealed by the well-behaved LSs in
AA-stacked graphite, including the intraband and interband inter-LS
vertical excitations, the Fermi-momenta-induced absorption peaks,
the non-uniform peak intensity, the multi-channel threshold peak,
intraband  two-channel peaks $\&$ interband double-peak structures
at distinct frequency ranges, the discontinuous $B_0$-dependence of
the initial interband channel, and the beating feature related to
the vertical Dirac cones. Magneto-absorption peaks agree with the
monolayer selection rule of ${\Delta\,n=\pm\,1}$ for AA-, ABC-, and
AB-stacked graphites. The second stacking can create the
monolayer-like characteristics, such as, the interband transitions,
non-composite symmetric peaks due to the same contribution of K and
H points, almost uniform intensity, and pure  ${\sqrt
B_0}$-dependence of absorption frequency. As for AB-stacked
graphite, there are four categories of interband inter-LS
excitations arising from the same or different groups. The strong
asymmetric peaks might appear at the identical frequency ranges and
thus exhibit the very complex absorption structures. Their main
features are rather different for the K- and H-point vertical
excitations, respectively, leading to the bilayer- and
monolayer-like absorption frequencies. Moreover, some extra peaks
come to exist under the LS anticrossings. The above-mentioned
differences could be examined by the experimental measurements of
ARPES, STS and optical spectroscopies on energy bands, DOS, and
absorption spectra, respectively.

The dimensional crossover of the essential properties occur in AA-
and AB-stacked layered graphenes as the layer number gradually
grows. The $N$-layer AA stacking has a vertical multi-Dirac cone
structure which is distributed within the $k_z$-dependent band width
of simple hexagonal graphite. It is a semi-metal under the overlap
of valence and conduction bands. However, an optical gap are induced
in a $N$-even system, since  absorption spectrum is only a
combination of $N$ intra-Dirac-cone vertical excitations. Under a
perpendicular magnetic field, the magneto-optical gap comes from a
forbidden transition region related to the intragroup LLs, and it is
greatly enhanced by the increasing field strength. The
magneto-threshold channels dramatically changes with the increasing
$B_0$, in which their intensities are about half of the others. The
low-lying Dirac-cone structure, magnetically quantized states,
special structures in DOS, and absorption peaks are expected to
approach  those of AA-stacked graphite for ${N>30}$. As for
AB-stacked graphene, band structure of a N-odd system resembles a
hybridization of massless Dirac cone and massive parabolic
dispersions, while that of a N-even system consists of only pairs of
parabolic subbands. When $N$ is very large, the monolayer- and
bilayer-like states are expected to correspond to the $H$ and $K$
point in Bernal graphite, respectively. The excitation channels are
only allowed between the respective monolayer-like subbands or
between the bilayer-like subbands; the magneto-optical selection
rule is also applicable to all the inter-LL transitions. With the
increase of layer number, the LL anticrossings happen more
frequently as compared to those in Bernal graphite. Electron-hole
asymmetry-induced twin-peak structures are revealed in both
multi-layer and bulk systems. However, in magneto-optical spectrum,
the measured profiles of the $B_0$-dominated peaks, including
threshold channel, intensity, spacing and frequency, could be used
to distinguish the stacking layer, configuration and dimension. On
the other hand, the dimensional crossover is hardly observed in ABC
stacking, mainly owing to the distinct lattice symmetries in 3D and
2D systems. For bulk graphite, a primitive unit cell, rhombohedron,
has a bi-particle lattice symmetry. The low-energy electronic
properties is described by the 3D anisotropic spiral Dirac cones.
Specially, the ABC-stacked graphene possess the surface-localized
and sombrero-shaped subbands, irrelevant to the rhombohedral
graphite. The zero-field absorption spectrum is contributed by the
${N\times\,N}$ excitation channels of the $N$ pairs of energy bands,
in which the low-frequency region is dominated by the
surface-localized states. The magneto-optical spectrum consists of
${N\times\,N}$ groups of inter-LL absorption peaks, each of which
displays the characteristic $B_0$-dependence regarding the
frequency, intensity and number. Furthermore, the frequent
anticrossing of LLs  due to the sombrero-shaped subbands lead to
extra peaks which have abnormal relations with the field strength.

The significant effects due to the lateral quantum confinement and
magnetic field clearly illustrate the dimension-diversified
essential properties. The periodical boundary condition induces the
decoupled angular-momentum states in carbon nanotube, but the open
one cannot create a transverse quantum number in graphene
nanoribbon. The former possesses the sine/cosine standing waves
along the azimuthal direction, regardless of radius and chirality.
However, the unusual standing waves in the latter depend on edge
structure, A and B sublattices; even zigzag and dimer lines. The
distinct characteristics of subenvelope functions dominate the
diverse selection rules, the conservation of angular momentum in
carbon nanotube, the same index of valence and conduction subbands
in armchair nanoribbons, and the index difference of odd integers in
zigzag nanoribbons. All the special structures in DOS and absorption
spectrum are presented in the asymmetric peaks of square-root form.
The magnetic quantization in cylindrical nanotubes is mainly
determined the field direction and strength. A parallel magnetic
field leads to the shift of angular momentum or the destruction of
double degeneracy, and the periodical Aharonov-Bohm effect in energy
dispersions, band gaps and absorption peaks.  However, the discrete
angular momenta are coupled with one another in the presence of a
perpendicular magnetic field, and QLLs hardly survive in cylindrical
systems except for very high field strength and large radius.
Apparently, optical spectra are dramatically altered by the magnetic
field, such as, more splitting peaks with the same rule and some
extra peaks without the optical rule under the parallel and
perpendicular ones, respectively. On the other hand, QLLs could
exist in graphene nanoribbon if the width is sufficient for the
localized oscillatory distribution. Their dispersionless
$k_x$-ranges can transform the asymmetric peaks of DOS and
absorption spectrum into the symmetric ones. Moreover, the latter
and the former, which correspond to the QLL- and edge-dependent
selection rules, respectively, appear at the lower and higher
frequencies.

Graphite, as well as graphite intercalation compounds, have been
extensively developed for various applications during a long tome as
a result of the unique properties. Pristine graphite is a suitable
precursor in the production of other carbon-related materials. It is
the most stable allotrope of carbon. The covalent bondings between C
atoms in the same layer can create graphite's high temperature
stability, and excellent electrical and thermal conductivity. Owing
to the superior mechanical properties, graphite fibers are
frequently utilized for reinforced composite
materials\cite{ASS401;79,Carbon114;59,CBM135;394,CST138;179}.
Graphite intercalation compounds, with the significant chemical
bondings between adatoms and carbons, exhibit wide applications in
electronics\cite{SSI300;169,SynMet222;351}, energy
storage\cite{JMCA5;4325} and
electrochemistry\cite{Carbon112;185,CeramInt43;4309}. The most
common use in industry serves as a low-cost
electrode.\cite{SABC241;455,JEC786;145} The high free electron
(hole) density is generated by the intercalation of metal atoms
(molecules).\cite{AdvPhys30;139} The metallic compounds could be
used as superconductors with high transition temperatures. e.g.,
11.5 K for C$_{6}$Ca and
C$_{6}$Yb.\cite{NatPhys1;39,NatPhys1;42,STAM9;044102} They have also
been applied for photo electro catalytic, electrochemical and
biomedical
sensor\cite{SAABC242;825,BB89;136,BB90;6,JCIS487;149,SABC239;325}.
Graphite composites display high conductivity for microbial fuel
cells applications in
electrochemistry\cite{IJHE38;15723,BB25;2167,MRB88;188}.
The field-effect transistor (FET) sensor based on graphite oxide
nanoparticle composites presents an advantage of low cost and
possesses high selectivity and excellent
stability.\cite{JAC694;1061,NanoLett12;1165}

Graphene-related materials have displayed the high potentials for
electronic, photonic and optoelectronic applications, such as
touch-screen panel device,\cite{NatNanotechnol5;574,MatLett107;247}
light-emitting diodes
(LEDs)\cite{ACSNano4;637,NatCom8;14560,APL95;203304,NanoLett4;911,
NanoLett6;1880}, solar
cells,\cite{small11;2963,MatLett192;84,APL88;233506,APL87;203511}
photo-detectors\cite{AOM3;989,OPTICA3;979,SciRep6;38569,ACSNano10;6963,SSA37;356,
NatNanotechnol4;839,NatPhoton4;297,ACSNano11;430} and
photo-modulators.\cite{Nature474;64,OptLett41;816} The direct
application of graphene in FETs is suppressed owing to its zero-gap
nature. However, bilayer graphene can open a sizable and tuneable
bandgap by applying a gate voltage, which is appropriate for making
large-area graphene FETs with extremely thin, shorter and higher
speeds
channels.\cite{NanoLett12;1324,IEEE30;261,IEEE56;2979,JAP118;244501}
The high transparency and flexibility of graphene could be utilized
to design thin, light and delicate
devices\cite{ACSNano4;637,NatCom8;14560}. Graphene-based optical
modulator possesses great advantages in low operation voltage, fast
modulation speed, small footprint and large optical bandwidth,
compared with semiconductor ones\cite{Nature474;64,OptLett41;816}.
By tuning the Fermi level or the free carrier concentration of
graphene sheets, the modulator is operated over a broad wavelength
range. As a result of the strong and rich interband optical
transitions, the graphene detectors are suitable for the
applications within a very wide energy spectrum, covering the
ultraviolet, visible, infrared and terahertz frequency
ranges\cite{Nature474;64,OptLett41;816}. When layered graphenes are
further doped with adatoms or molecules, they might change into
gap-modulated semiconductors or metals. Specifically, graphene
oxides and hydrogenated graphenes possess the adatom-modulated
energy gaps in a wide range of ${0<E_g<4.0}$ eV, depending on the
concentration and distribution of
adatoms.\cite{Carbon93;967,RSCAdv6;24458} The tunable and
controllable electronic properties make them serve as potential
candidates, such as
FETs,\cite{AFM24;117124,CAP14;738,Nanotech21;165202};
supercapacitors,\cite{NatNanotechnol6;496,Carbon49;573}
sensors,\cite{Carbon50;4228,NanoLett8;3137}
photovoltaic\cite{JPCL2;3006,ACSNano4;5263,RSCAdv4;35493} and
light-emitting devices.\cite{RSCAdv4;35493,JMC22;2929} The alkali-
and Al-induced high free carrier density might have high potentials
in future technological applications, e.g., high-capacity
batteries,\cite{Nature520;324,JES160;A1781} and energy
storages.\cite{PRB81;205406,JAP105;4307}

The wide-range applications in electronic devices have been made
with carbon nanomaterials, such as 1D carbon
nanotubes\cite{APL106;213503,APL73;2447,APL108;163104,
AdvMater21;2586} and graphene
nanoribbons,\cite{AdvMater21;2586,Science319;1229} 2D few-layer
graphene\cite{AdvMater21;2586} and 3D graphites.\cite{SSI300;169,SynMet222;351,JMCA5;4325,SABC241;455,JEC786;145,
JCIS487;149,SABC239;325,SAABC242;825,BB89;136,NatPhys1;39,
NatPhys1;42,STAM9;044102,BB90;6} Up to now, FET
based on semiconducting nanotubes and nanoribbons are widely
developed, mainly owing to the advantage of high mobility under the
low scatterings. In carbon nanotubes, the intrinsic 1D electronic
structures, with the decoupled states of angular momenta, dominate
the 1D quantized electrical properties, including the
radius-dependent resistance, capacitance and inductance, which are
diameter dependent and responsible for the non-monotonic dependence
of the electrical mobility.\cite{NanoLett4;35} The first FETs made
of carbon nanotubes were reported in
1998,\cite{APL73;2447,APL108;163104} which have superior electrical
properties of the conducting channels by the gate-voltage
modulation. The nano-scaled carbon nanotubes are responsible for low
scatterings and allows the gate's ability to control the potential
of the channel in the ultimate thin FETs, while suppressing
short-channel effects.\cite{APL108;163104,APL106;213503} Moreover,
semiconducting nanotubes present wide applications on optoelectronic
devices, such as electroluminescent light
emitters,\cite{PRB73;085421,Science300;783}
supercapacitors\cite{ACSApplMater6;15434} and
photodetectors.\cite{ACSNano10;6963,SciRep6;38569,AOM3;989}. Both
electrons and holes confined on the cylindrical surface are driven
towards each other by applying the appropriate biases on source and
drain of the FET. The recombination due to the two types of excited
carriers emits electroluminescence, and the photon-emission process
is extensively utilized in
LEDs.\cite{NanoLett11;23,NatPho10;420,NatNanotech5;27} The
application of photodetector is based on the electric current
generated by the resonant excitations. On the other side, metallic
carbon nanotubes could serve as high-performance interconnects in
integrated electronic devices.\cite{JMS52;643,JEM44;4825}


The seminconducting graphene nanoribbons could be directly used in
applications of
FETs,\cite{Science319;1229,NatNanotech6;45,APLMat3;011101} since
electronic states are confined in a narrow width and have obvious
energy spacings (gaps).\cite{PRL98;206805,Science312;1191} Edge
structures and ribbon widths lead to different 1D electronic and
optical properties, e.g., the strong dependence of wave function on
the edge or center position. With the decreasing ribbon width, the
carrier mobility is degraded by the edge boundary, while the
potential barrier is enhanced for the conducting channels. The
on/off ratio is improved for narrow ribbons when the temperature is
sufficiently
lowered.\cite{PRL100;1229,Science319;1229,NatNanotech6;45} Another
promising application of graphene nanoribbons is polymer composite
and electrode material for
batteries\cite{ACSAMI6;9590,ACSAMI7;26549,AdvMater25;6298} and
supercapacitors.\cite{JMCA2;7484,JMCA3;4931,NatChem8;718} The
synthesis of the graphene nanoribbon composite has produced an
effective component to improve the electrochemical stability and
enhanced specific capacity of the electrode materials.







\section{Concluding remarks}

This work presents a systematic review of essential properties for
simple hexagonal, Bernal and rhombohedral graphites. The generalized
tight-binding model and the gradient approximation are developed to
explore the electronic and optical properties under the magnetic
quantization. Furthermore, the effective-mass approximation can
provide the qualitative pictures and the semi-quantitative results.
A thorough comparison is made among 3D graphite, 2D graphenes, 1D
graphene nanoribbons and carbon nanotubes by covering the dependence
on the layer number, stacking configuration, dimension,
width/radius, edge/chirality, and boundary condition. This is useful
in understanding the dimensional crossover behavior. The calculated
results agree with those from other theoretical calculations and are
validated by the experimental measurements, while most of
predictions require further detailed examinations. The theoretical
framework is useful in promoting the future studies on other layered
materials, e.g., Si-,\cite{PRB94;205427,NatNanotech10;227} Ge-,\cite{AdvMater26;4820,PRL108;155501} Sn-,\cite{PRL111;136804}, P-,\cite{PRL111;057005,PRB94;045410} and
Bi-related 2D and 3D materials.\cite{NanoLett12;4674} Specifically, the
generalized tight-binding model is suitable for solving the critical
Hamiltoians with the multi-orbital bondings, the spin-orbital
couplings, the interlayer atomic interactions; the external electric
and magnetic fields.\cite{PRB83;195405,JVSTB28;386,OptExp22;7473,PRB94;205427,
RSCAdv5;51912,SciRep7;40600,PRB94;045410} This model could combine with the
single- and many-particle theories to comprehend the other physical
properties, such as Coulomb
excitations\cite{PLA352;446,PRB74;085406,ACSNano5;1026} and
transport properties.\cite{arXiv170401313}

The intralayer and interlayer atomic interactions of ${2p_z}$
orbitals account for the diverse essential properties in layered
graphites. The AA-stacked graphite has the highest density of free
electrons and holes (the largest band overlap), the widest energy
dispersions along ${\hat k_z}$, the vertical Dirac-cone structures,
and many LSs cross the Fermi level. These directly reflect the
highest stacking symmetry, or the strongest interlayer interactions.
The optical spectrum presents a shoulder structure and a prominent
plateau structure at the low and middle frequencies, respectively.
Due to the unusual magnetic quantization, one group of valence and
conduction LSs can create the multi-channel threshold peak, some
intraband two-channel peaks, and a lot of interband double-peak
structures with the beating phenomena. The unique oscillational
magneto-absorption spectra are never predicted or identified in the
previous studies on any materials. On the other hand, the layered
graphenes exhibit more low-frequency shoulder structures, and
optical gaps in $N$-even cases. The finite-layer confinement effect
is almost vanishing for ${N>30}$; that is, the AA-stacked graphenes
and graphite possess the same optical spectra there. The
magneto-absorption peaks have a symmetric structure and a uniform
intensity. The threshold intraband peak is non-well-behaved in the
$B_0$-dependence, and some initial interband peaks are absent. These
are closely related to the quantized LLs from the multi-Dirac-cones.
It is also noticed that all absorption peaks due to LLs and LSs
agree with the selection rule of ${\Delta\,n=\pm,1}$, and they
possess the ${\sqrt B_0}$-dependences except for the multi-channel
ones.


Bernal graphite is intriguing for the studies of massless and
massive Dirac-quasi-particles. The research interest in 2D
graphenes is based on the properties of bulk graphite, which are
deduced to represent the coexistence of the 2D monolayer and bilayer essential properties at different $k_{z}$ wave vectors.
Both Bernal graphene and graphite exhibit the optical response
of Dirac fermions regardless of external fields. The band structure
displays a massless-Dirac-like behavior in the vicinity of the H
point, where the in-plane dispersion is linear and doublely degenerate
to reflect the two isolated graphene sheets in the primitive unit
cell. On the other hand, the in-plane energy dispersions near the K
point resemble the massive-Dirac-like behavior which is specified
to the AB-stacked bilayer graphene. These monolayer-like and
bilayer-like energy dispersions have indeed been observed by ARPES\cite{PRL100;037601,ASS354;229,PhyB407;827,PRB79;125438,NatPhy2;595}.
The field evolution of the 1D LSs is depicted in the Pierels
tight-binding model. Under a strong magnetic field, the crossing of
the low-lying LSs near the Fermi level gives rise to the change of
semimetallic Bernal graphite into a zero-gap semiconductor.
In the scale of $B_{0}$, two series of square-root divergent peaks
with linear and square-root dependences are verified by STS, and
these peaks account for the monolayer-like and bilayer-like Landau states
that accumulate at the band edges of the LSs at the K and H points,
respectively. Furthermore, depending on the curvatures of the LSs,
the measured intensities of their own respective peaks are
consistent with the theoretical calculations. The magneto-optical
properties elucidated in the framework of tight-binding model reveal
far more significant results than those results derived from a
simplified effective-mass approximation. This provides clarity to
the information of the grephene-like properties in graphites and the
true epitaxial graphenes.

In the vicinity of the H and K points, the inter-LS channels are the
dominant contributions to the magneto-absorption spectrum, of which
the spectral intensity is determined by the DOS intensity and the
dipole transition probability. The monolayer-like and bilayer-like
absorption spectra are predicted to coexist in the bulk spectrum
following the characteristic magnetic field frequency dependences
$\propto \sqrt{B_{0}}$ and $\propto B_{0}$.
The measured Fermi velocity can be used to interpret $\gamma_{0}$
and $\gamma_{1}$ from the H-point and K-point channels; the deduced
values in graphite match those in few-layer graphenes. Infrared
magneto-absorption spectroscopies have confirmed the splitting of
the absorption peaks for the optical transition channels near the K
and H points. However, the observability of the peak splittings in
optical spectroscopy depends on the competition among the magnetic
field, ambient temperature and experiment resolution. These main
features are very useful in identifying the stacking configurations
and the dimensionality of systems from experimental measurements. The
splitting at the H point is attributed to electron-hole asymmetry,
reflecting the inherent complexity of the full interactions in SWM
model. However, the latter is still under debate, which doesn't
result from Dirac-cone asymmetry but might originate from
spin-orbital coupling, anticrossing of LSs or parallel magnetic
flux. These results require a more elaborated model and better
experimental verifications. In addition, the inconspicuous peaks,
coming from the band-edge states of the anticrossing LSs near the
the H point, could be possibly observed with the extra peak
intensities enhanced by the degree of the hybridization of the LSs.


ABC stacking configuration has different point-group symmetries for the
corresponding 2D and 3D structures, leading to distinct
characteristics of electronic properties and optical spectra. For
example, the massless-Dirac quasi-particles are preserved in
ABC-stacked graphite, and the bulk stack is topologically nontrivial
for the existence of surface-localized states. The massless-Dirac
characteristics are even more obvious than those of AA- and
AB-stacked graphites, because the energy dispersion dependence on
$k_{z}$ is weaker than those in Bernal graphite and simple hexagonal
graphite by one or two orders of magnitudes. The low-energy
electronic and optical properties are reviewed in both rhombohedral
and hexagonal unit cells, which are, respectively, built from 2 and
8 sublattices. The former with $p_{3}m$ symmetry is the primitive
unit cell of ABC-stacked graphite, whereas the latter with $p_{3}$
symmetry is chosen to represent AA- and AB-stacked graphite for the sake
of convenience. They provide the same physical results, while due to
the zone-folding effect in the latter, the use of the former is more
appropriate to comprehend the evolution of the Dirac cone and the
magnetic quantization under the influence of different
$\beta_{i}^{,}$s.
In the minimal model only with $\beta_{0}$ and $\beta_{1}$, the
Dirac points rotate in a circular path with a constant radius of $\beta_{1}(v_{0}\hbar)^{-1}$ at the Fermi level. Each Dirac cone behaves as in monolayer graphene with a Fermi velocity of $3\beta_{0}b/2$, giving rise to linearly increased intensity of $\omega$ in the absorption spectrum. Under a magnetic field, the corresponding LSs
are totally reduced to 0D dispersionless LLs that are classified to
one group just like monolayer graphene. Such LLs induce 2D
delta-function-like peaks in the DOS that are intensity-equal and
followed by a simple square-root energy relationship
$E^{c,v}(n^{c,v})\propto\sqrt{n^{c,v}B_{0}}$. Also, the
magneto-absorption spectrum is identical to that of monolayer
graphene in which the 2D spectral peaks are in the sequence,
$E(n^{c}\rightarrow
n^{v})\propto\sqrt{B_{0}}(\sqrt{n^{c}}+\sqrt{n^{v}})$, where
$n^{c}-n^{v}=\pm 1$. These 2D characteristics based on the minimal
model implies that the stacking effect is not demonstrated in the
minimal model because of the lack of the consideration of all
interlayer atomic interactions.

Considering the additional $\beta_{i}^{,}$s more than the minimal model,
the Dirac cone changes into tilted and anisotropic, and furthermore
it spirals around the corners of the 1st BZ with a varying radius. These behaviors are deduced to be caused by the influences of $\beta_{3}$ and $\beta_{4}$. The distortion of the isoenergy surfaces causes a deviation of linear absorption intensity in the low-energy region. With the
knowledge of the anisotropic energy dispersions, the
magneto-electronic and magneto-optical properties are reviewed
within the semi-classical Onsager quantization and generalized
Peierls tight-binding schemes. Both schemes are consistent in the low-energy region. The 1D $k_{z}$-dependent LSs are symmetric
about the $n^{c,v}=0$ ones in the 1st BZ; however, the electron-hole
symmetry is broken down because the dispersion of the $n^{c,v}=0$
LSs moves according to the Dirac-point spiral. Based on the magneto
selection rule $\Delta n=\pm 1$, the vertical transitions of a
single channel are found to have approximately the same energy along
K-H. Moreover, the energy deviation from the monolayer energy
dependence directly indicates the distortion of Dirac cones. As the case for the experimental verification on the theoretically predicted optical and electronic properties of AB-stacked graphite, the spiral
Dirac-cone structure can be verified by using the same experimental
techniques, such as ARPES, STS, and magneto-optical spectroscopy.

The reduced dimension in the transverse $y$-direction can greatly
diversify electronic properties and optical spectra of
carbon-related systems, especially for carbon nanotubes and graphene
nanoribbons . The essential properties are very sensitive to the
radius/width, chirality/edge and periodical/open boundary condition.
As a result of the cylindrical symmetry, each carbon nanotube has
the angular-momentum-dependent electronic states, being revealed as
the sine/cosine-form standing waves. The band-edge state energies
and 1D energy dispersions strongly depend on radius and chirality,
and so do the frequency, number and intensity of 1D asymmetric
absorption peaks. The ${\Delta\,J=0}$ selection rule, which comes
from the specific standing waves, represents the conservation of
angular momentum during the vertical excitations. This is
independent of geometric structures. A parallel magnetic field
induces the splitting of double degeneracy, the metal-semiconductor
transition and the periodical Aharonov-Bohm effect, while a
perpendicular one creates the coupling of distinct angular-momentum
components and thus the destruction of selection rule or the extra
absorption peaks. However, it is very difficult to observe the
QLL-dominated essential properties except for very high
perpendicular  magnetic fields or large carbon nanotubes.

As a result of the open boundary condition, graphene nanoribbons
quite differ from carbon nanotubes, covering the absence of a
transverse quantum number, the edge-dominated standing waves,  the
edge-dependent selection rules, and the coexistence with the
QLL-induced selection rule. The edge structure plays an important
role in the existence of edge-localized states, the uniform or
non-uniform energy spacings, and the state degeneracy. Specifically,
the subenvelope functions strongly rely on A and B two sublattices,
the zigzag/dimer lines, state energies, and wave vectors.  For
zigzag and armchair nanoribbons,  one and two special relations are,
respectively, presented in valence and conduction subbands, being
responsible for the ${\Delta\,J\,=2I +1}$ and 0 selection rules of
the zero-field optical spectra. The magnetic QLLs, being similar to
LLs in monolayer graphene, are mainly determined by the competition
between the width and magneto-length. They could exhibit the
lower-frequency symmetric absorption peaks with the
${\Delta\,J=\pm\,1}$ selection rule. Furthermore, the asymmetric
absorption peaks associated with the edge-dependent selection rule
(the specific parabolic subbands) could survive at higher frequency.
The transformation between these two types of absorption peaks are
clearly revealed in the $B_0$-dependent magneto-optical spectra. It
is relatively observed in zigzag nanoribbons, compared with armchair
systems. The latter could present  very complicated peak structures
because of the strong competition of ${\Delta\,J=0}$ and ${\pm\,1}$
rules.

Part of theoretical calculations agree with the experimental
measurements. ARPES has identified the 3D energy bands of Bernal
graphite,\cite{PRL100;037601,ASS354;229,PhyB407;827,PRB79;125438,NatPhy2;595}
Dirac cone structure in monolayer graphene,\cite{PRL110;146802}
two/three linear valence bands in bilayer/trilayer AA
stacking,\cite{NatMater12;887,NanoLett8;1564} two pairs of parabolic
bands in bilayer AB stacking,\cite{PRL98;206802,Science313;951}
linear and parabolic bands in tri-layer ABA
stacking,\cite{PRB88;155439,PRL98;206802} partially flat,
sombrero-shaped and linear bands in tri-layer ABC
stacking,\cite{PRB88;155439} and 1D parabolic subbands and energy
gaps in graphene nanoribbons.\cite{PRB73;045124,ACSNano6;6930} The
similar ARPES examinations could be done for electronic structures
of simple hexagonal and rhombohedral graphites, and carbon
nanotubes. The STS confirmations on the DOS characteristics cover a
finite value at the Fermi level (the semi-metallic behavior) and the
bilayer- and monolayer-like LS energy spectra in Bernal
graphite,\cite{PhyB407;827,PRB79;125438,PRL100;037601,ASS354;229,NatPhy2;595}
the V-shaped structure vanishing at ${E=0}$ and the ${\sqrt
{B_0}}$-dependent LL energies in monolayer graphene, a special
structure at ${E\sim\,0.3}$ eV and the linear $B_0$ dependence
(linear and square-root dependences) for LL energies in bilayer
(trilayer) AB stacking,\cite{PRL106;126802,NatPhys6;109,PRB91;155428,
PRB87;165102,APL107;263101} a prominent peak near $E_F$ arising from flat bands in tri-layer ABC stacking,\cite{APL107;263101,ACSNano9;5432,PRB91;035410} the radius-
and chirality-enriched energy gaps and asymmetric prominent peaks in
carbon nanotubes,\cite{Science292;702,Nature391;59,Nature391;62} and
the confinement-induced band gaps in graphene
nanoribbons.\cite{APL105;123116,SciRep2;983,PRB91;045429,ACSNano7;6123} The other magneto-electronic prominent structures in DOS, which are presented
by the AA- and ABC-stacked graphene/graphite, carbon nanotubes, and
graphene nanoribbons, deserve closer experimental verifications. As
to the geometry-diversified electronic excitations, the
optical/magneto-optical measurements have confirmed the
$\pi$-electronic strong peak at middle frequency and the K- and
H-dominated magneto-absorption peaks of the inter-LS transitions in
AB-stacked
graphite,\cite{PRL100;136403,PRB80;161410,PRL102;166401,PRB86;155409,JAP117;112803,
PR138;A197} the low-frequency shoulder structure $\&$ the $\pi$ peaks and the monolayer- and bilayer-like inter-LL absorption frequencies in
few-layer AB stackings,\cite{PRL106;046401,PRB81;155413,PRB83;125302,
PRB85;245410,PRL111;096802,PRL104;176404} the two low-frequency characteristic peaks in trilayer ABC stacking,\cite{PRL104;176404,NatPhys7;944} and the radius-, chirality-, and magnetic-field-dependent absorption peaks in carbon nanotubes.\cite{Science304;1129} Furthermore, converted absorption frequencies for the lowest sombrero-shaped band have been observed by magneto-Raman spectroscopy in ABC-stacked graphene of up to 15 layers,\cite{NanoLett16;3710}. Optical
spectroscopies could be further used to check the intra- and
inter-LS absorption peaks and the beating spectra in
AA-stacked graphite, the transitions of intra-Dirac cone and
intra-LL-group in few-layer AA stackings, the monotonic/complex
magneto-excitation spectra in ABC-stacked graphite/graphenes, and
the edge- and QLL-dependent selection rules in graphene nanoribbons.

The various geometric structures and the diverse intrinsic
properties clearly indicate that the graphite-related systems are
suitable for the development of basic and applied sciences.
Furthermore, the chemical doping of atoms and molecules could
greatly enhance the application ranges. Pristine graphite is the
suitable precursor for the other carbon-related materials. Graphite
fibers are frequently used as reinforced materials because of the
super-excellent mechanical
properties.\cite{ASS401;79,Carbon114;59,CBM135;394,CST138;179}
Graphite intercalation compounds, with tunable carrier densities,
could serve as electrodes,\cite{SSI300;169,SynMet222;351}
superconductors,\cite{NatPhys1;39,NatPhys1;42,STAM9;044102} photo
electro catalytic, electrochemical and biomedical
sensors,\cite{SAABC242;825,BB89;136,BB90;6,JCIS487;149,SABC239;325}
and microbial fuel cells.\cite{IJHE38;15723,BB25;2167,MRB88;188}
Layered graphenes and their compounds, which possess the rich and
controllable electronic and optical properties, are expected to
present the wide-range applications, such as
FETs,\cite{NanoLett12;1324,IEEE30;261,IEEE56;2979,JAP118;244501}
photodetectors,\cite{Nature474;64,OptLett41;816} optical
modulators,\cite{Nature474;64,OptLett41;816} solar
cells,\cite{small11;2963,MatLett192;84,APL88;233506,APL87;203511}
touch-screen panel device,\cite{NatNanotechnol5;574,MatLett107;247}
various sensors,\cite{Carbon50;4228,NanoLett8;3137} high-capacity
batteries,\cite{Nature520;324,JES160;A1781}
supercapacitors,\cite{NatNanotechnol6;496,Carbon49;573} and energy
storages.\cite{PRB81;205406,JAP105;4307} The dimension-enriched
essential properties in 1D carbon nanotubes have the high potentials
in FETs,\cite{APL73;2447,APL108;163104,APL106;213503}
LEDs,\cite{NanoLett11;23,NatPho10;420,NatNanotech5;27}
electroluminescent light emitters,\cite{PRB73;085421,Science300;783}
photodetectors,\cite{ACSNano10;6963,SciRep6;38569,AOM3;989} and
high-performance interconnects.\cite{JMS52;643,JEM44;4825} Moreover,
the finite-size effects of graphene nanoribbons are available in
developing the nanoscaled devices, e.g.,
FETs.\cite{PRL100;1229,Science319;1229,NatNanotech6;45,APLMat3;011101}

The current work is closely related to the layered materials, with
various lattice symmetries, planar/curved structures, stacking
configurations, layer numbers, and dimensions. The emergent group-IV
2D materials, which cover graphene,
silicene,\cite{PRB94;205427,NatNanotech10;227}
germanene,\cite{AdvMater26;4820,PRL108;155501}
tinene,\cite{NatMater14;1020} and monolayer Pb,\cite{PRL111;057005}
are high potential candidates in studying the rich and unique
physical, chemical and material phenomena. Such systems possess a
lot of intrinsic properties in terms of lattice symmetries, planar
or buckled structures, intra- and inter-layer atomic interactions,
single- or multi-orbital chemical bondings,\cite{PRB94;045410}
distinct site energies, and spin-orbital couplings. The complicated
relations among the significantly important interactions are
expected to create the critical Hamiltonians and thus greatly
diversify the essential properties. The generalized tight-binding
model, which is reliable under the uniform/nonuniform magnetic and
electric fields,\cite{PRB83;195405,JVSTB28;386,OptExp22;7473,PRB94;205427,
RSCAdv5;51912,SciRep7;40600,PRB94;045410} deserves further developments to make
thorough and systematic investigations, especially for the
diversities among five layered systems. All the atomic interactions
and external fields could be included in the calculations on
electronic structures and optical properties simultaneously. The
diverse phenomena in group-IV layered systems might become the
main-stream research topics in the near future, such as, orbital-,
spin- and valley-dominated magnetic quantizations, optical and
magneto-optical selection rules, dimensional crossovers,
adatom/molecule doping-induced energy gaps or free carrier
densities, stacking-modulated Dirac-cone structures and quantum Hall
conductivities, and element-dependent  plasmon modes and Landau
dampings.

The combination of the generalized tight-binding model with the
static Kubo formula is suitable for studying the quantum Hall effect
(QHE) in layered materials. It could provide the reliable LL energy
spectra and wave functions even under the complicated anti-crossing
behaviors. As a result, the available inter-LL transitions for the
QHE, the selection rules, are obtained exactly. The study on the
bilayer and trilayer graphenes shows that the various stacking
configurations greatly diversify the quantum transport
properties.\cite{arXiv170401313} The diverse features cover the
non-integer conductivities, the integer conductivities with the
distinct heights, the LL-splitting-induced reduction and complexity
of quantum conductivity, a zero or finite conductivity at the
neutral point, and the well-like, staircase, composite, and abnormal
plateau structures in the magnetic-field-dependencies. Similar
studies on other 2D systems are expected to present more quantum
phenomena.

As for the electronic Coulomb excitations, the delicate random-phase
approximation has been successfully developed for 2D graphene
systems, according to the layer-dependent subenvelope
functions.\cite{PRB74;085406,PLA352;446} This point of view is same
that used in the generalized tight-binding model under various
external fields. That is, their combination could include the
intralyer and interlayer atomic interactions, the intralayer and
interlayer Coulomb interactions, and the magnetic and electric
fields
simultaneously.\cite{PLA352;446,PRB74;085406,ACSNano5;1026,PRB89;165407,
PRB86;125434,
SciRep3;1368,AOP339;298,NJP16;125002,RSCAdv5;51912,PRB94;205427,SciRep7;40600}
Up to now, the systematic studies on single-particle and collective
excitations (electron-hole pairs and plasmon modes) are made for
bilayer AA and AB stackings without/with magnetic
quantization,\cite{PRB74;085406,PRB89;165407} monolayer graphene
under a magnetic field,\cite{ACSNano5;1026} few-layer graphenes in
the presence of an electric
field,\cite{PRB86;125434,SciRep3;1368,AOP339;298} silicene
without/with gate voltage or magnetic
quantization,\cite{NJP16;125002,RSCAdv5;51912,PRB94;205427} and
germanene.\cite{SciRep7;40600} Such systems might exhibit the
unusual excitation phase diagrams associated with transferred
momenta and energies, being never revealed in 2D electron gas
systems. The many-particle phenomena arising from the
electron-electron interactions in emergent 2D materials are worthy
of thorough investigations.

\renewcommand{\baselinestretch}{0.2}

\begin{figure}
\centering
\includegraphics[width=0.9\linewidth]{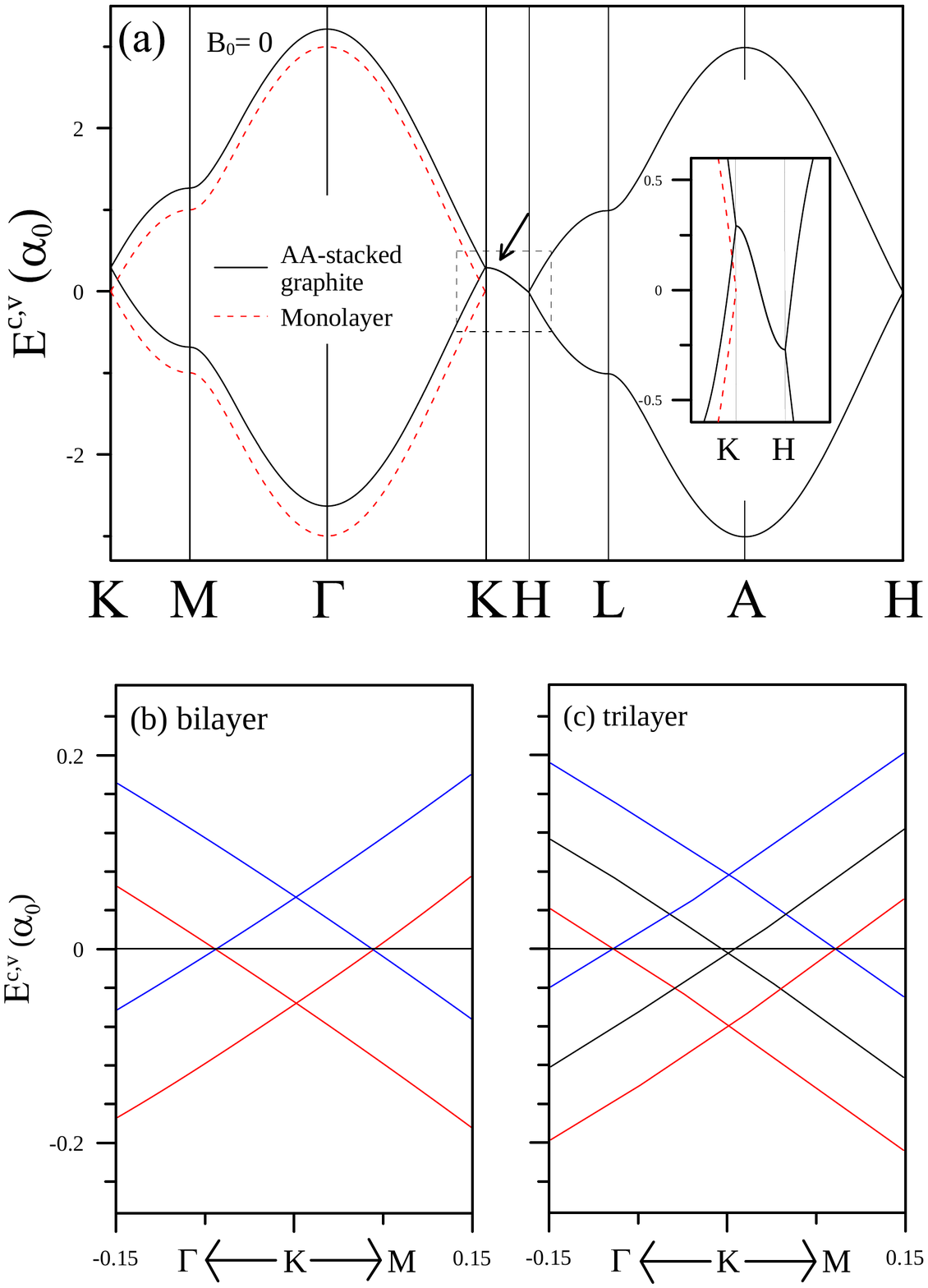}
\caption{Band structures for AA-stacked (a) graphite and monolayer
graphene; (b) bilayer and (c) trilayer systems.}
\label{fig:graph}
\end{figure}

\begin{figure}
\centering
\includegraphics[width=0.9\linewidth]{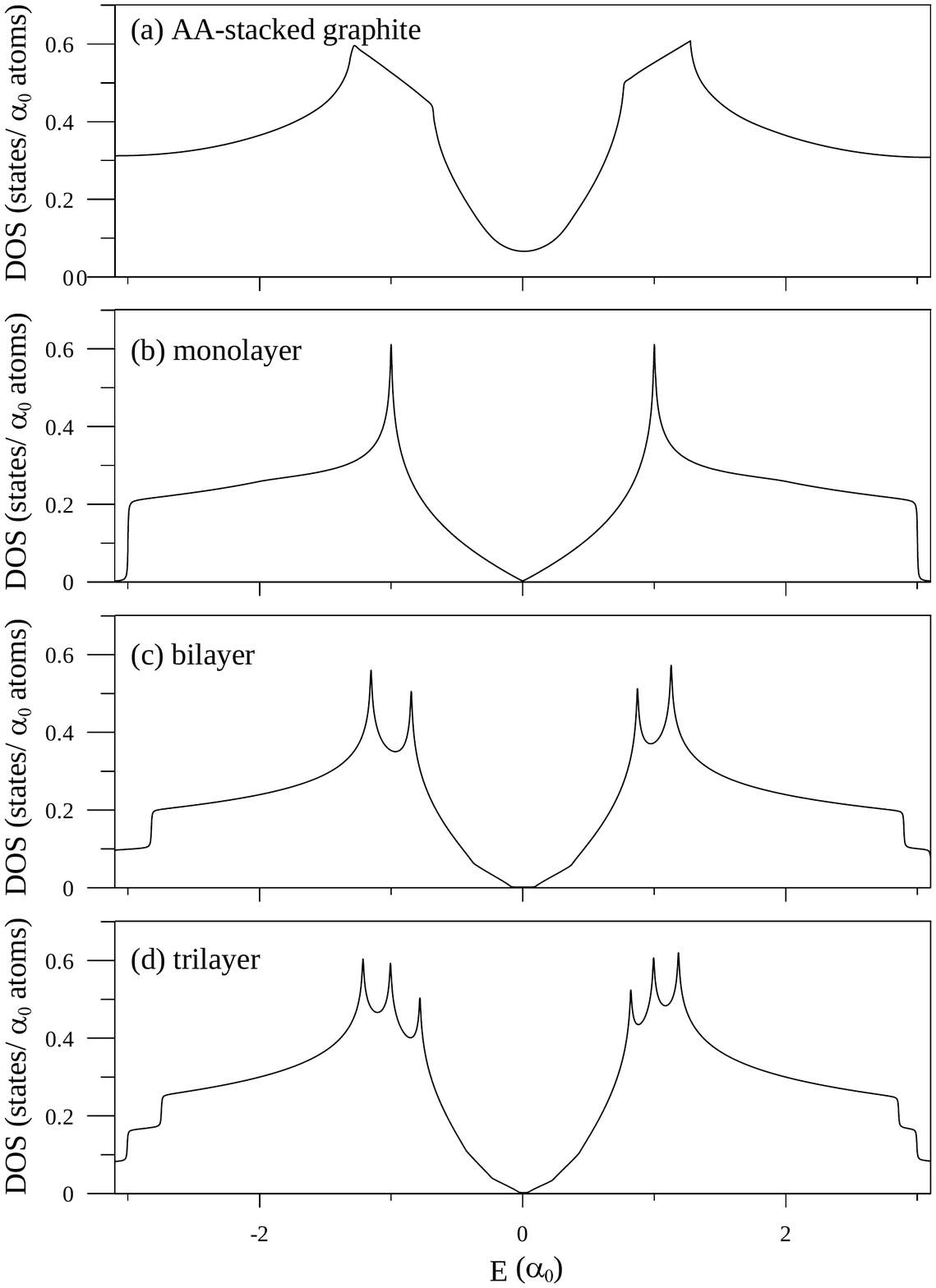}
\caption{Density of states for (a) simple hexagonal graphite, and (b)
monolayer, (c) bilayer and (d) trilayer  graphenes.}
\label{fig:graph}
\end{figure}

\begin{figure}
\centering
\includegraphics[width=0.9\linewidth]{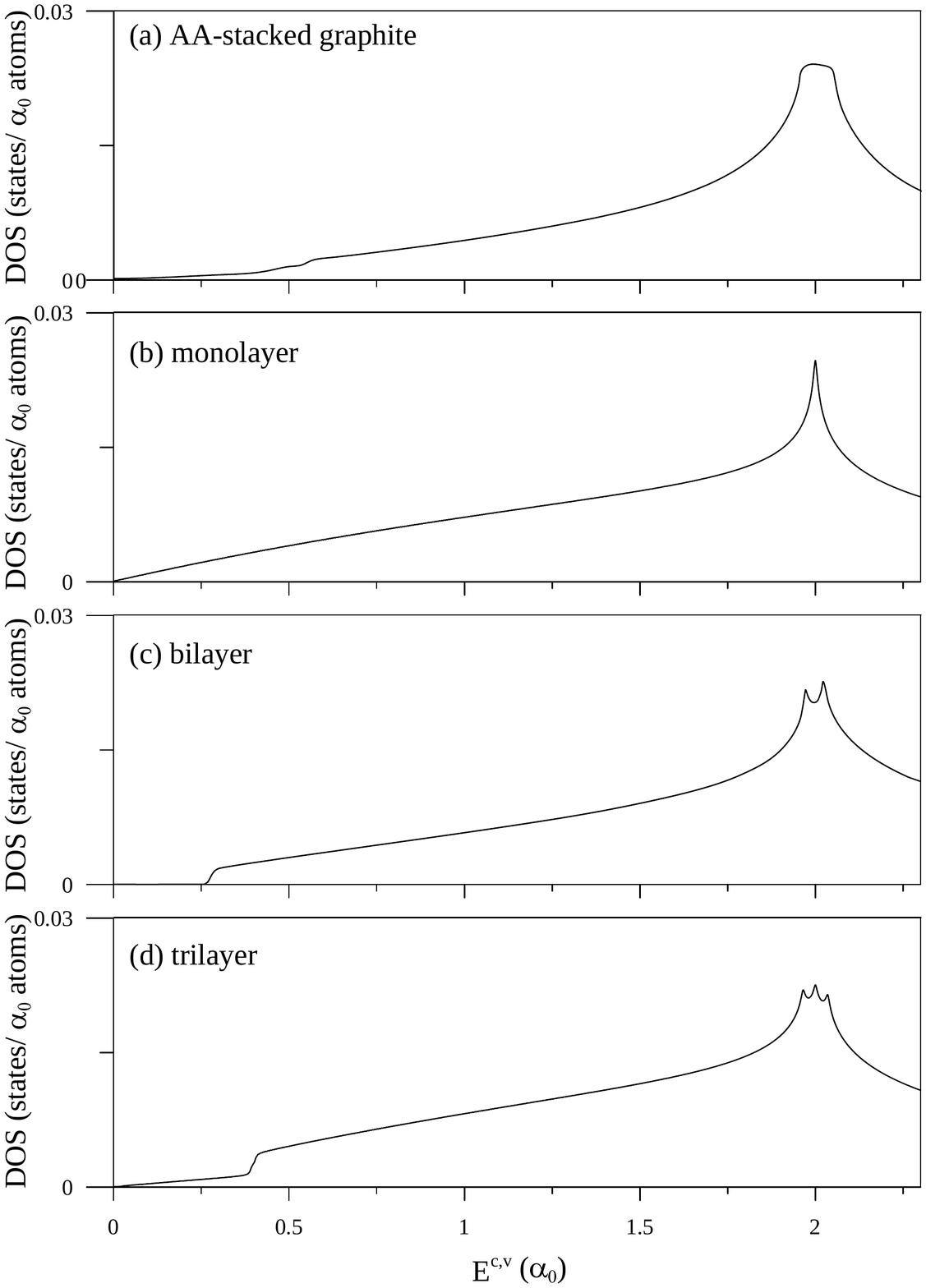}
\caption{Optical absorption spectra of the AA-stacked (a) graphite,
(b) monolayer, (c) bilayer, and (d) trilayer graphenes.}
\label{fig:graph}
\end{figure}

\begin{figure}
\centering
\includegraphics[width=0.9\linewidth]{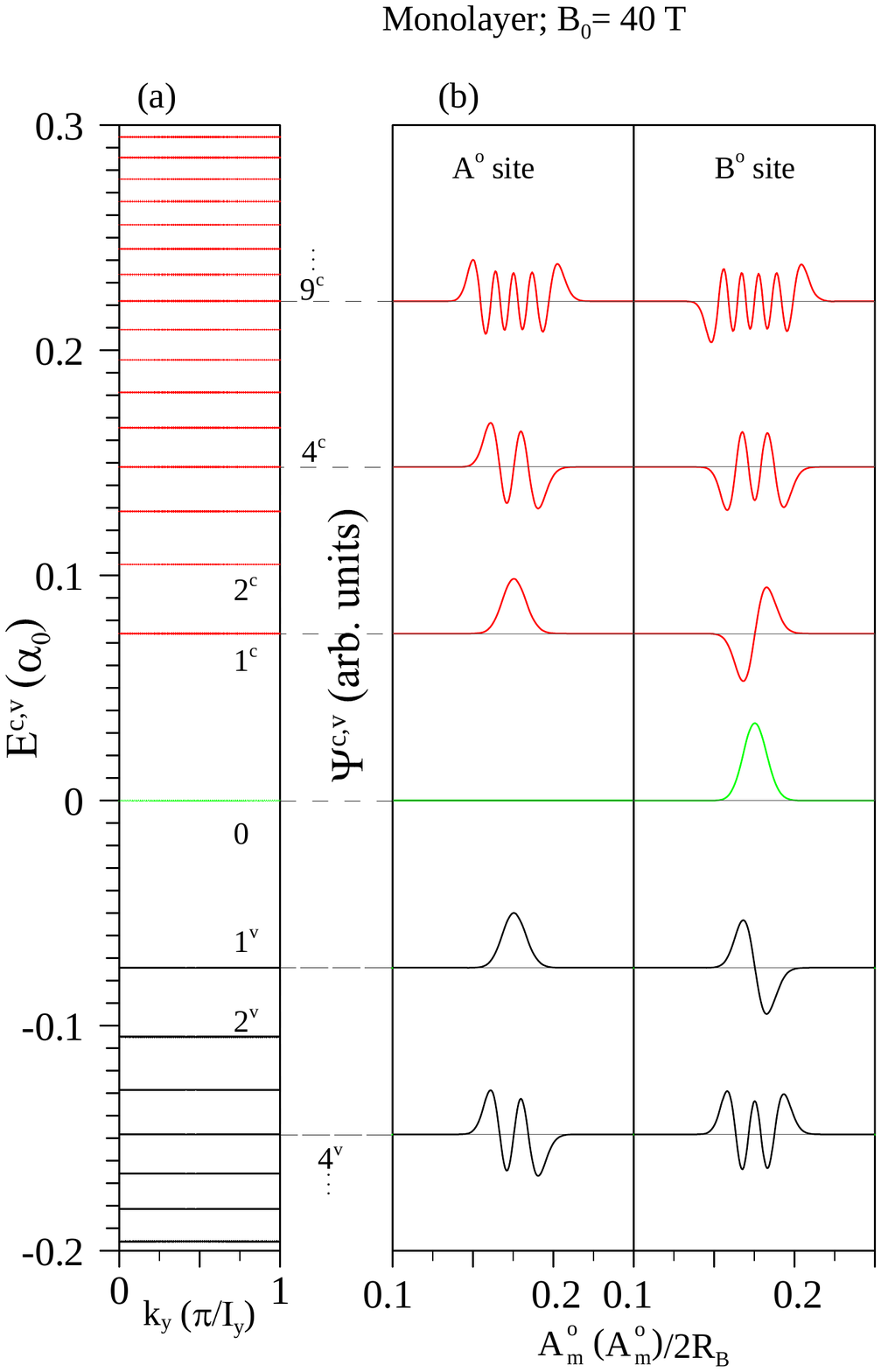}
\caption{The Landau levels of monolayer graphene at ${B_0=40}$ T: (a)
energy spectrum, and (b) amplitudes of subenvelope functions at two
sublattices.}
\label{fig:graph}
\end{figure}

\begin{figure}
\centering
\includegraphics[width=0.9\linewidth]{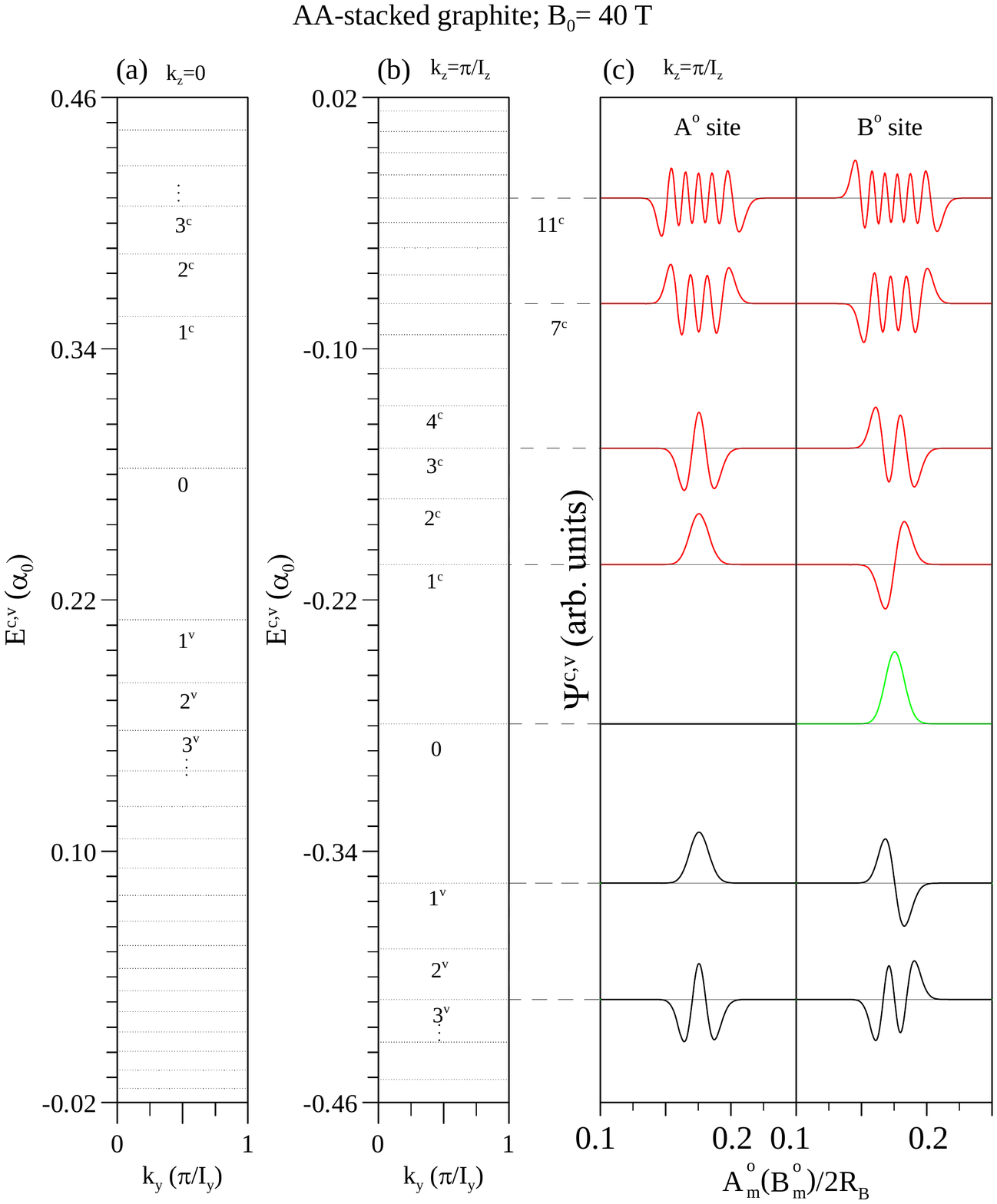}
\caption{The Landau subbands of simple hexagonal graphite at
${B_0=40}$ T:  energy spectrum corresponding (a) ${k_z=0}$ $\&$ (b)
${k_z=\pi\,/I_z}$, and  (c) the amplitudes of subenvelope functions
of the former.}
\label{fig:graph}
\end{figure}

\begin{figure}
\centering
\includegraphics[width=0.9\linewidth]{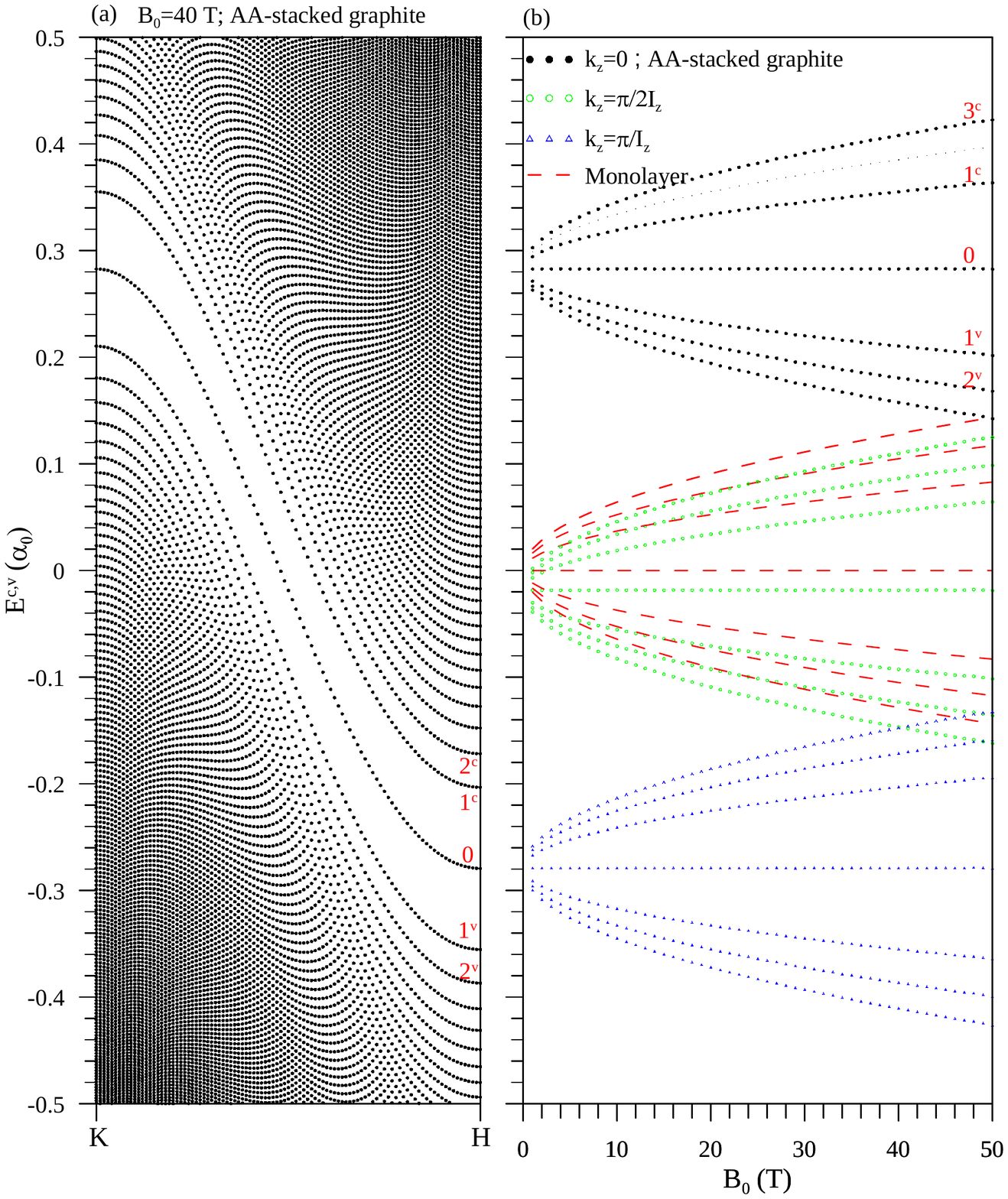}
\caption{The LS energy spectra of the AA-stacked graphite (a) along
the KH dircetion at  ${B_0=40}$ T and (b) for the $B_0$-dependence
at various $k_z$s. Also shown in (b) is that of monolayer graphene.}
\label{fig:graph}
\end{figure}

\begin{figure}
\centering
\includegraphics[width=0.9\linewidth]{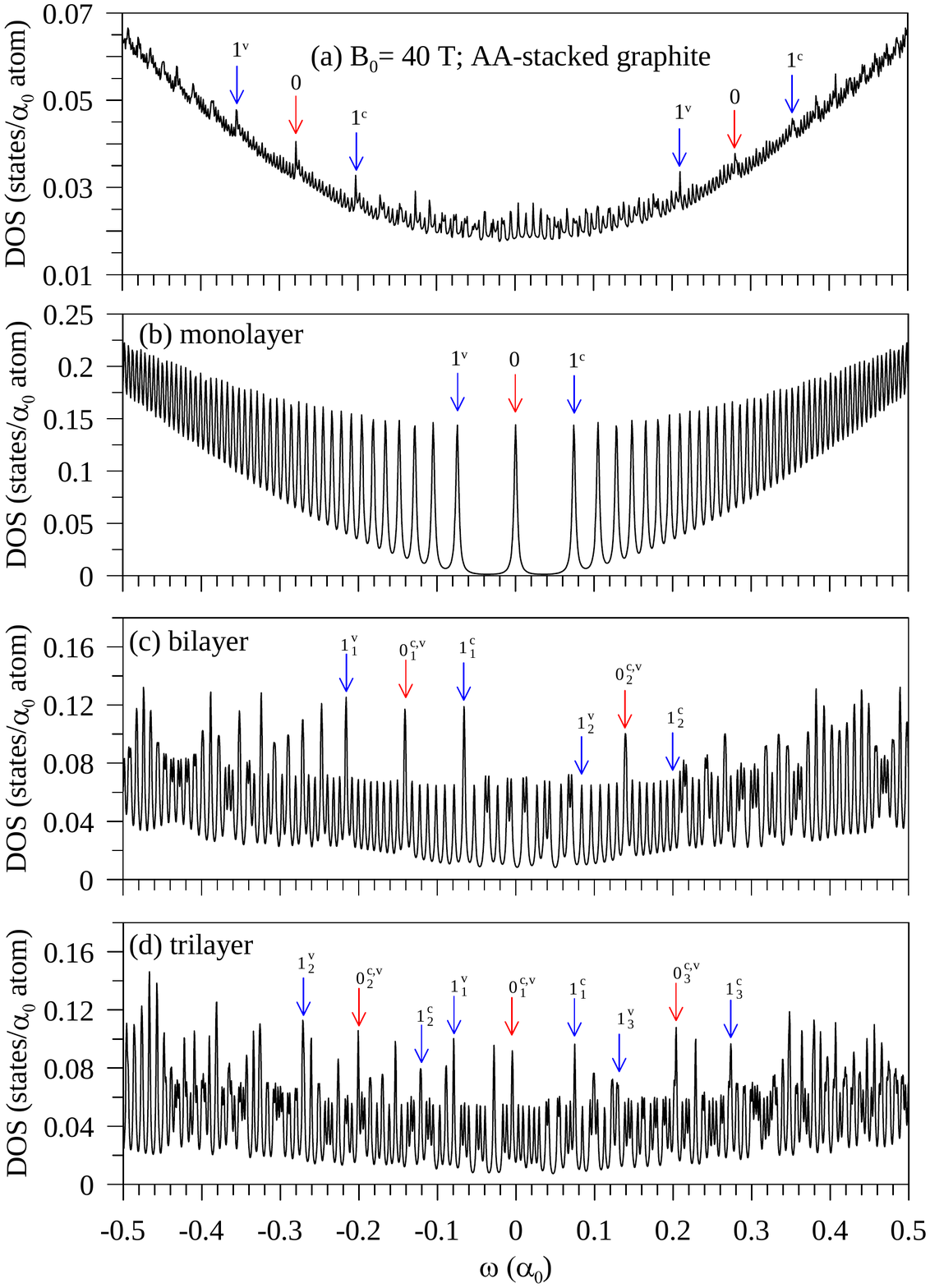}
\caption{Magneto-electronic density of states at ${B_0=40}$ T for (a)
simple hexagonal graphite, and (b) monolayer, (c) bilayer and (d)
trilayer graphenes.}
\label{fig:graph}
\end{figure}

\begin{figure}
\centering
\includegraphics[width=0.9\linewidth]{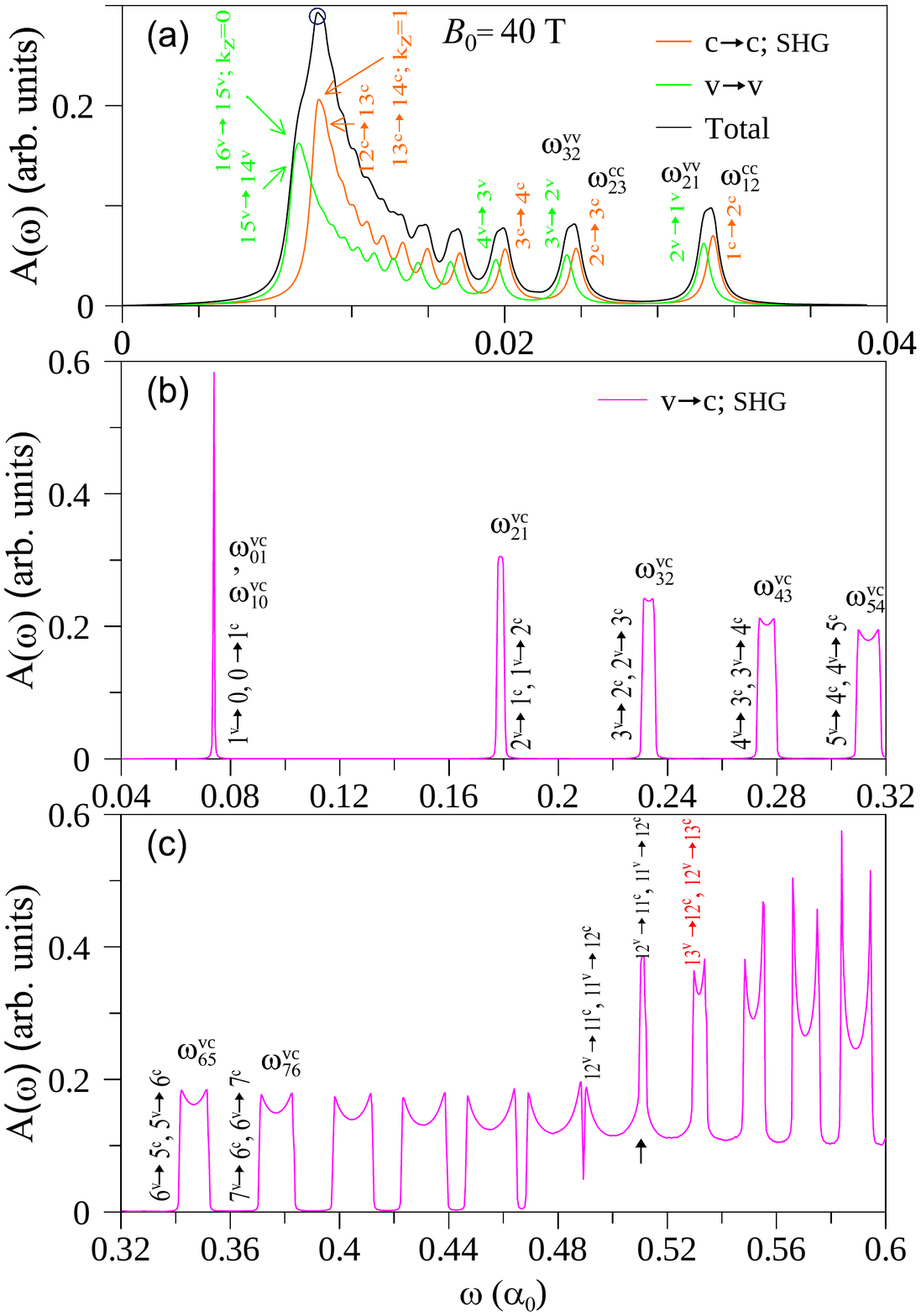}
\caption{Magneto-optical absorption spectrum of simple hexagonal
graphite in (a) to (c) within different frequency ranges at
${B_0=40}$ T.}
\label{fig:graph}
\end{figure}

\begin{figure}
\centering
\includegraphics[width=0.9\linewidth]{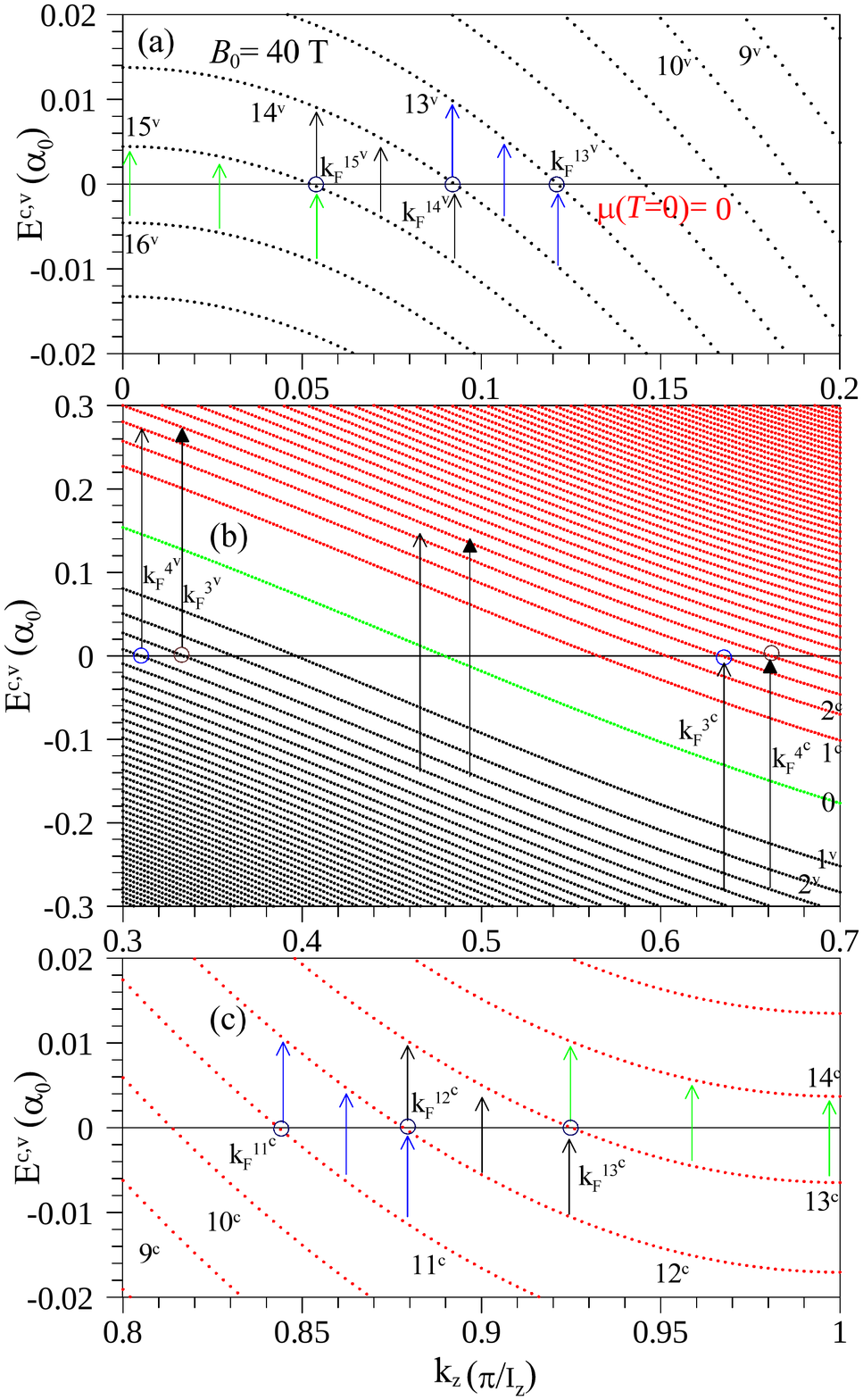}
\caption{Vertical optical transitions due to the Landau subbands
near (a) ${k_z=0}$, (b) ${k_z=\pi\,/2I_z}$, and (c)
${k_z=\pi\,/I_z}$.}
\label{fig:graph}
\end{figure}

\begin{figure}
\centering
\includegraphics[width=0.9\linewidth]{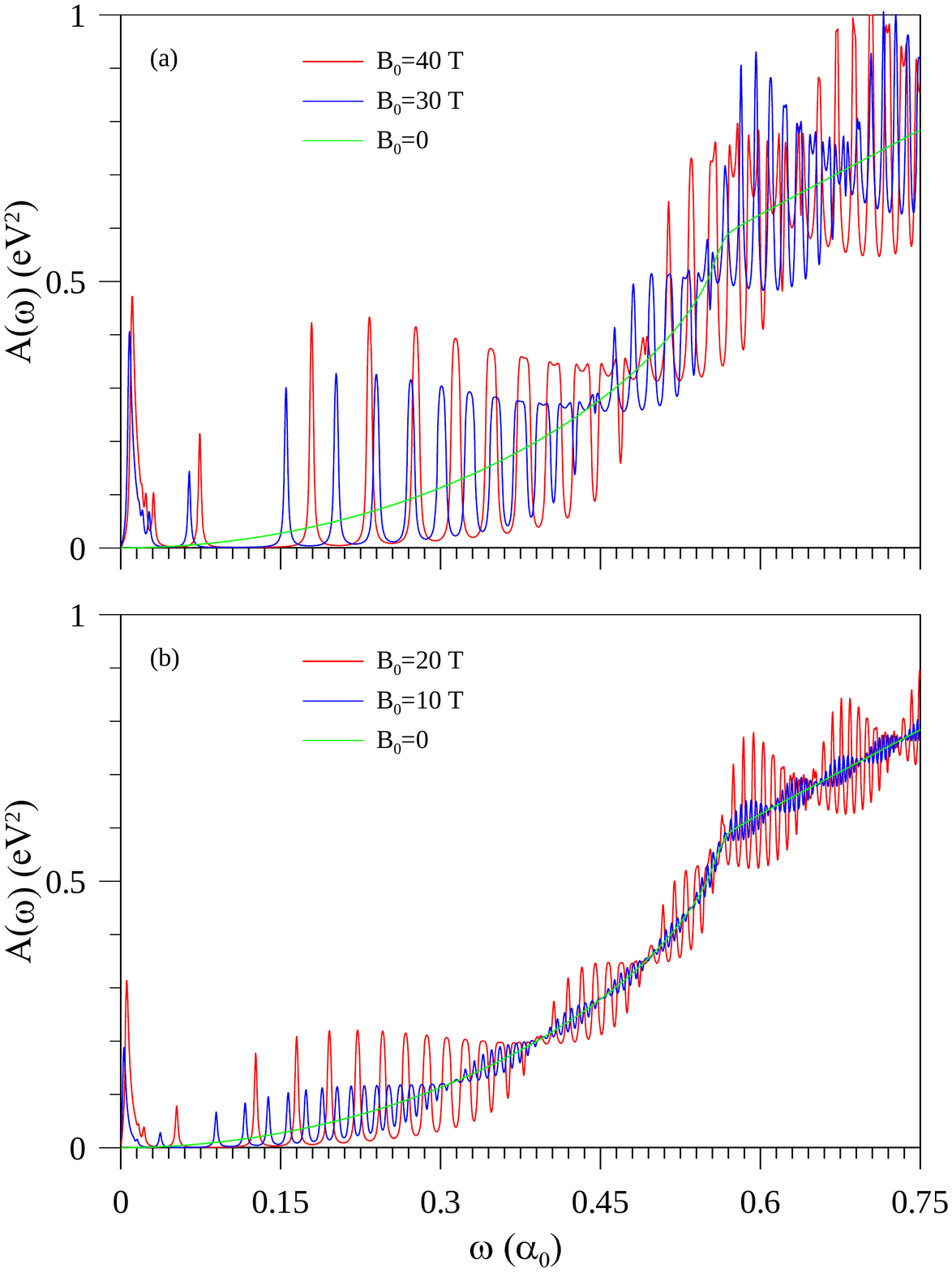}
\caption{The beating magneto-absorption spectra of simple hexagonal
graphite for various field strengths in (a) and (b).}
\label{fig:graph}
\end{figure}

\begin{figure}
\centering
\includegraphics[width=0.9\linewidth]{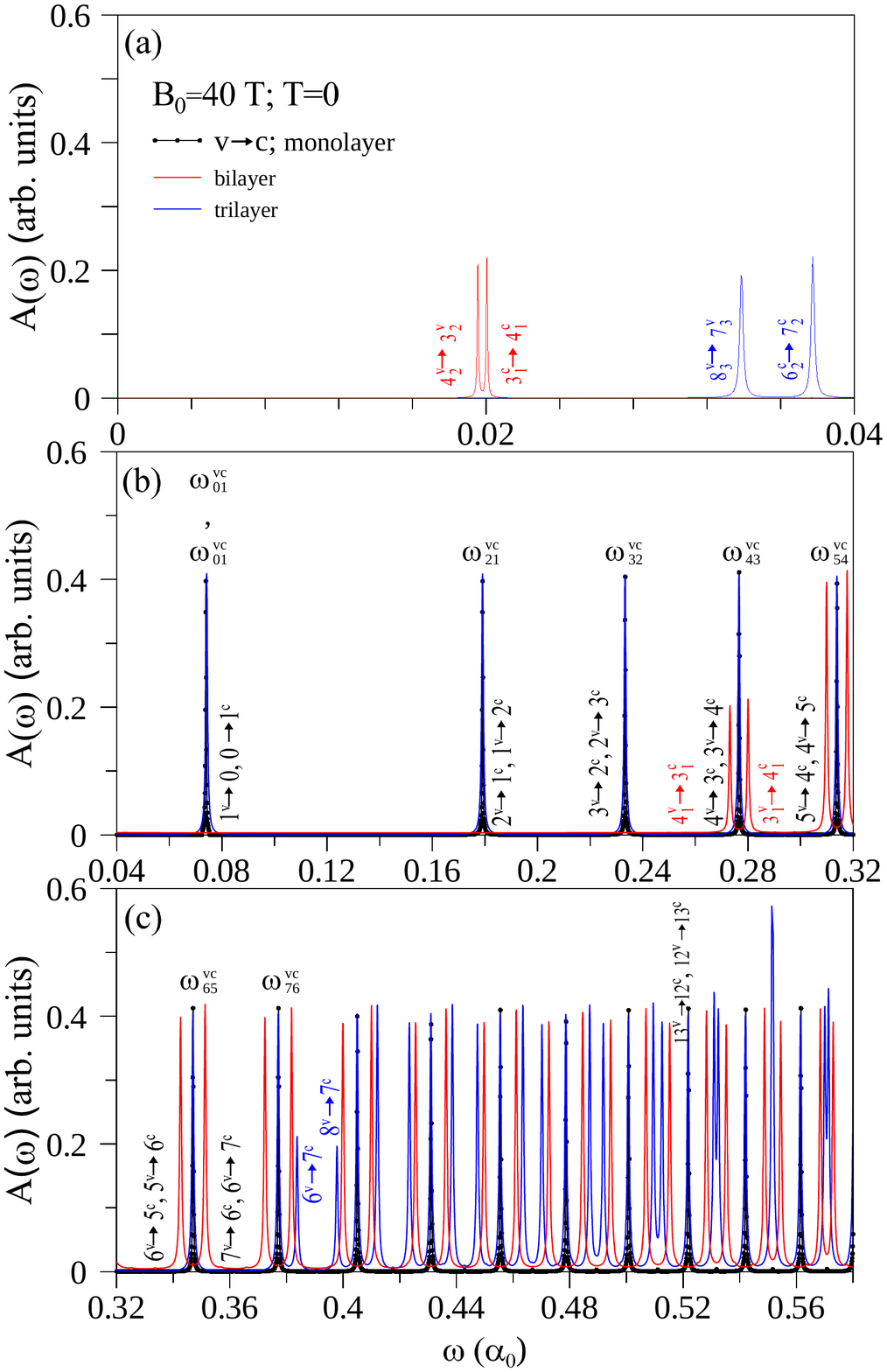}
\caption{Magneto-optical absorption spectra of monolayer, bilayer
and trilayer graphenes in (a) to (c) within different frequency
ranges at ${B_0=40}$ T.}
\label{fig:graph}
\end{figure}

\begin{figure}
\centering
\includegraphics[width=0.9\linewidth]{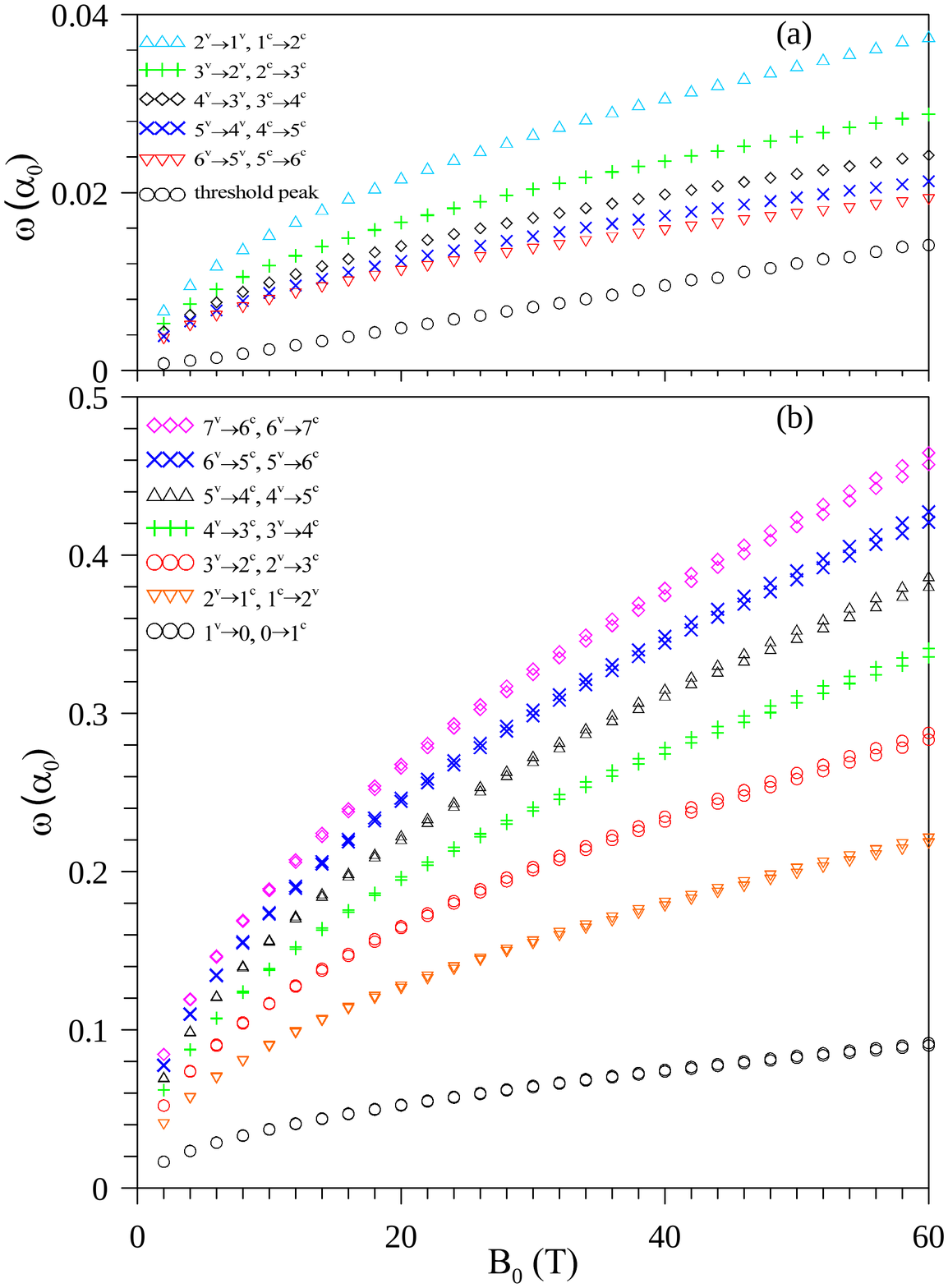}
\caption{The $B_0$-dependent absorption frequencies of simple
hexagonal graphite corresponding to (a) intraband and (b) interband
excitation channels.}
\label{fig:graph}
\end{figure}

\begin{figure}
\centering
\includegraphics[width=0.9\linewidth]{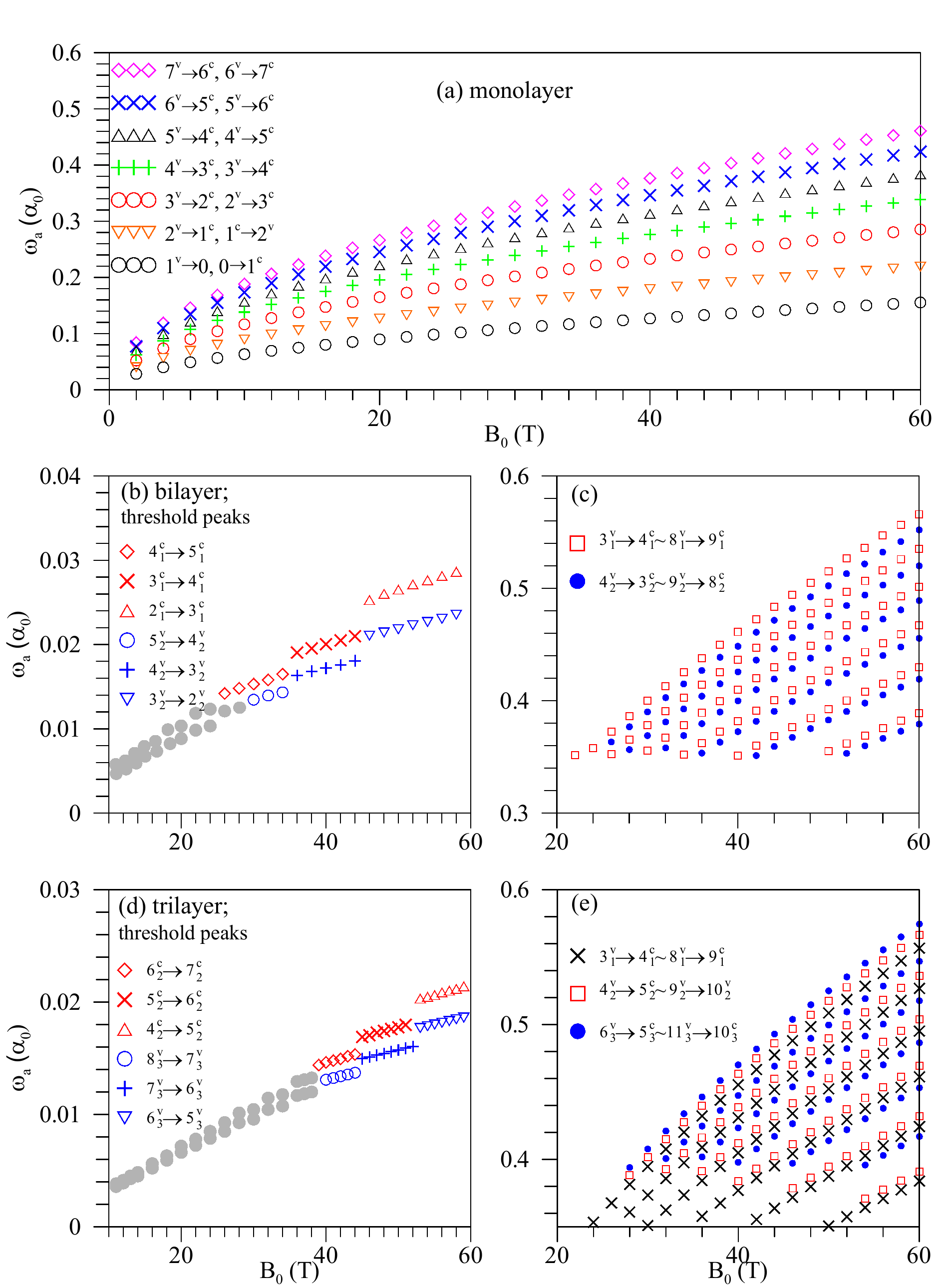}
\caption{The magneto-absorption frequencies for (a) monolayer,
(b)-(c) bilayer and (d)-(e) trilayer graphenes. The discontinuous
$B_0$-dependence of the threshold channel is shown in (b) and (d).}
\label{fig:graph}
\end{figure}

\begin{figure}
\centering
\includegraphics[width=0.9\linewidth]{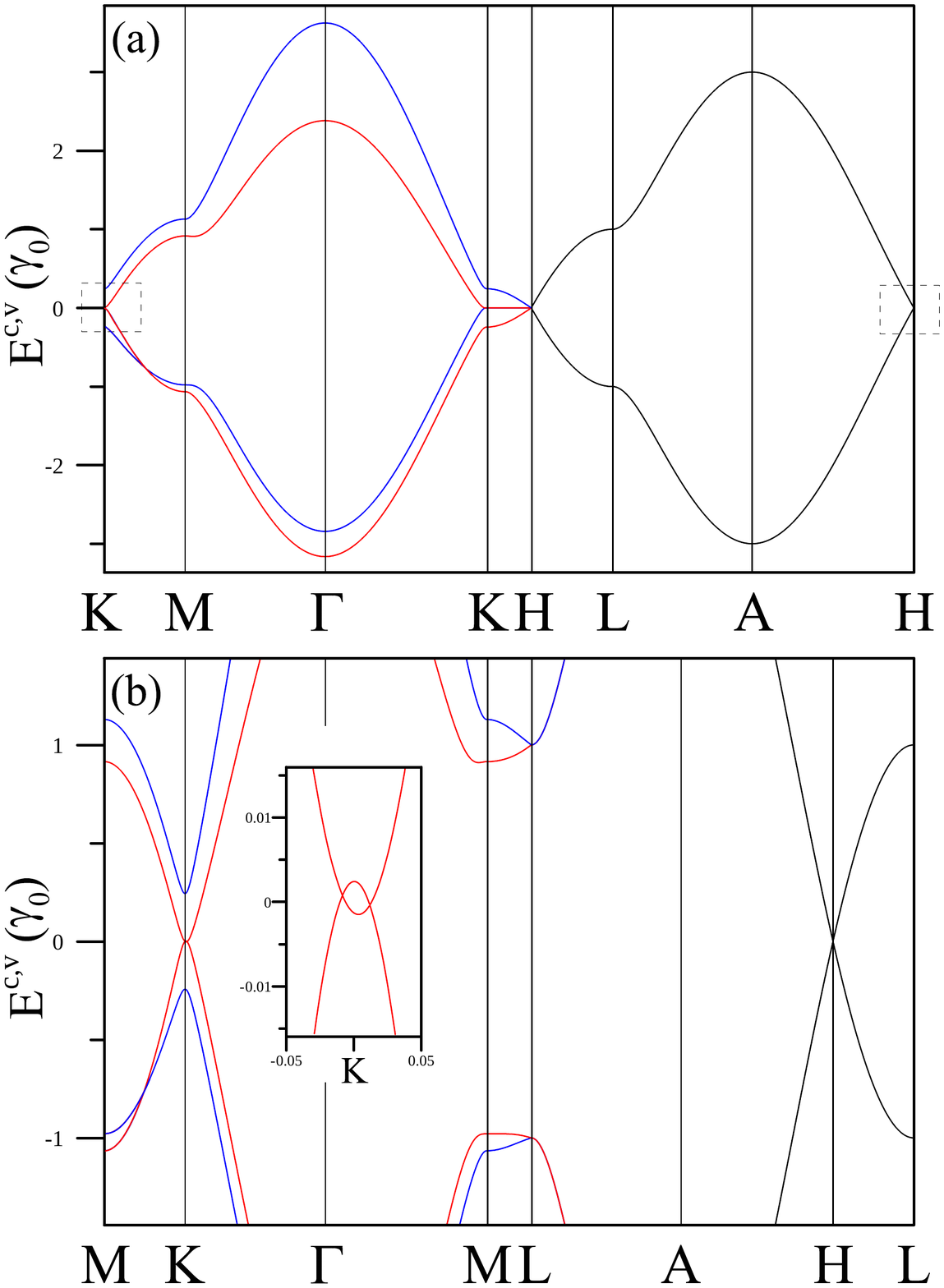}
\caption{Band structures for AB-stacked (a) graphite and (b) the
zoomed-in view at low energies near the K and H points.}
\label{fig:graph}
\end{figure}

\begin{figure}
\centering
\includegraphics[width=0.9\linewidth]{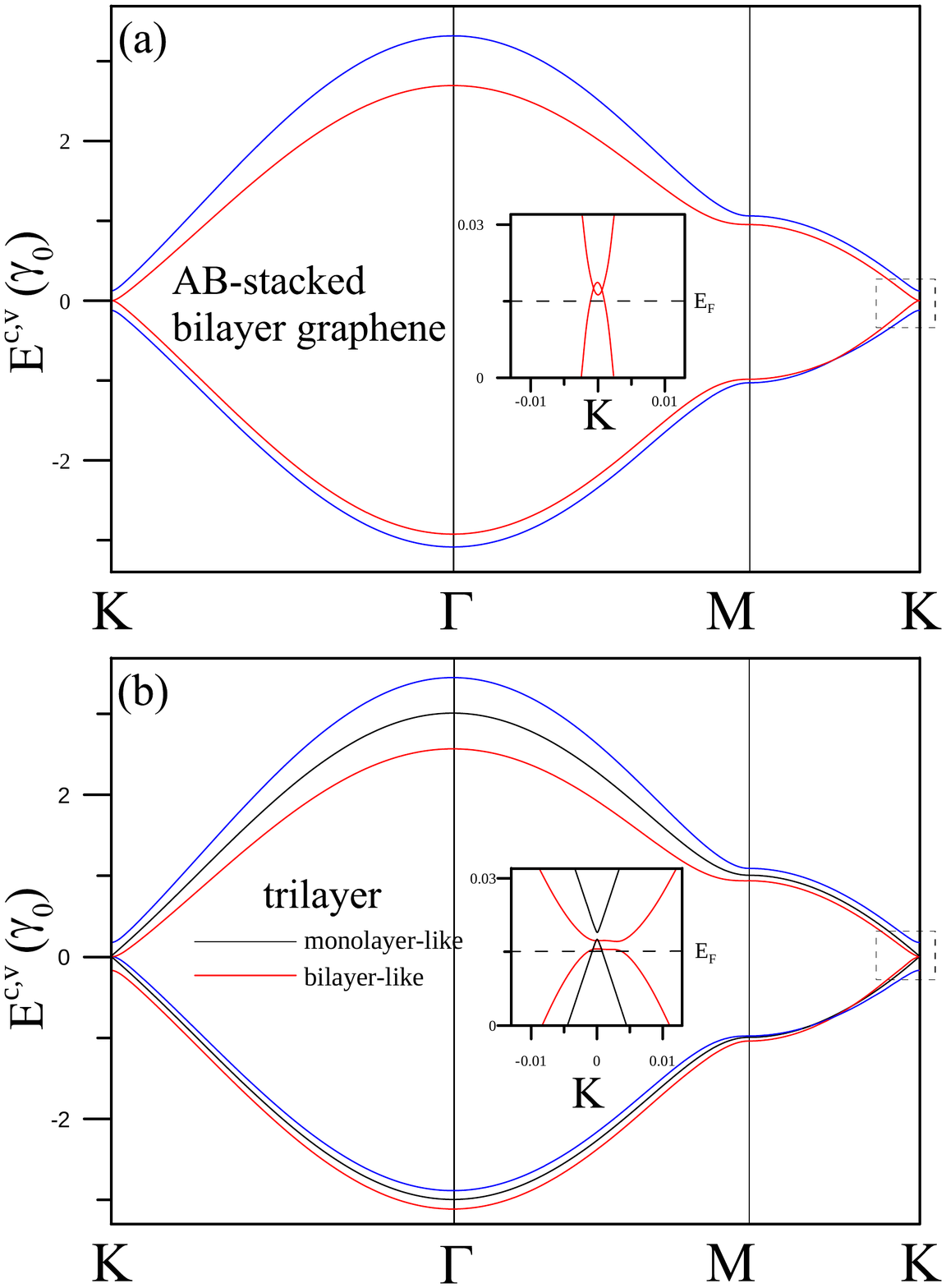}
\caption{Band structures of AB-stacked (a) bilayer and (b) trilayer
graphenes.}
\label{fig:graph}
\end{figure}

\begin{figure}
\centering
\includegraphics[width=0.9\linewidth]{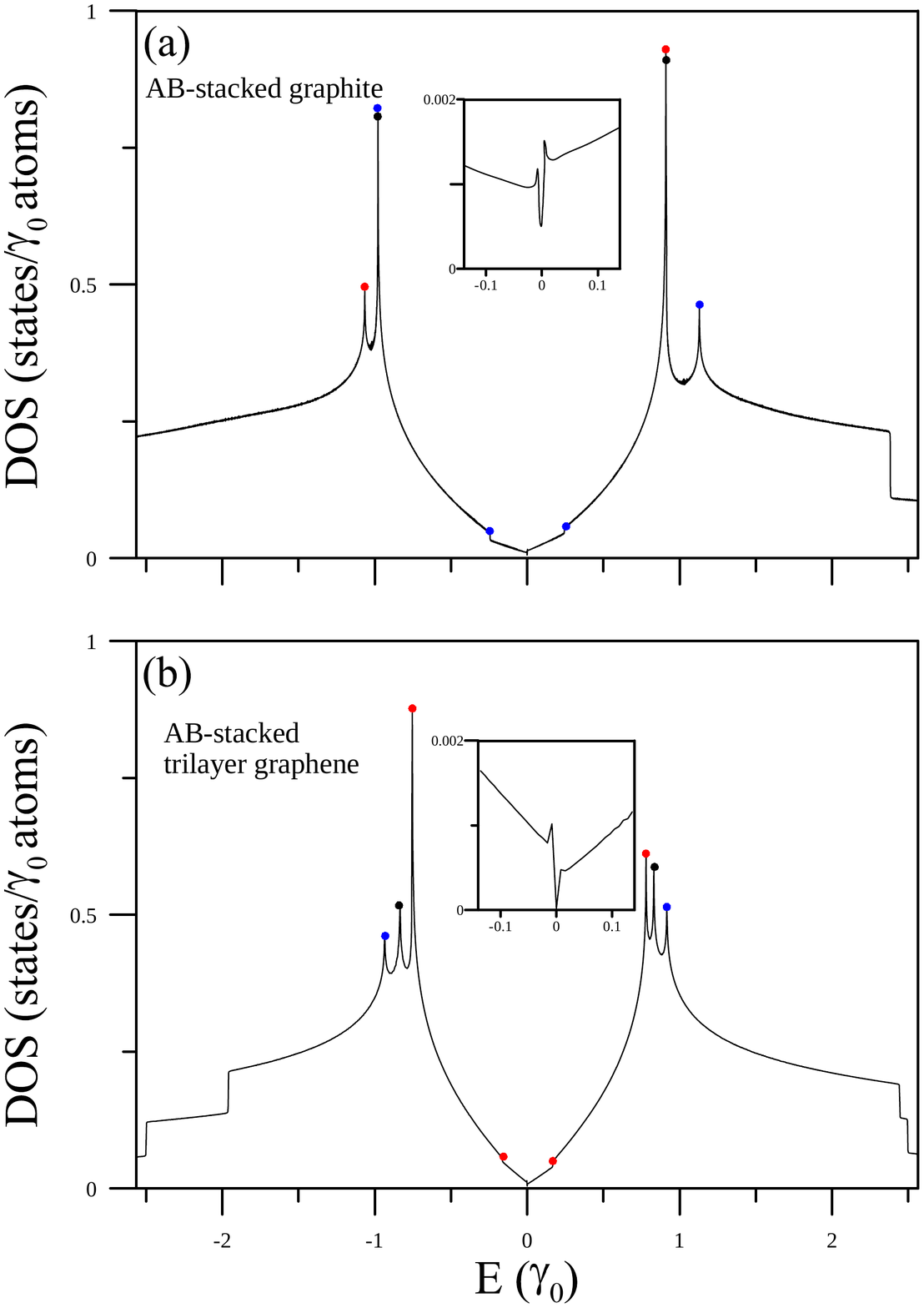}
\caption{The DOS of AB-stacked (a) graphite and (b) trilayer graphenes.}
\label{fig:graph}
\end{figure}

\clearpage

\begin{figure}
\centering
\includegraphics[width=0.9\linewidth]{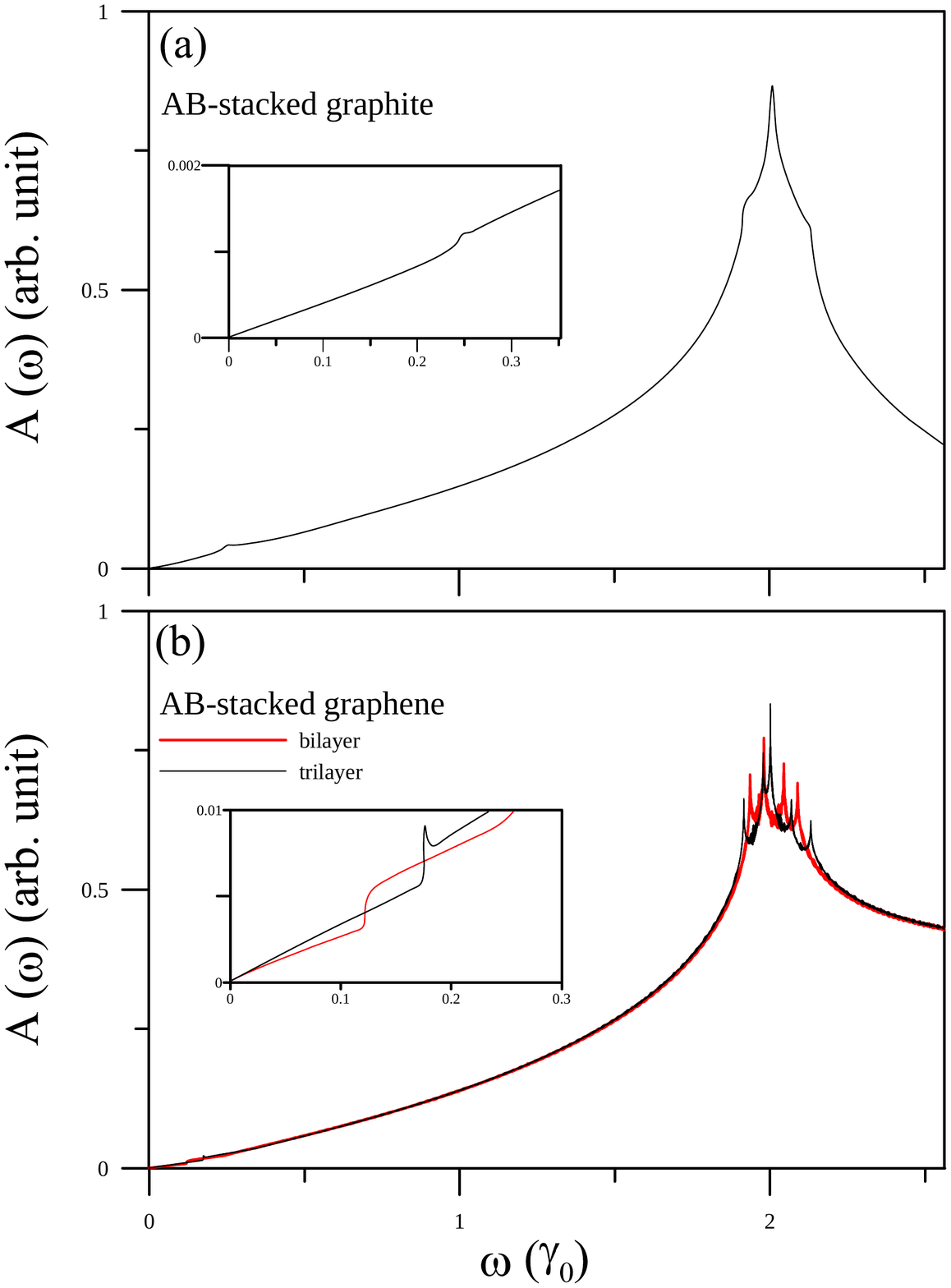}
\caption{Optical absorption spectra of Bernal (a) graphite and (b) graphene.}
\label{fig:graph}
\end{figure}

\begin{figure}
\centering
\includegraphics[width=0.9\linewidth]{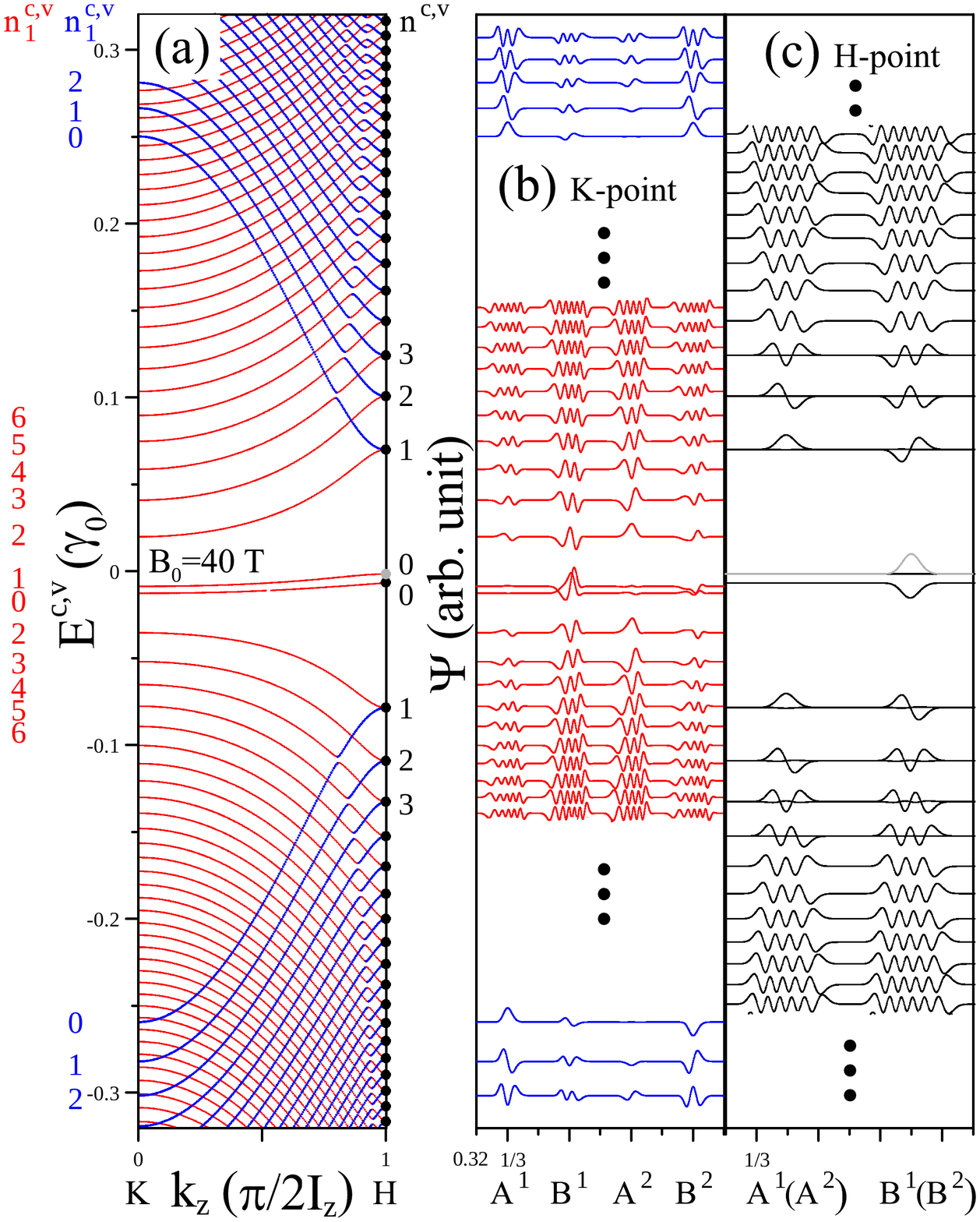}
\caption{(a) Landau subbands of Bernal graphite. The subenvelope
functions are shown for the (b) K and (c) H points.}
\label{fig:graph}
\end{figure}

\begin{figure}
\centering
\includegraphics[width=0.9\linewidth]{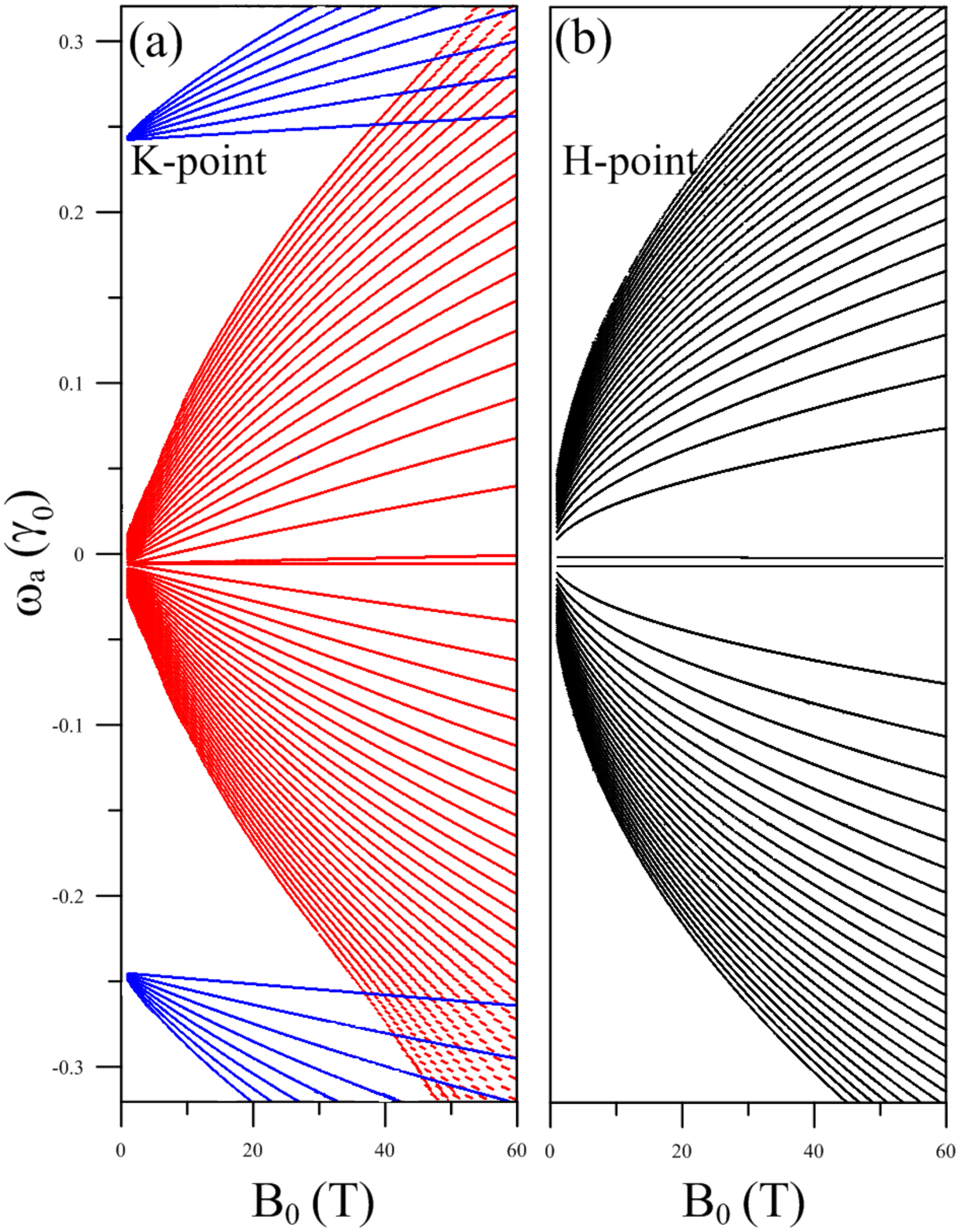}
\caption{$B_{0}$-dependent energies of the (a) K-point and (b)
H-point LSs.}
\label{fig:graph}
\end{figure}

\begin{figure}
\centering
\includegraphics[width=0.9\linewidth]{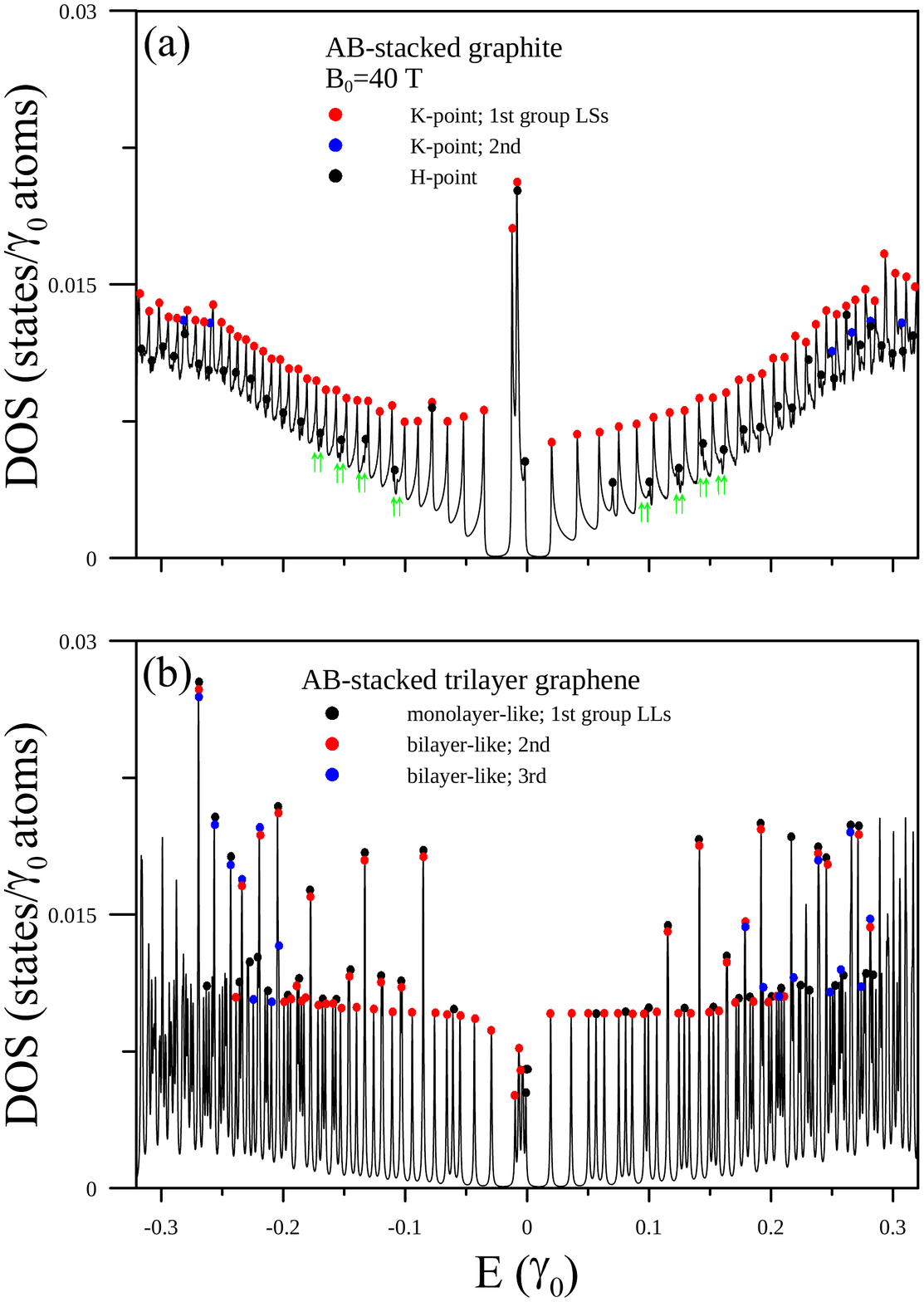}
\caption{DOS of Bernal graphite at $B_{0}=40$ T. Also shown is
that of Bernal trilayer graphene.}
\label{fig:graph}
\end{figure}

\begin{figure}
\centering
\includegraphics[width=0.9\linewidth]{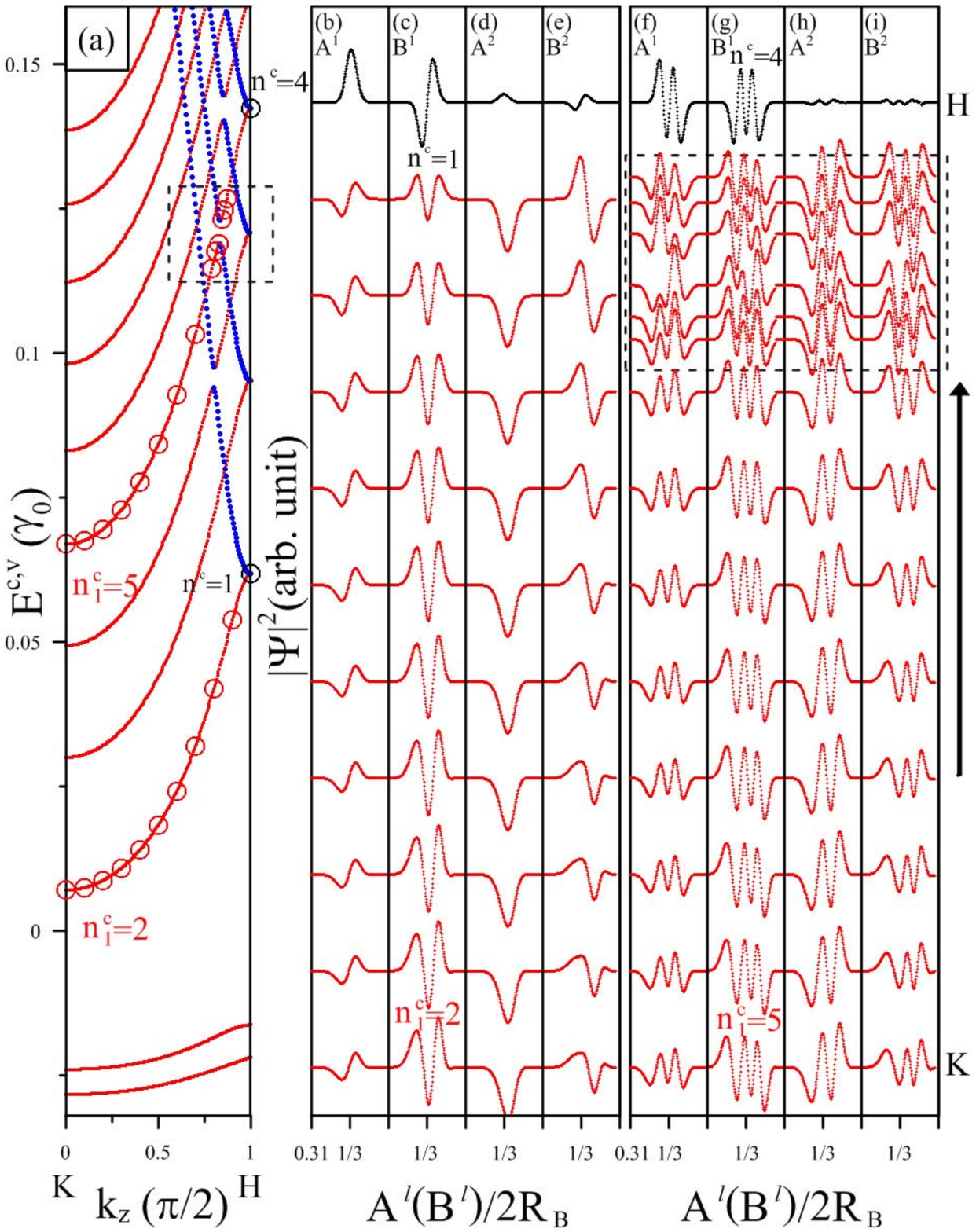}
\caption{(a) The LS anticrossing pattern of Bernal graphite. (b)-(i)
he evolution of the subenvelope functions along $k_{z}$ for the
low-lying LSs. The dashed rectangular region indicates the
hybridized subenvelope functions of the anticrossing LSs.}
\label{fig:graph}
\end{figure}

\begin{figure}
\centering
\includegraphics[width=0.9\linewidth]{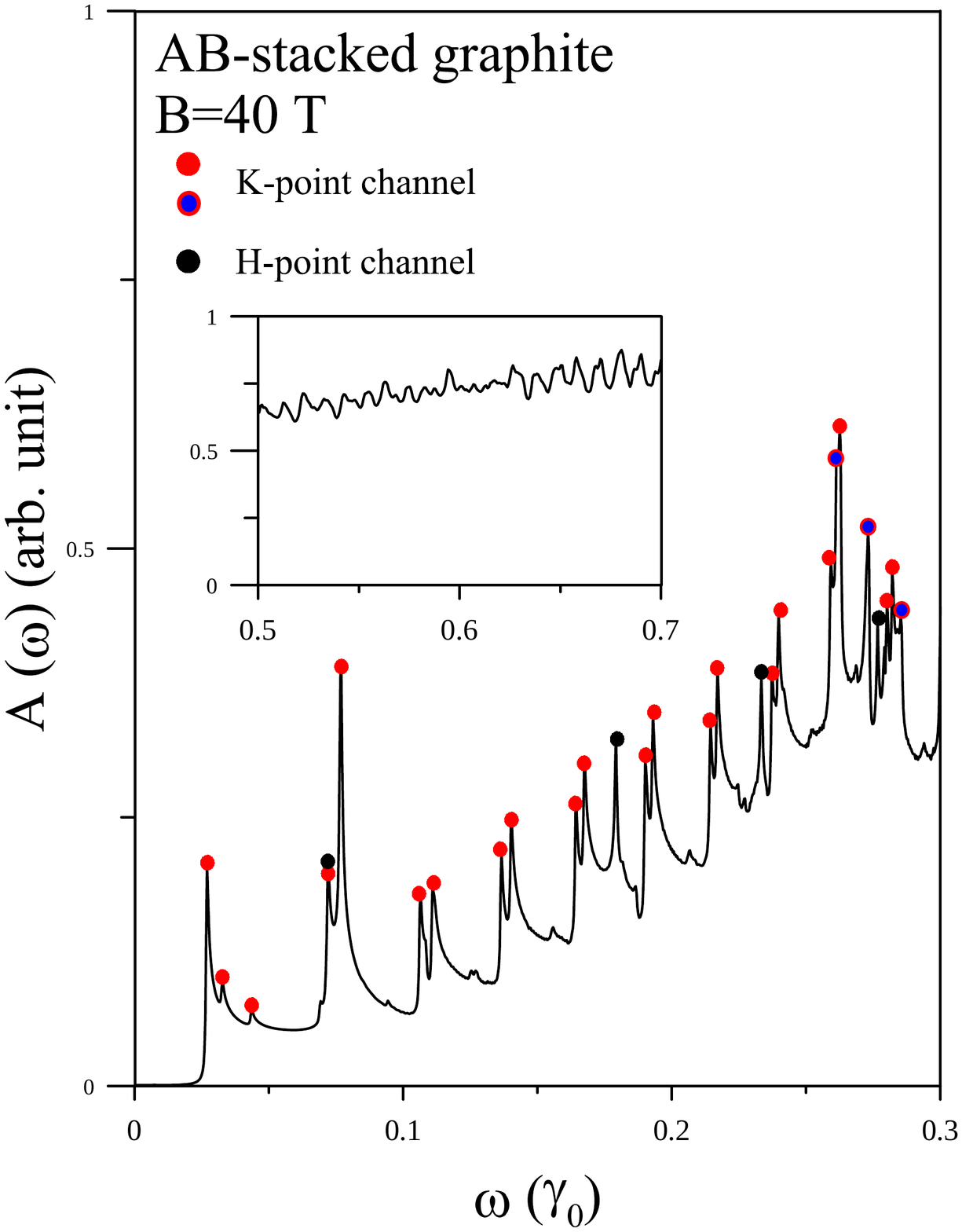}
\caption{Magneto-absorption spectra of Bernal graphite. The
absorption peaks corresponding to the transitions of massive
(massless) Dirac fermions near the K (H) point are marked by
red (black) dots.}
\label{fig:graph}
\end{figure}

\begin{figure}
\centering
\includegraphics[width=0.9\linewidth]{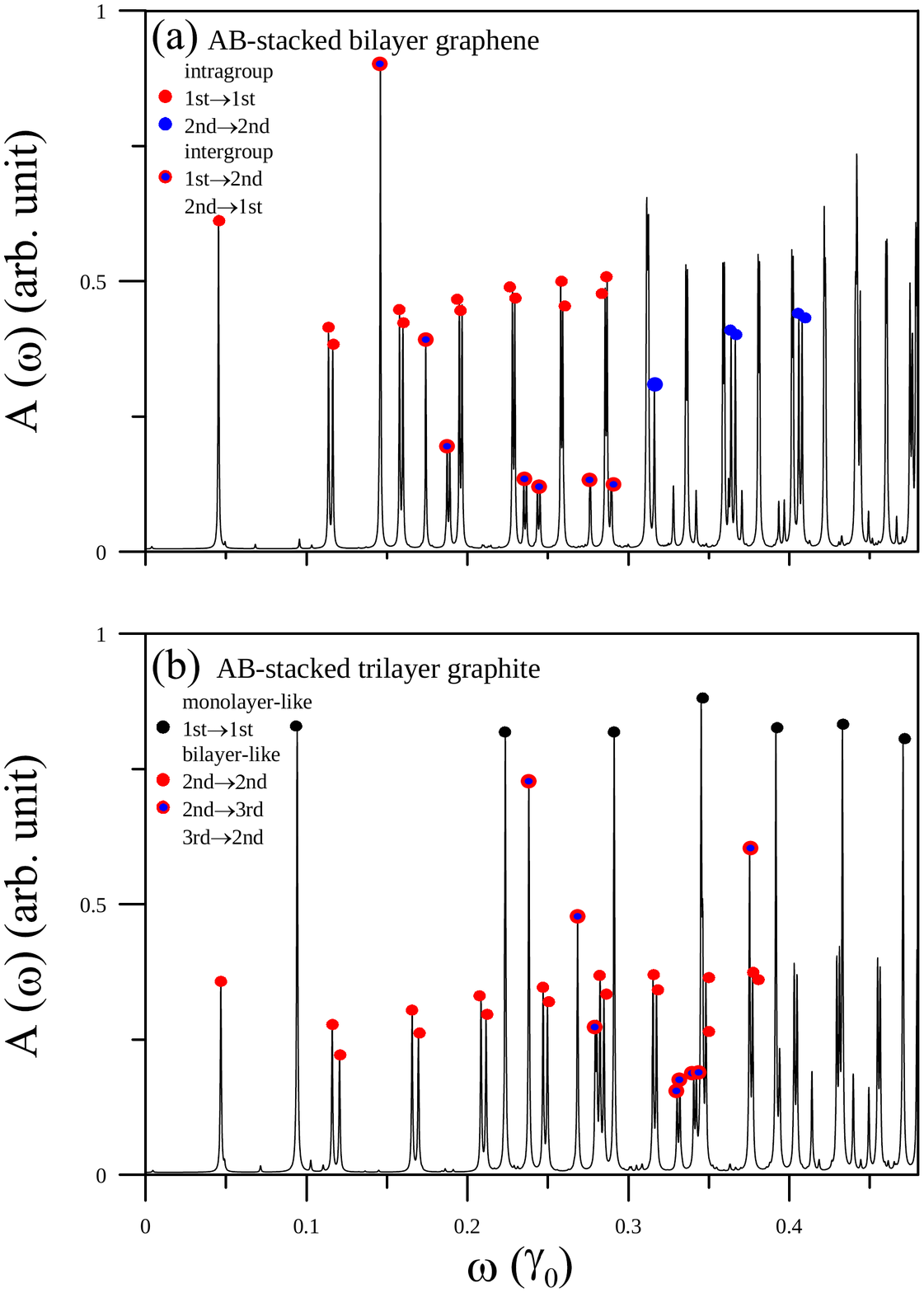}
\caption{Magneto-absorption spectra of AB-stacked (a) bilayer and
(b) trilayer graphenes.}
\label{fig:graph}
\end{figure}

\begin{figure}
\centering
\includegraphics[width=0.9\linewidth]{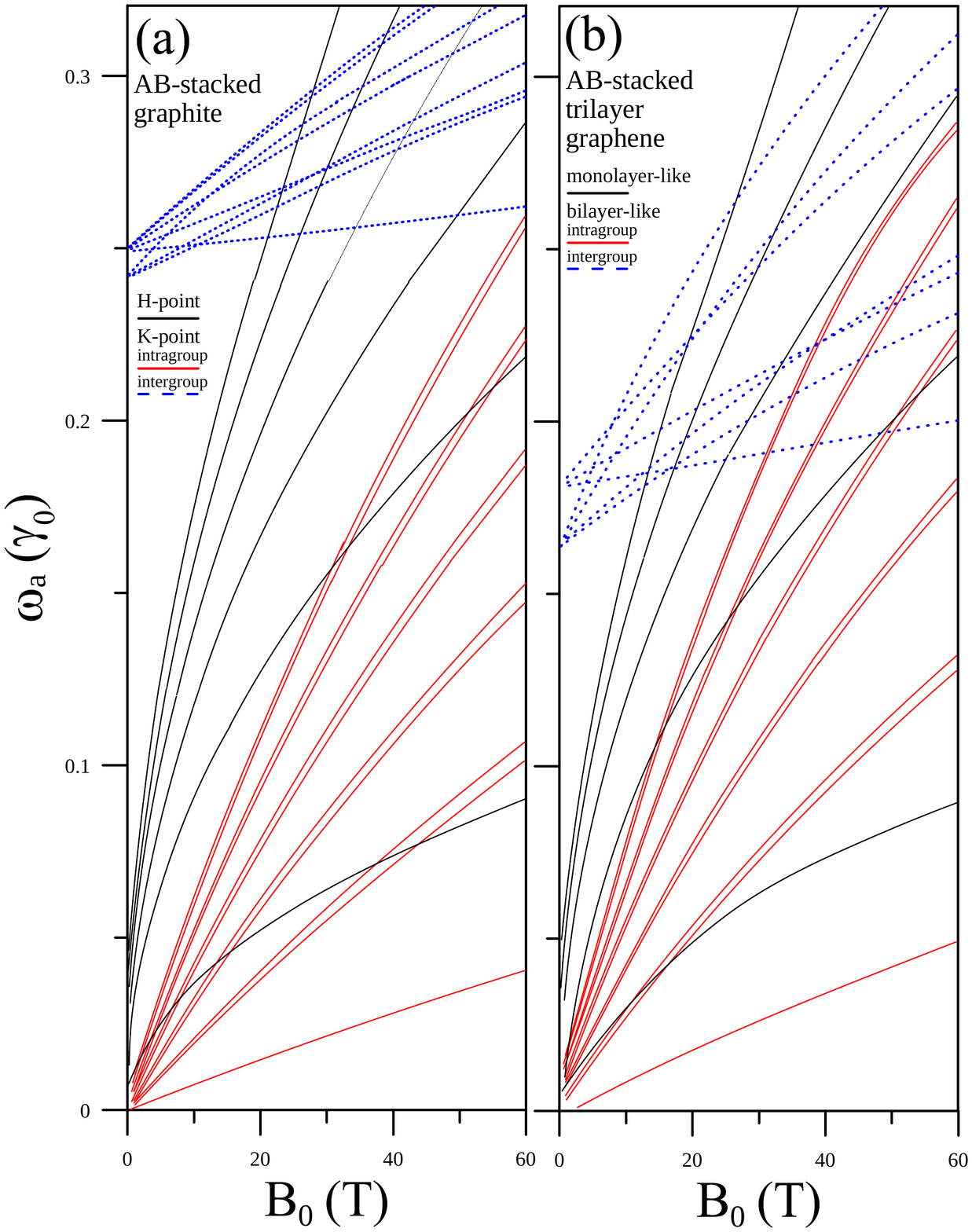}
\caption{$B_{0}$-dependent absorption frequencies of Bernal
(a) graphite and (b) bilayer and trilayer graphenes.}
\label{fig:graph}
\end{figure}

\begin{figure}
\centering
\includegraphics[width=0.9\linewidth]{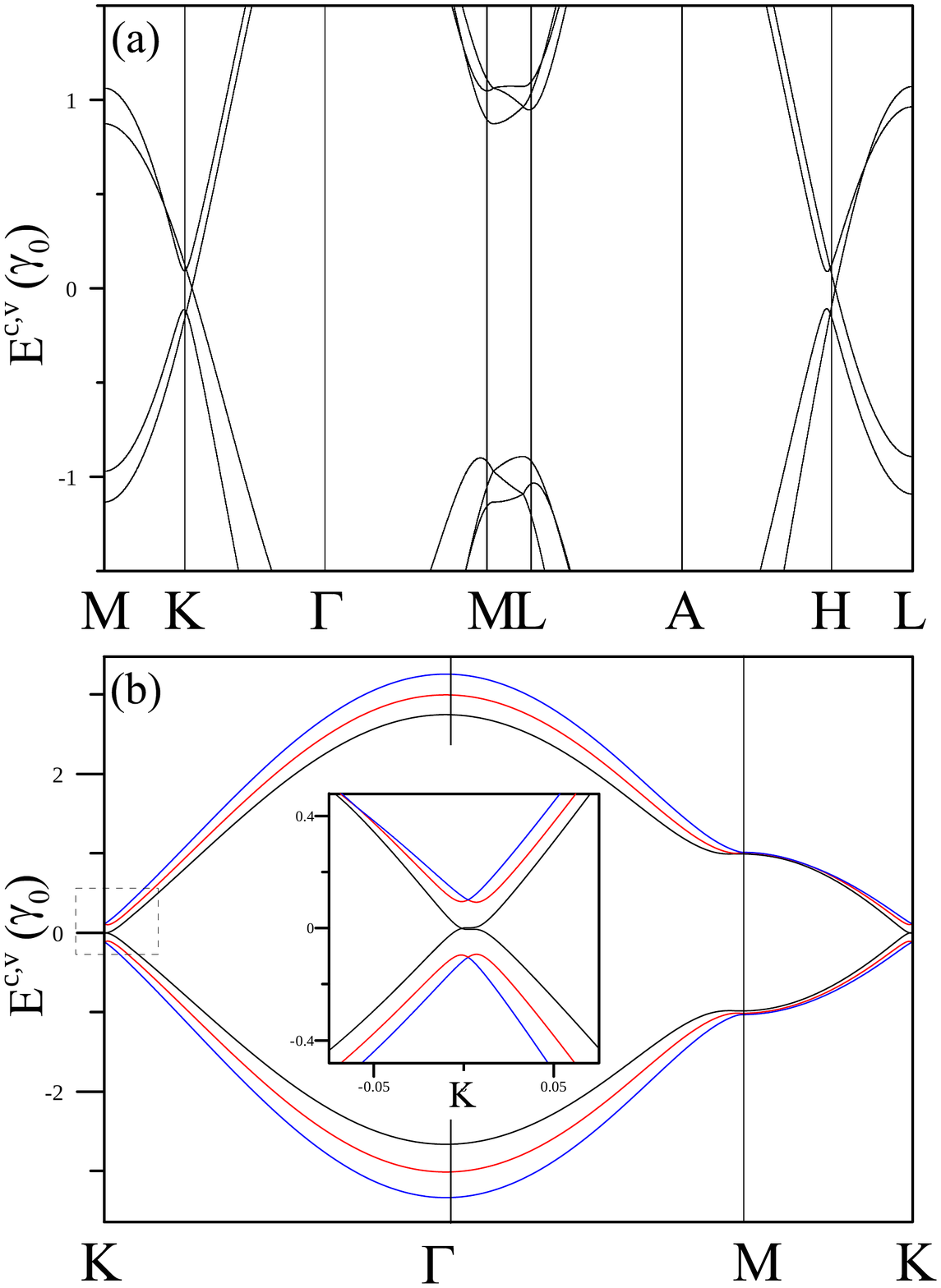}
\caption{Band structure of rhombohedral (a) graphite and (b) trilayer graphene.}
\label{fig:graph}
\end{figure}

\begin{figure}
\centering
\includegraphics[width=0.9\linewidth]{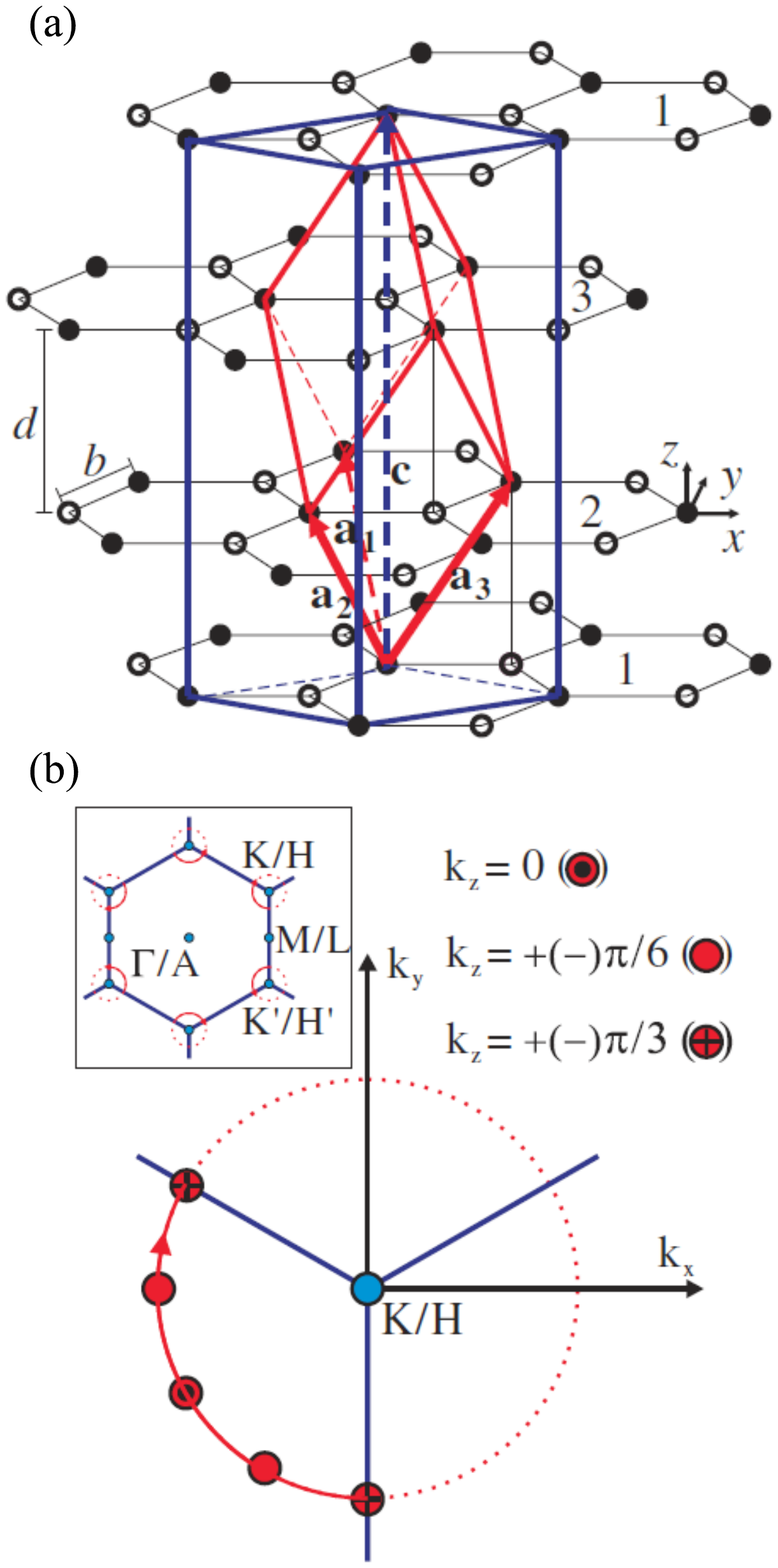}
\caption{(a) Rhombohedral primitive unit cell (red color) and triple
hexagonal unit cell (blue color). (b) ($k_{x},k_{y}$)-projection of
the Dirac-point spiral at $k_{z}=0$, $k_{z}=\pm\pi/6$ and
$k_{z}=\pm\pi/3$.}
\label{fig:graph}
\end{figure}

\begin{figure}
\centering
\includegraphics[width=0.9\linewidth]{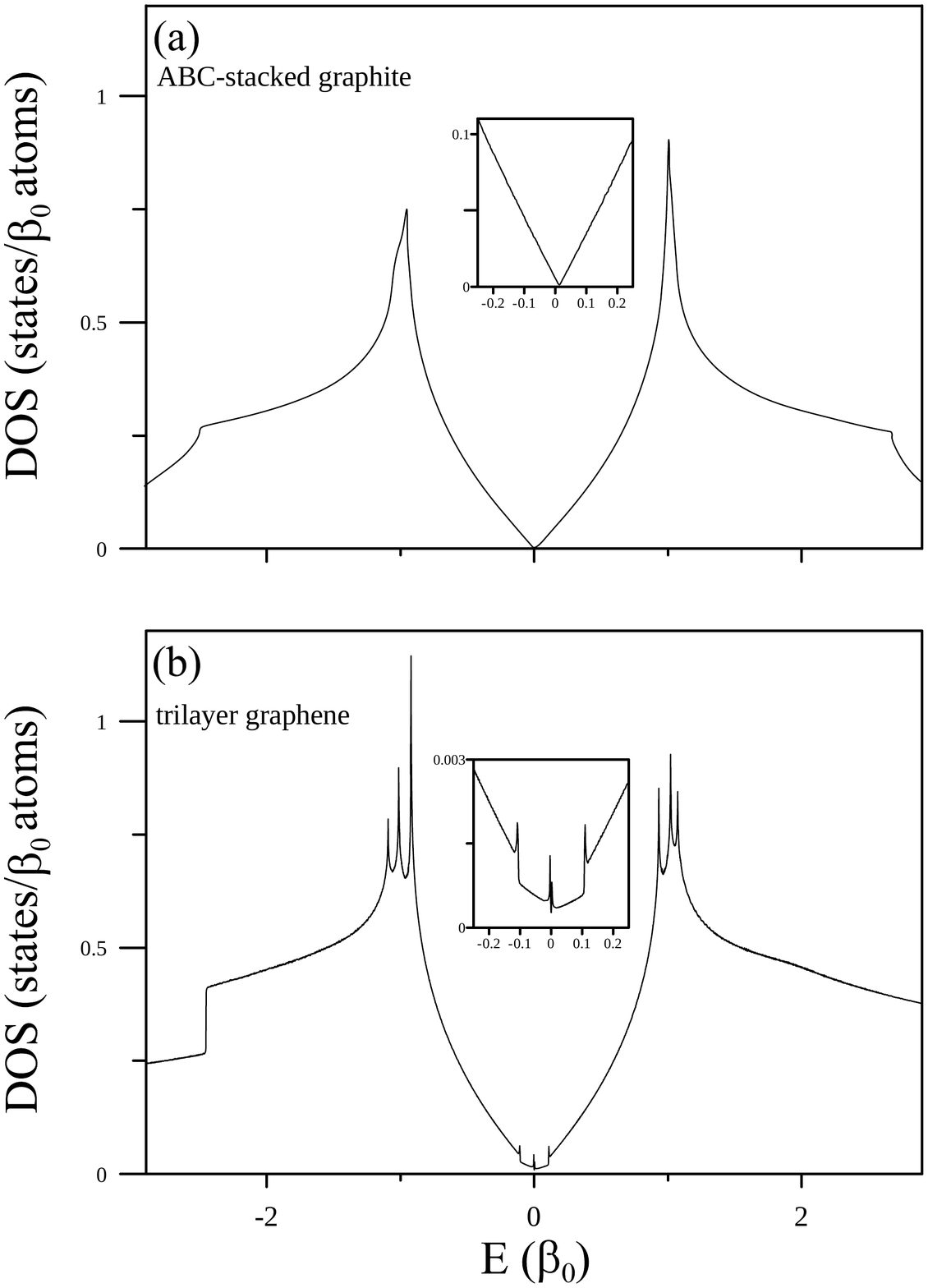}
\caption{DOSs of ABC-stacked (a) graphite and (b) trilayer
graphene.}
\label{fig:graph}
\end{figure}

\begin{figure}
\centering
\includegraphics[width=0.9\linewidth]{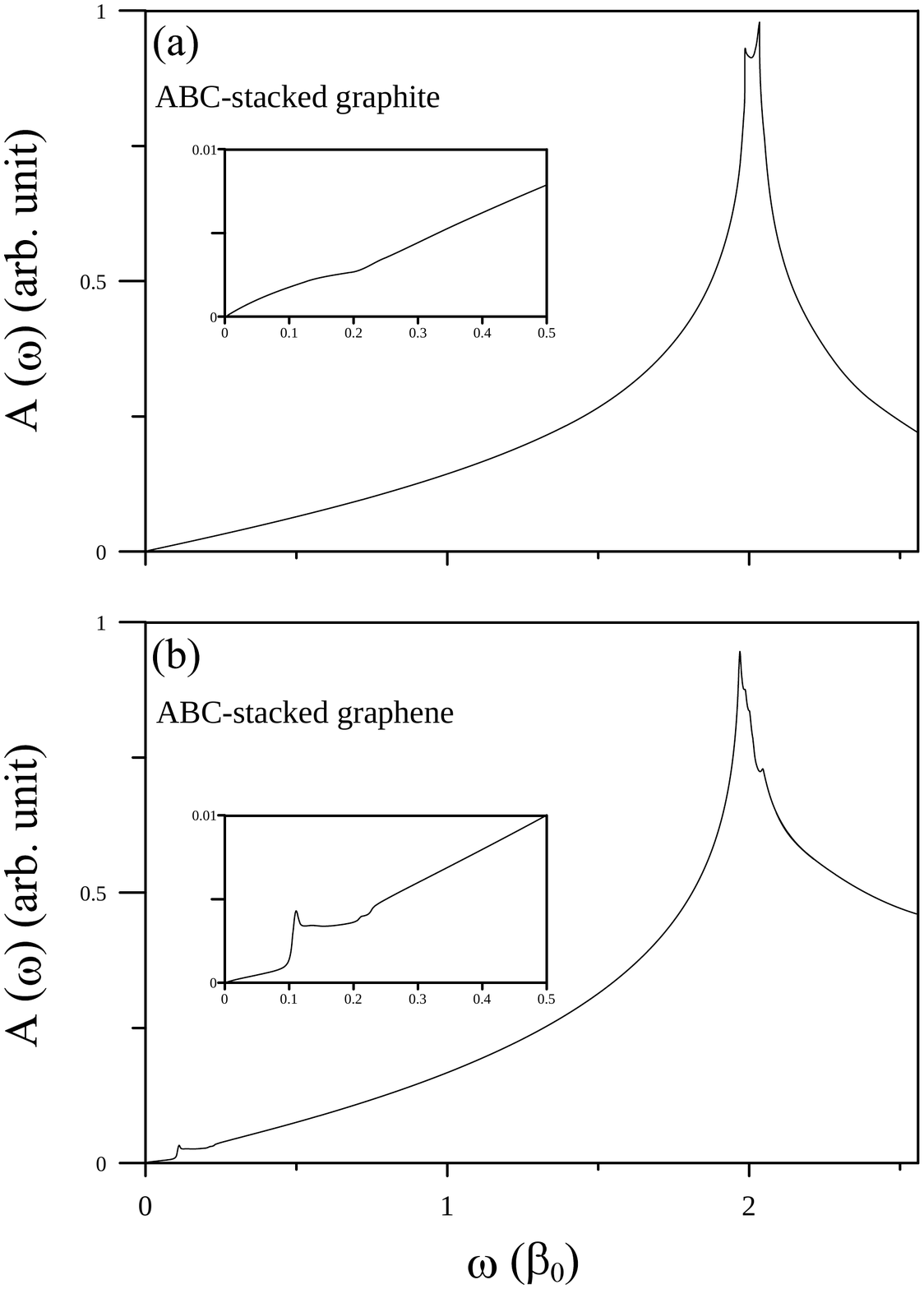}
\caption{Absorption spectra of ABC-stacked (a) graphite and (b)
trilayer graphene. The insets of (a) and (b) show the zoomed-in view
at low frequencies.}
\label{fig:graph}
\end{figure}

\begin{figure}
\centering
\includegraphics[width=0.9\linewidth]{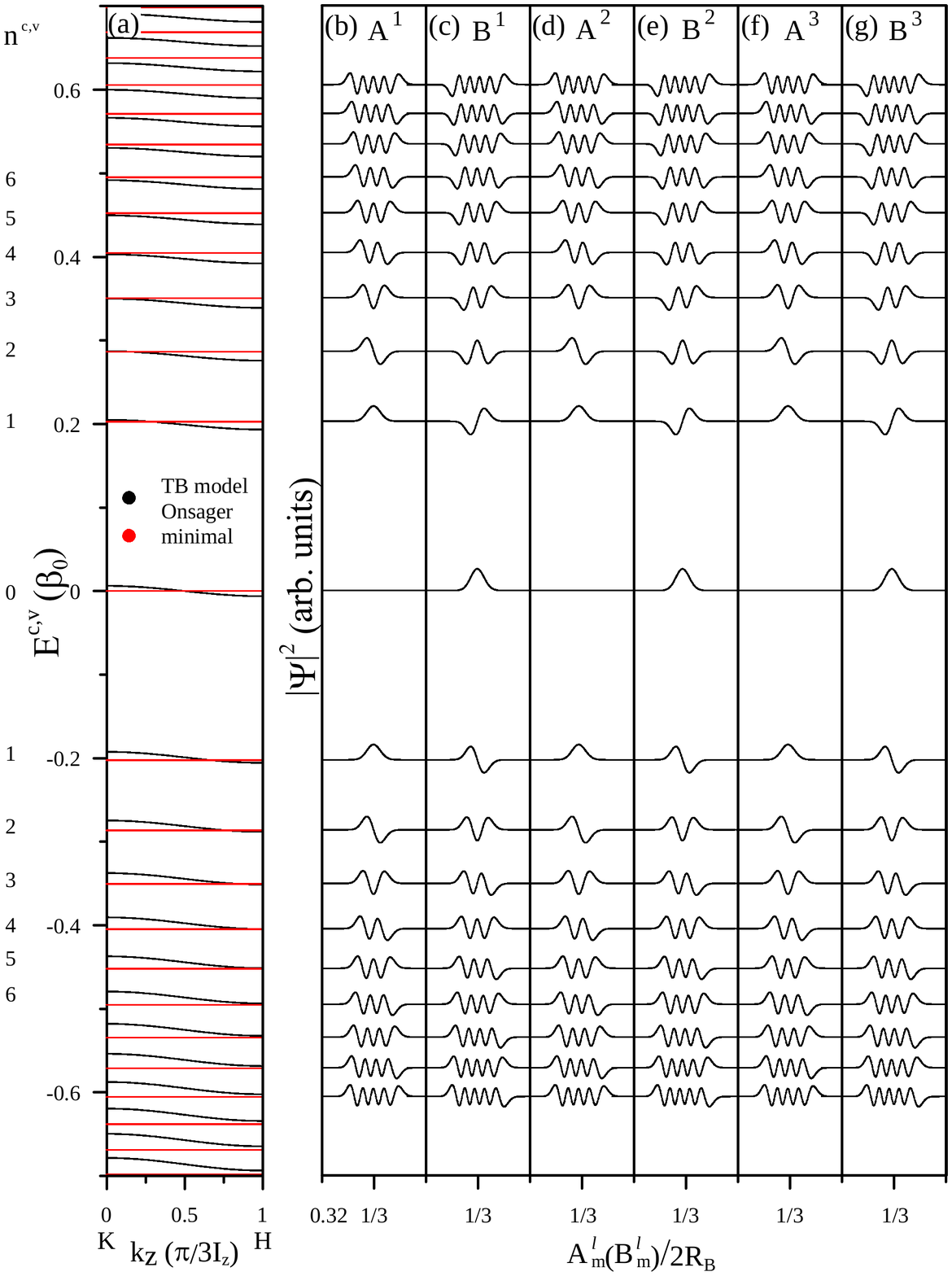}
\caption{(a) The $k_{z}$-dependent Landau subbands at $B_{0}=40$ T.
The black and blue colors, respectively, represent the calculations
from the full tight-binding model and the minimal model. (b)-(g) The
subenvelope functions at the K point.}
\label{fig:graph}
\end{figure}

\begin{figure}
\centering
\includegraphics[width=0.9\linewidth]{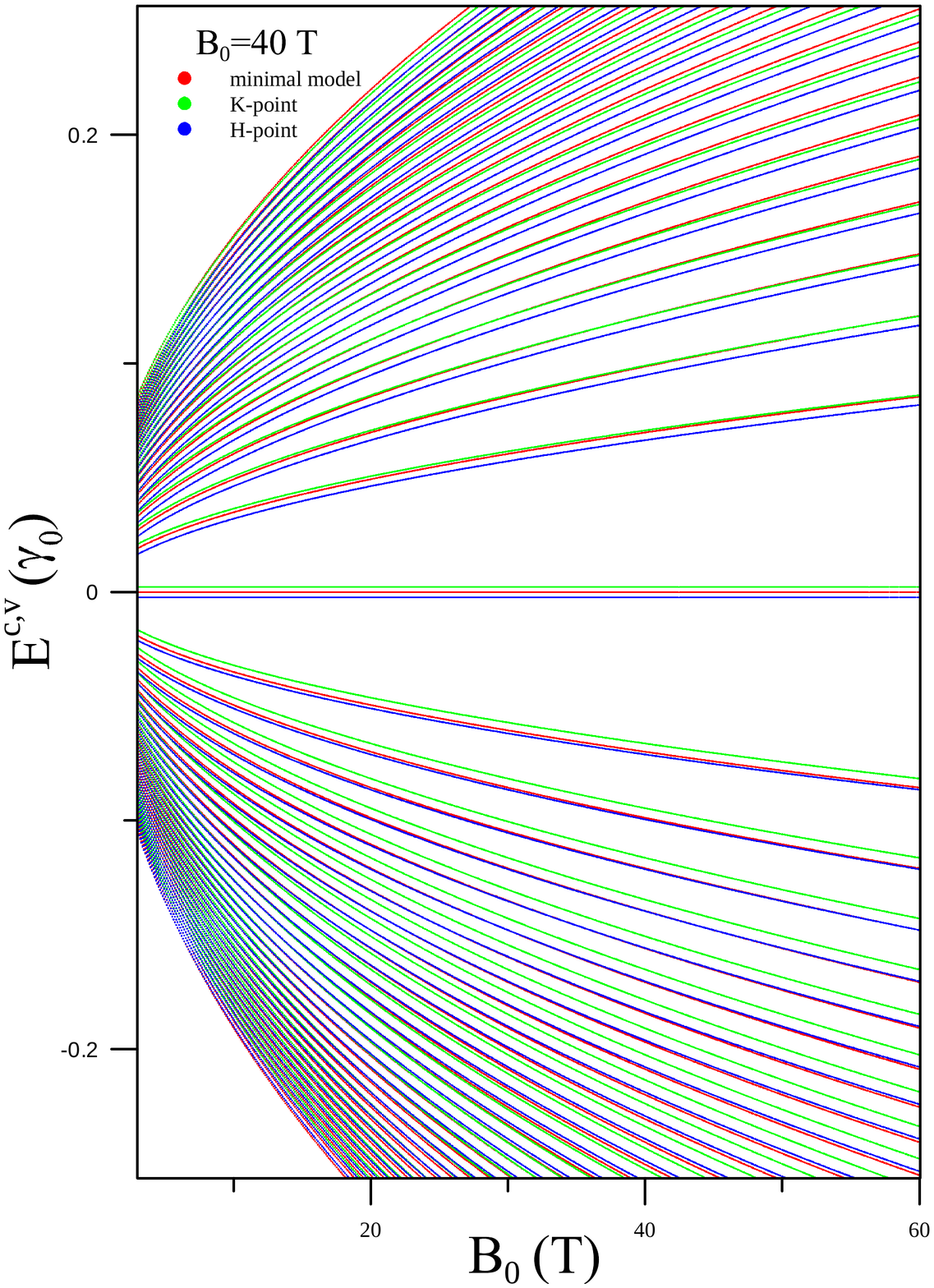}
\caption{$B_0$-dependent Landau subband energies of rhombohedral graphite.}
\label{fig:graph}
\end{figure}

\begin{figure}
\centering
\includegraphics[width=0.9\linewidth]{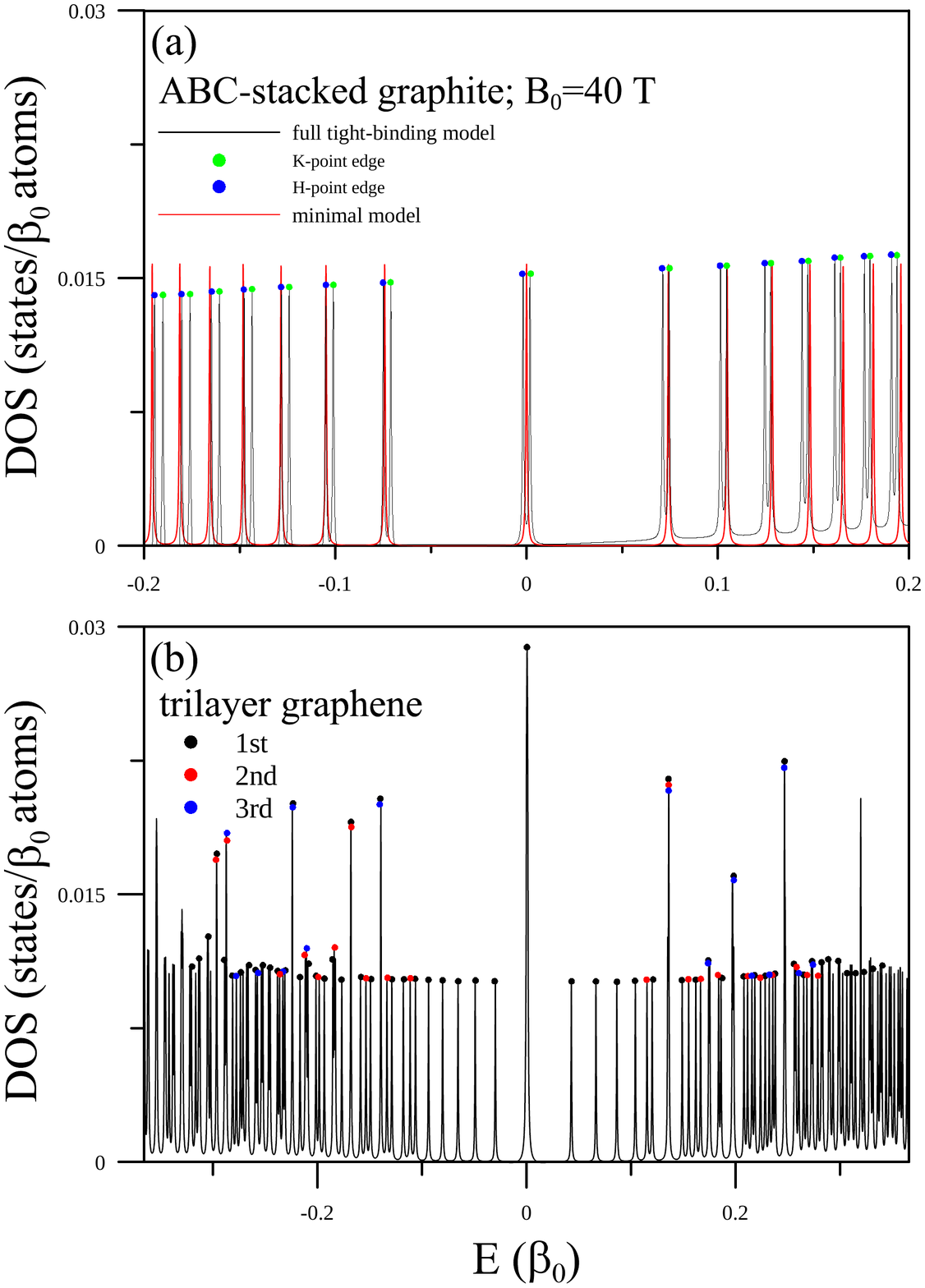}
\caption{DOS of ABC-stacked (a) graphite and (b) trilayer graphene
under $B_{0}=40$ T. The black and red curves in (a) represent the
calculations from the tight-binding model and minimal model,
respectively.}
\label{fig:graph}
\end{figure}

\begin{figure}
\centering
\includegraphics[width=0.9\linewidth]{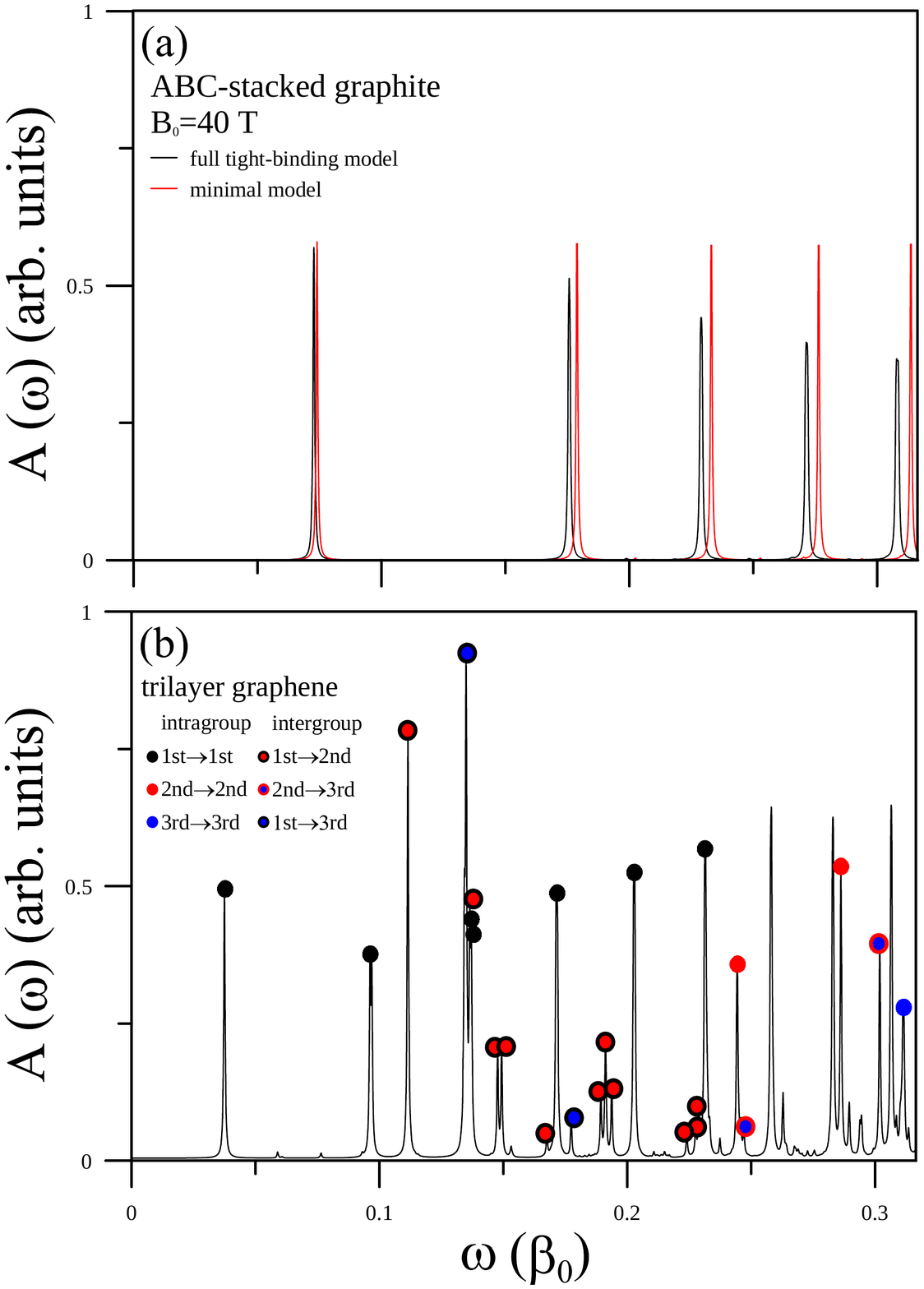}
\caption{(a) Absorption spectra of rhombohedral graphite and (b)
trilayer ABC-stacked graphene under $B_{0}=40$ T.}
\label{fig:graph}
\end{figure}

\begin{figure}
\centering
\includegraphics[width=0.9\linewidth]{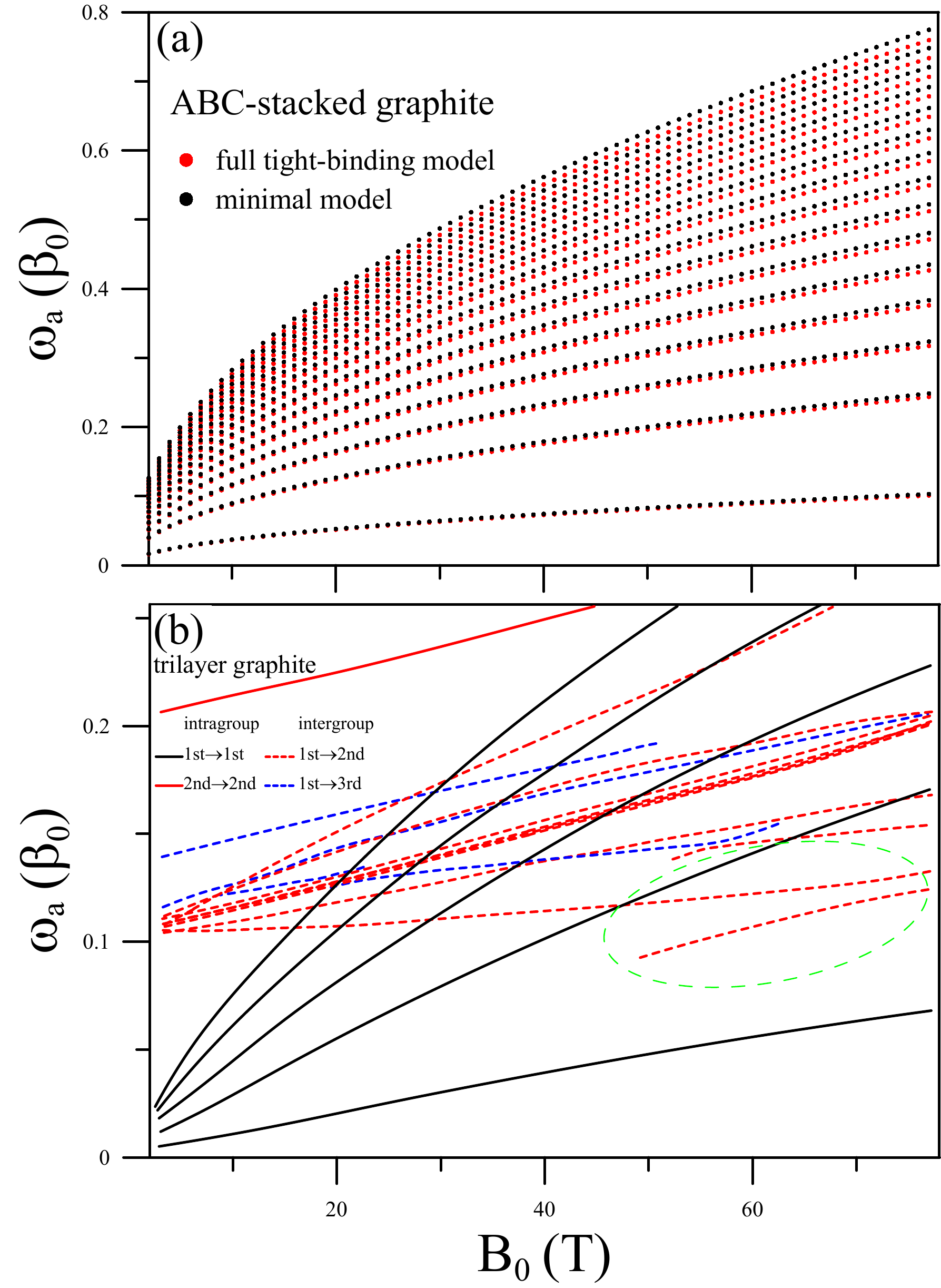}
\caption{$B_0$-dependent absorption frequencies for the ABC-stacked
(a) graphite and (b) graphene. In (b), the solid and dashed curves represent the intragroup and intergroup absorption frequencies, respectively.}
\label{fig:graph}
\end{figure}

\clearpage

\begin{figure}
\centering
\includegraphics[width=0.9\linewidth]{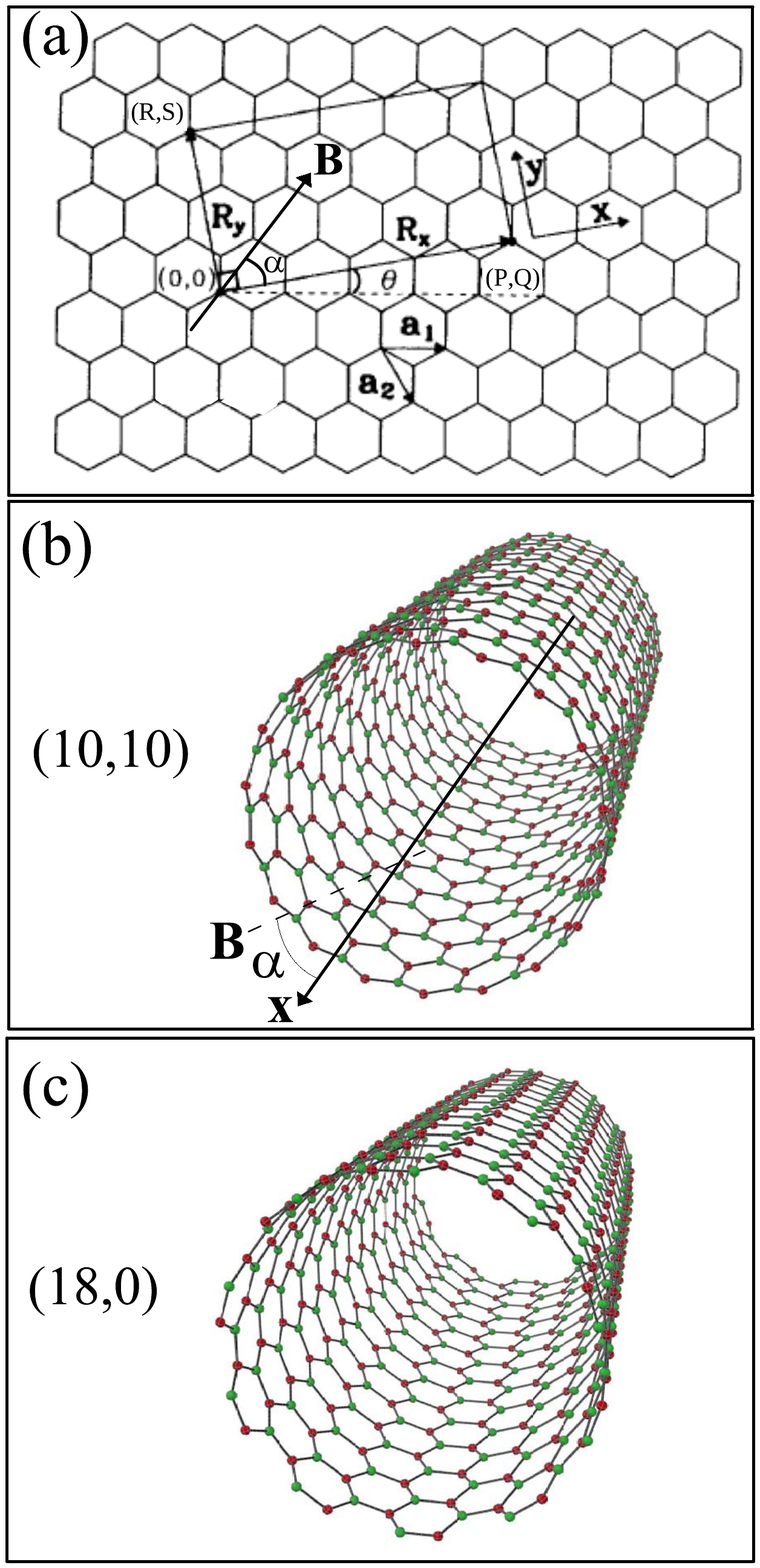}
\caption{Geometric structures for (a) graphene $\&$ carbon
nanotubes;  (b) (10,10) $\&$ (18,0) nanotubes. Also shown in (b) is
the relation between nanotube axis and magnetic field.}
\label{fig:graph}
\end{figure}

\begin{figure}
\centering
\includegraphics[width=0.9\linewidth]{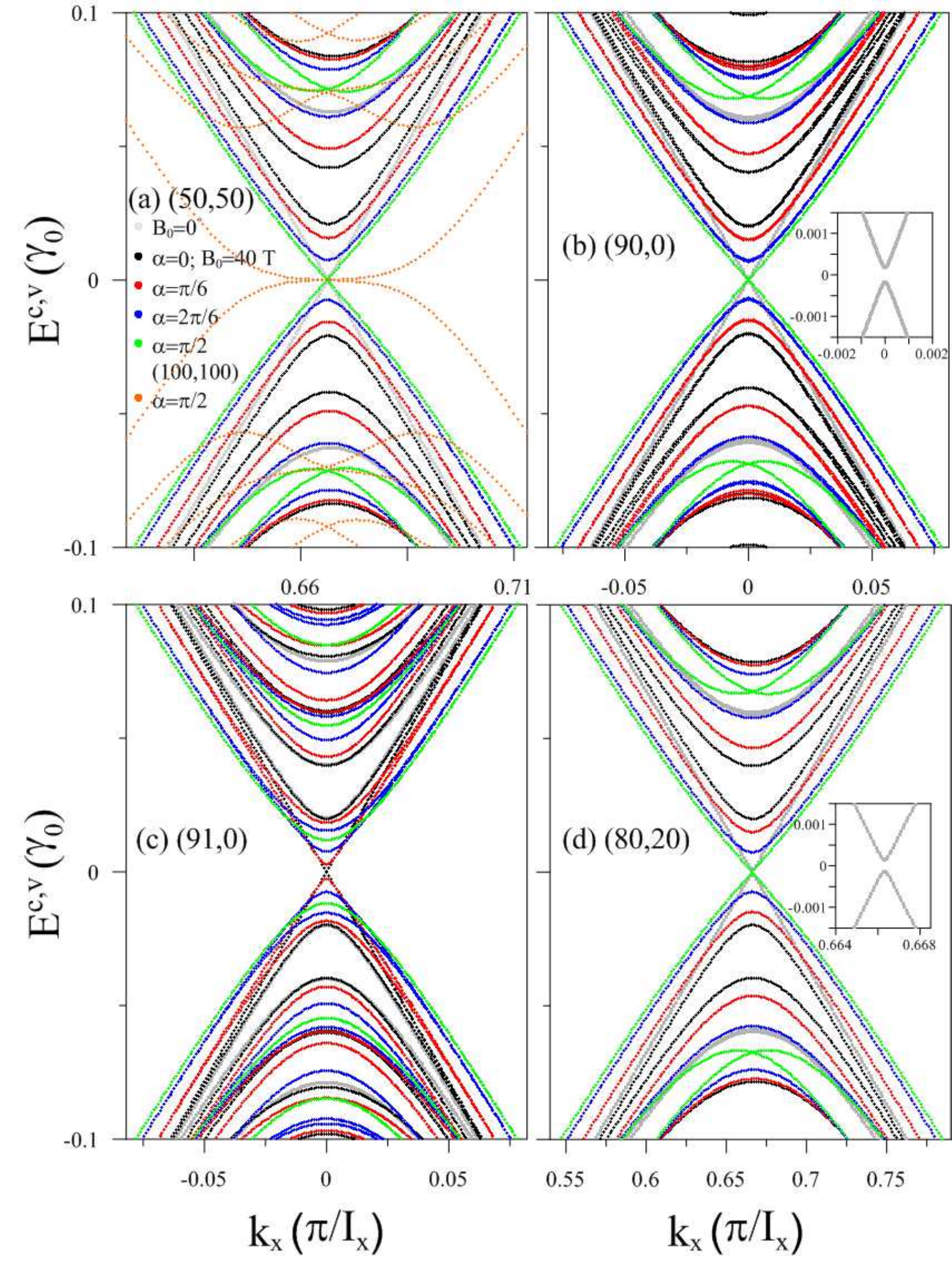}
\caption{Band structures of (a) (50,50), (b) (90,0), (c) (91,0) and (d)
(80,20) carbon nanotubes at various
magnetic-field and ${B_0=0}$ and 40 T. Also shown in (a) are that of (100,100) nanotube at ${\alpha\,=\pi\,/2}$ (orange curves), and in the insets of (b) and (d) are narrow energy gaps.}
\label{fig:graph}
\end{figure}

\begin{figure}
\centering
\includegraphics[width=0.9\linewidth]{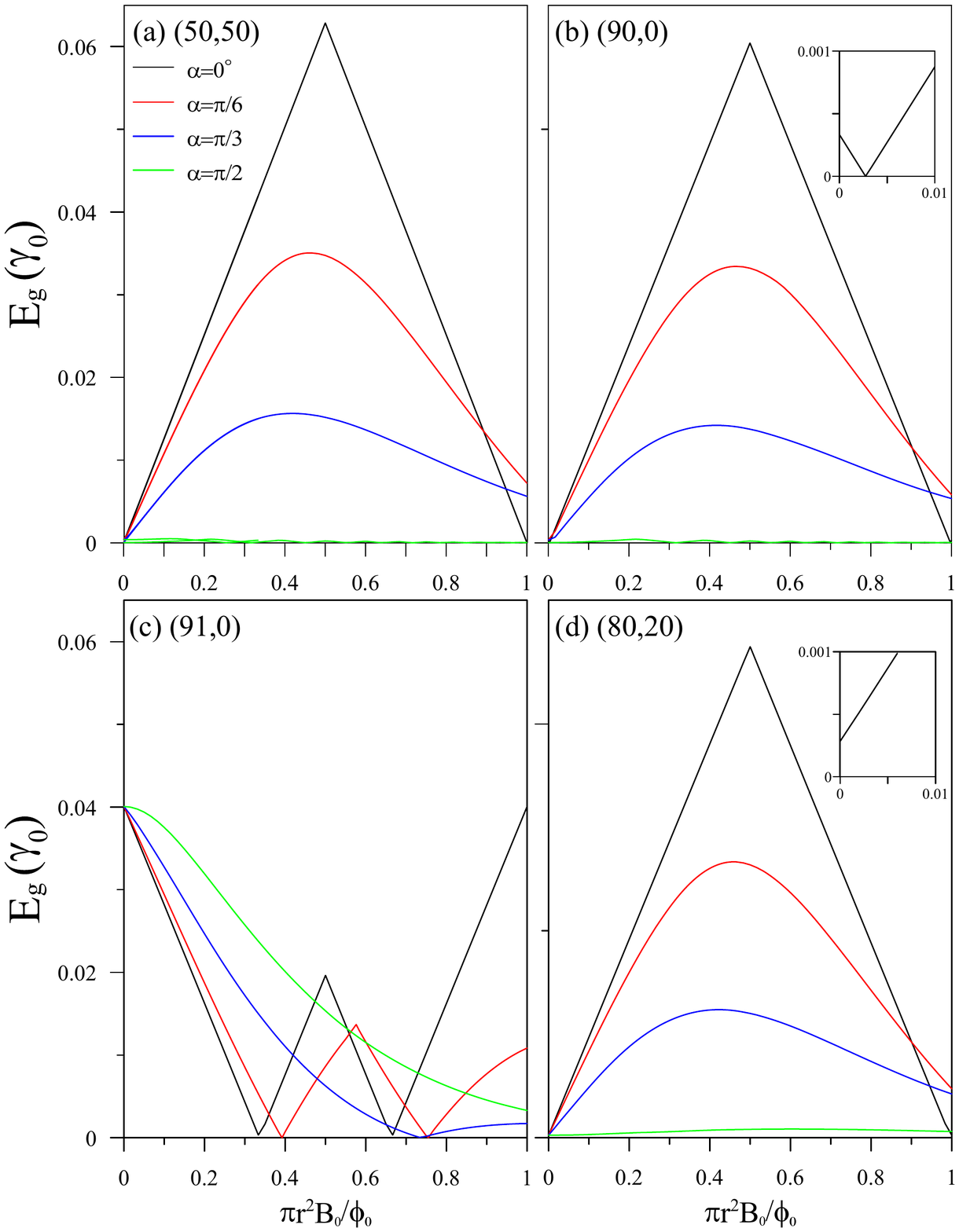}
\caption{Magnetic-flux-dependent energy gaps for (a) (50,50), (b)
(90,0), (c) (91,0) and (d) (65,35) carbon nanotubes at various
magnetic-field directions.}
\label{fig:graph}
\end{figure}

\begin{figure}
\centering
\includegraphics[width=0.9\linewidth]{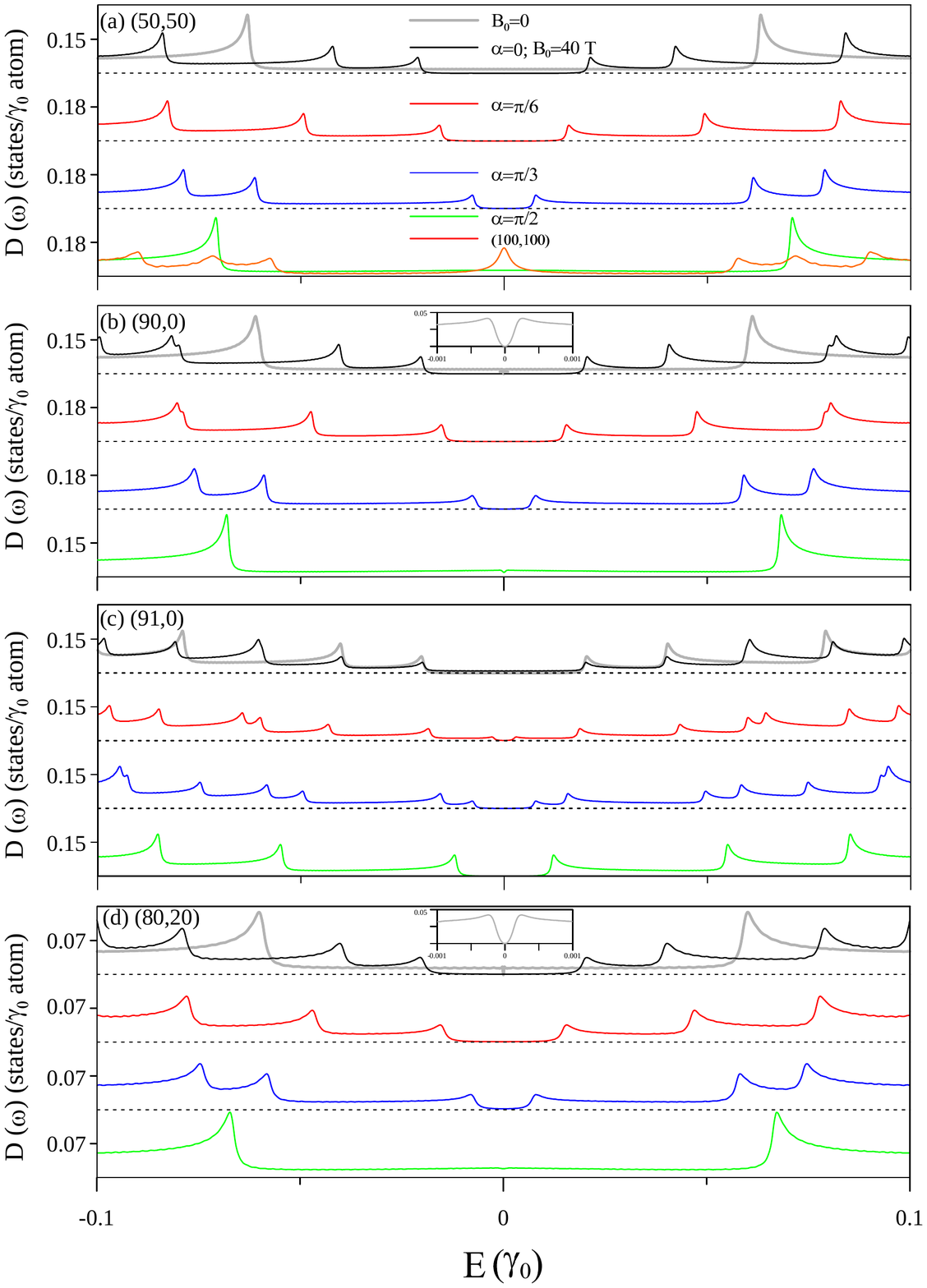}
\caption{Density of states for (a) (50,50), (b) (90,0), (c) (91,0)
and (d) (65,35) carbon nanotubes at various magnetic-field
directions.}
\label{fig:graph}
\end{figure}

\begin{figure}
\centering
\includegraphics[width=0.9\linewidth]{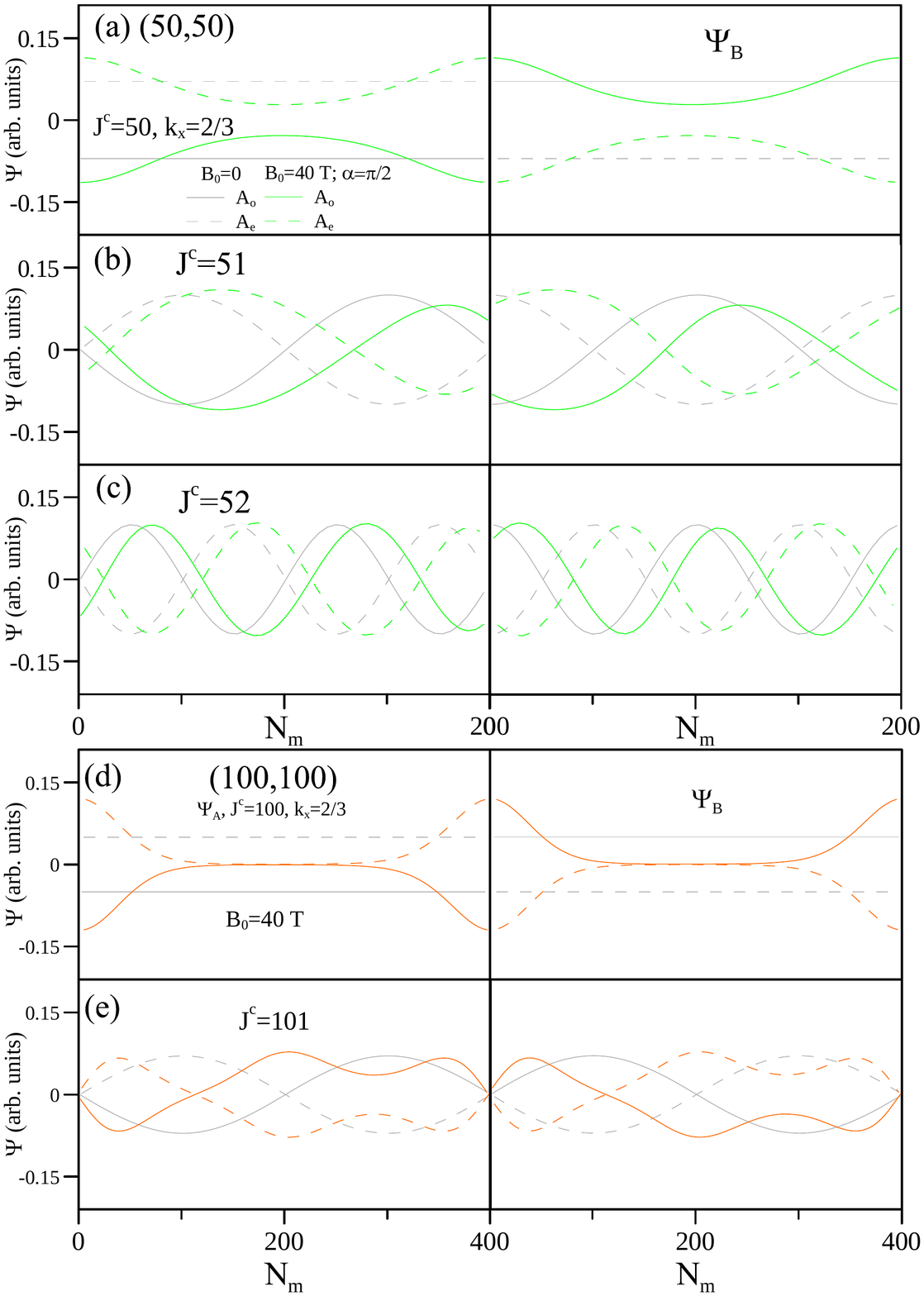}
\caption{The subenvelope functions of the (50,50) nanotube at
${k_x=2/3}$ for (a) ${J^{c}=50}$, (b) ${J^{c}=51}$ and (c)
${J^{c}=52}$. ${B_0=0}$ and ($B_0$=40 T,${\alpha\,=\pi\,/2}$) are,
respectively, shown by the black and red curves. Also shown are
those of the (100,100) nanotube for (d) ${J^{c}=100}$ and (e)
${J^{c}=101}$.}
\label{fig:graph}
\end{figure}

\begin{figure}
\centering
\includegraphics[width=0.9\linewidth]{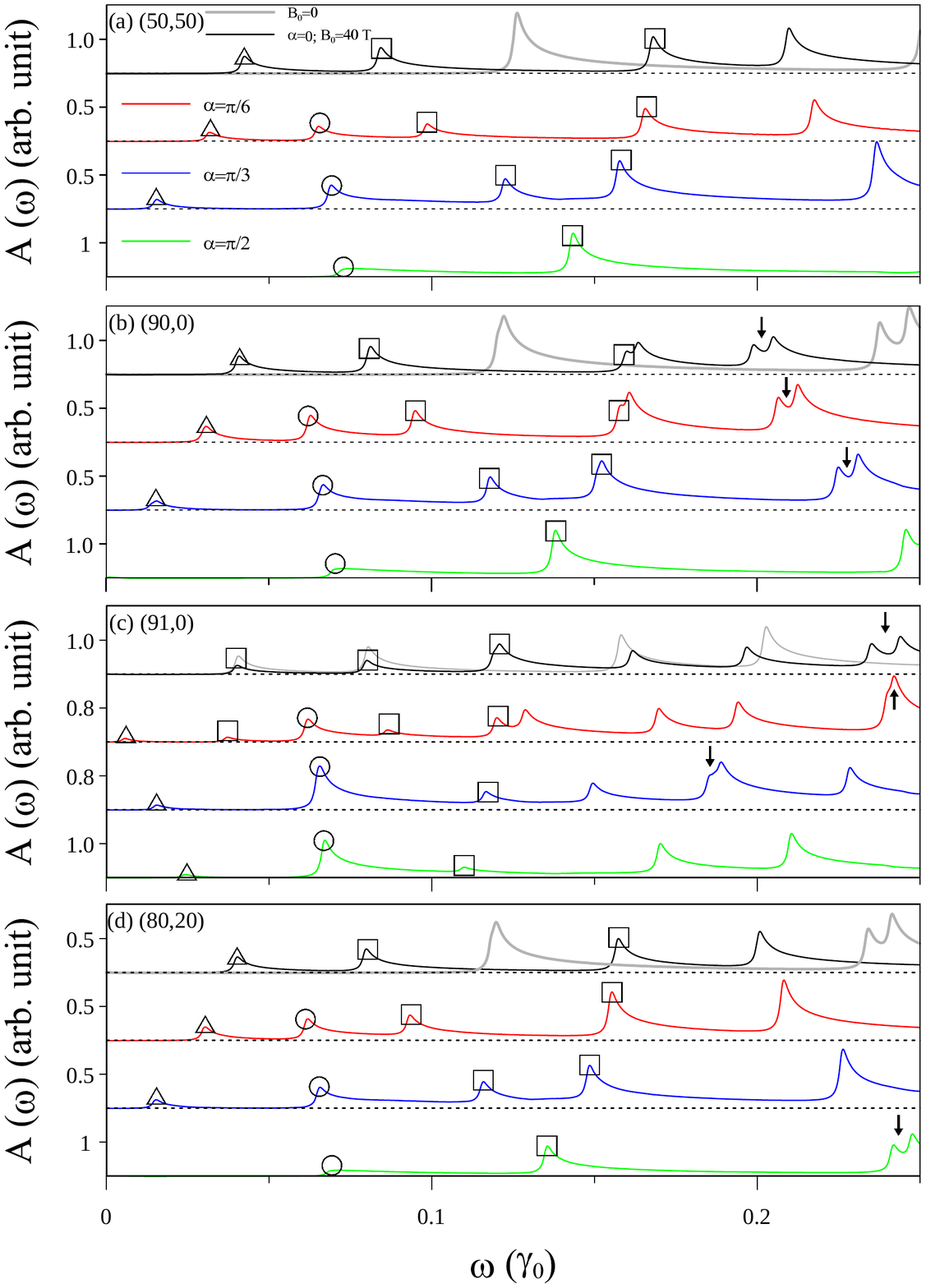}
\caption{Optical absorption spectra of (a) (50,50), (b) (90,0), (c)
(91,0) and (d) (65,35) carbon nanotubes under zero field and various
magnetic-field directions.}
\label{fig:graph}
\end{figure}

\begin{figure}
\centering
\includegraphics[width=0.9\linewidth]{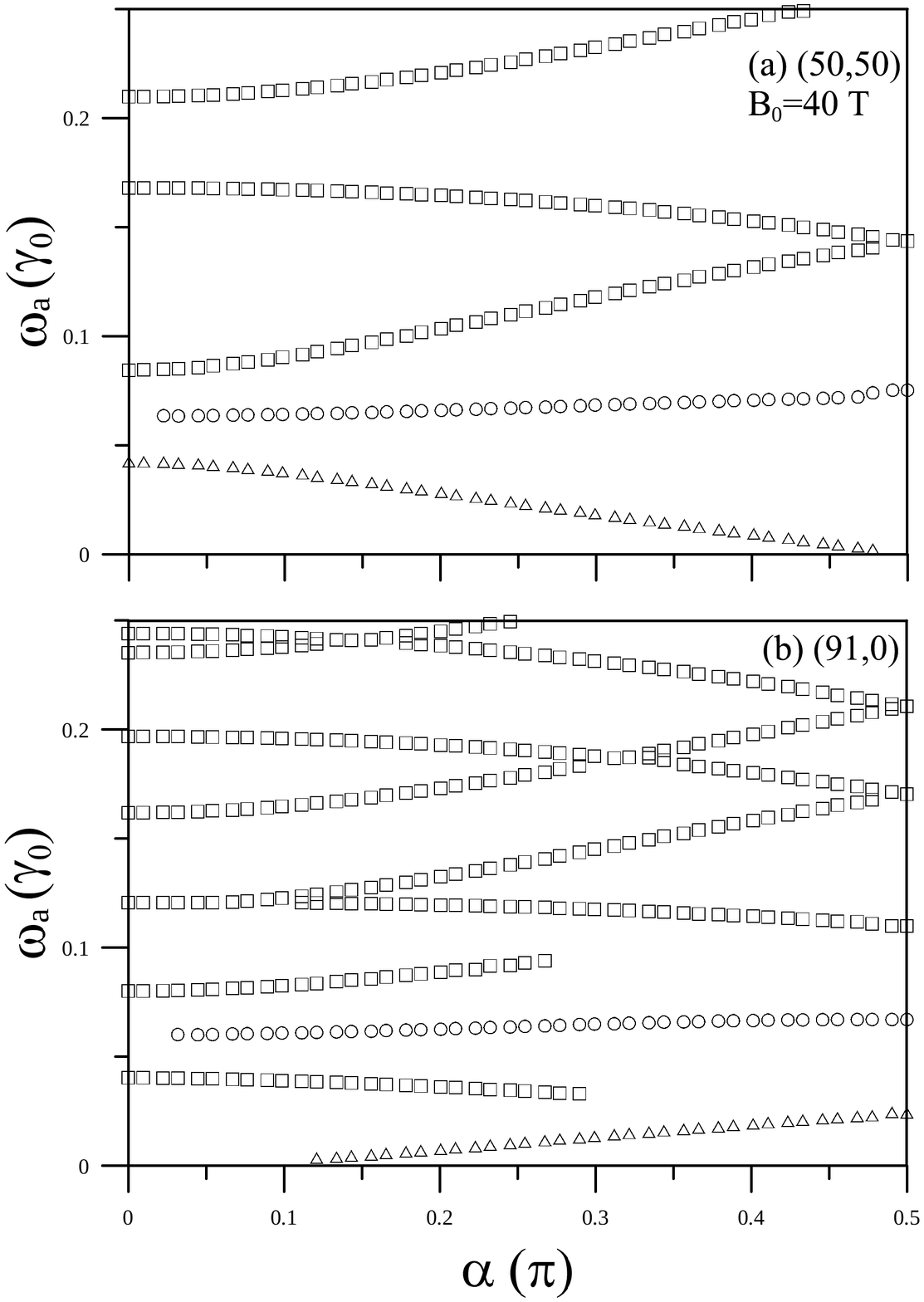}
\caption{Optical absorption frequencies of (a) (50,50), and (b) (91,0) carbon nanotubes under various magnetic-field directions at $B_{0}$=40 T.}
\label{fig:graph}
\end{figure}

\begin{figure}
\centering
\includegraphics[width=0.9\linewidth]{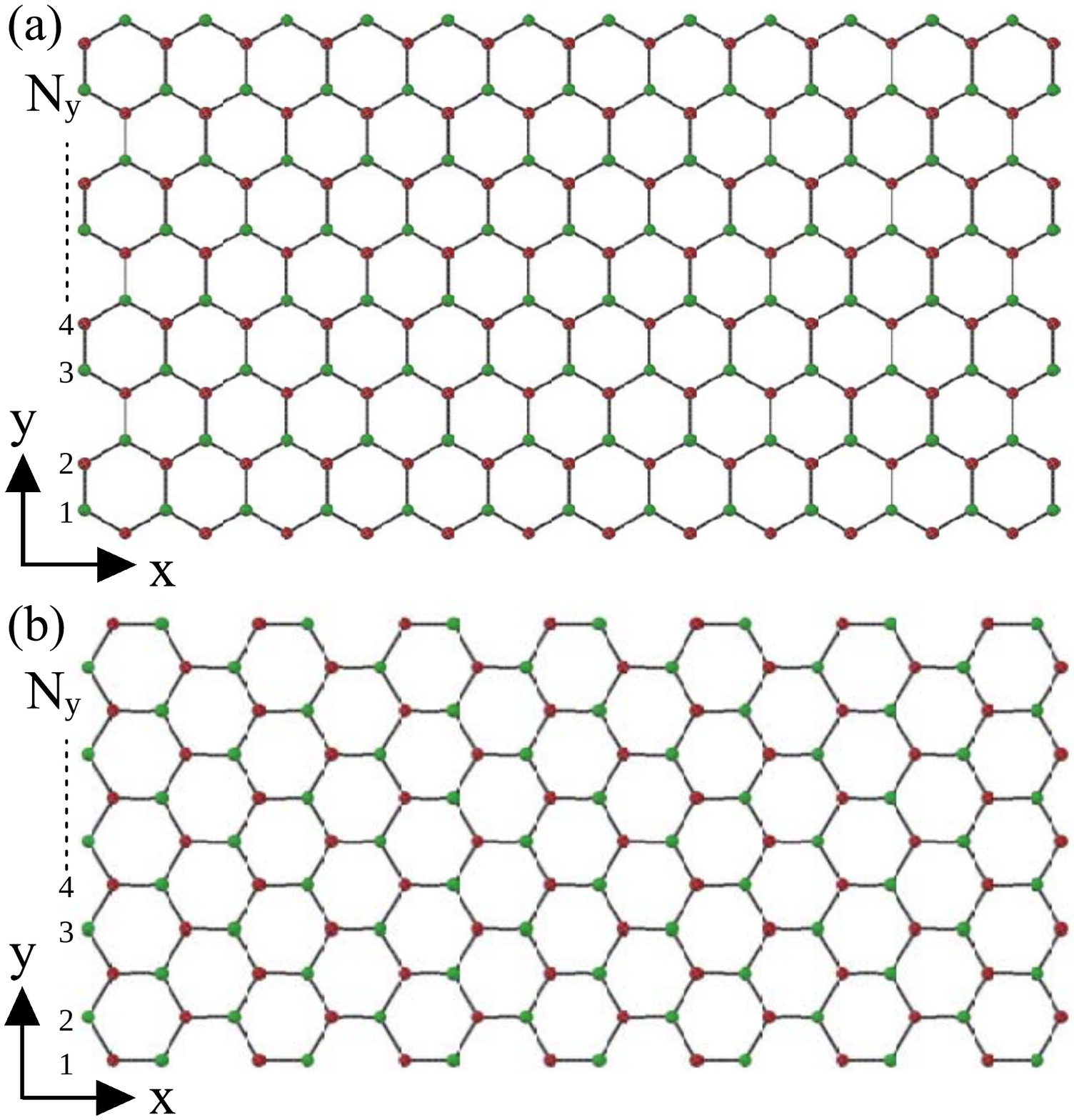}
\caption{Geometric structures of (a) zigzag and (b) armchair
graphene nanoribbons. $N_y$ is the number of zigzag or dimer lines.}
\label{fig:graph}
\end{figure}

\begin{figure}
\centering
\includegraphics[width=0.9\linewidth]{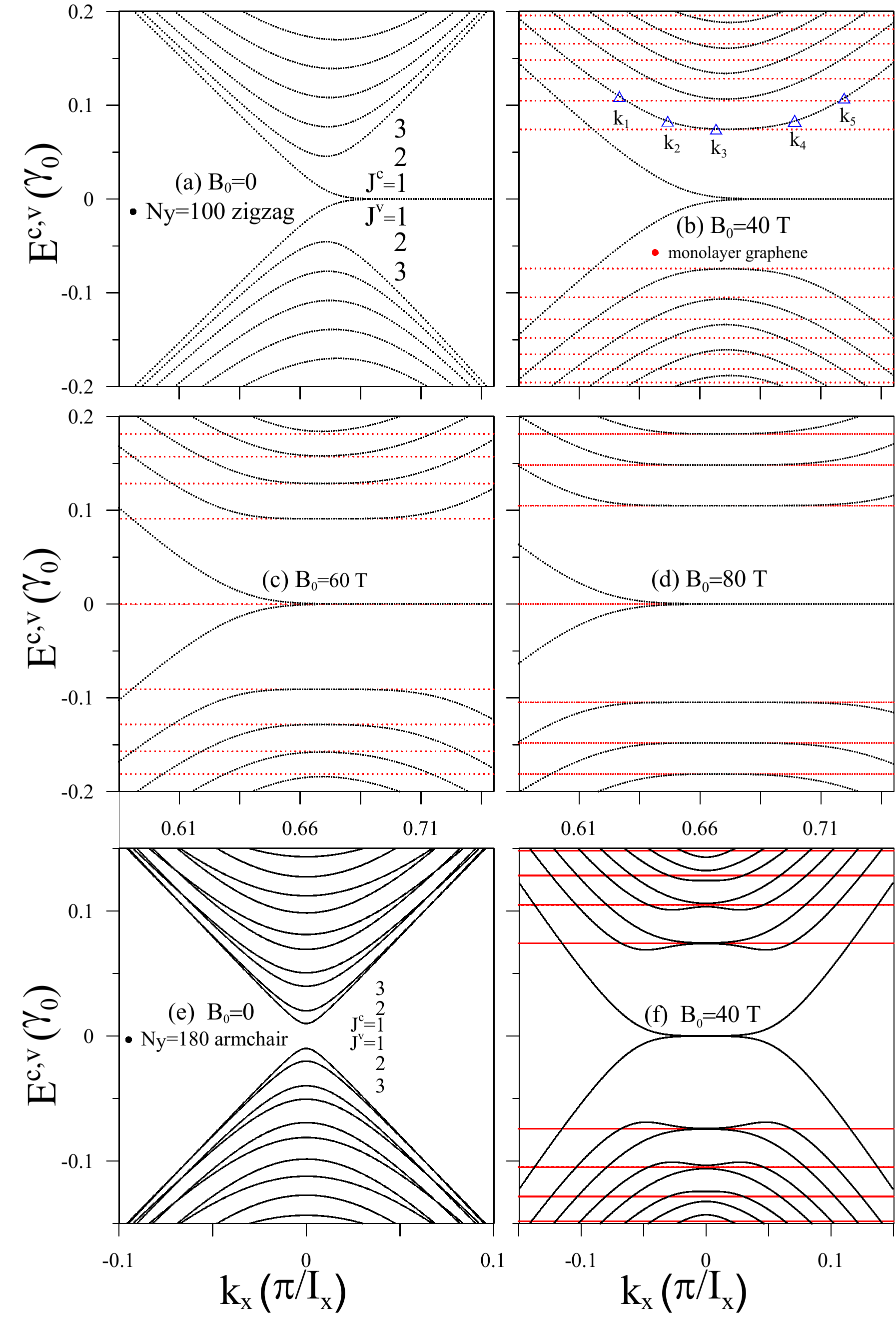}
\caption{Band structures of a ${N_y=100}$ zigzag nanoribbon at (a)
${B_0=0}$, (b) ${B_0=40}$ T, (c) ${B_0=60}$ T; (d) ${B_0=80}$ T, and
those of a ${N_y=180}$ armchair nanoribbon at (e) ${B_0=0}$; (b)
${B_0=40}$ T. The red lines represent the LL energies of monolayer graphene. ${J^{c,v}}$ is the subband index, as measured from the Fermi level.}
\label{fig:graph}
\end{figure}

\begin{figure}
\centering
\includegraphics[width=0.9\linewidth]{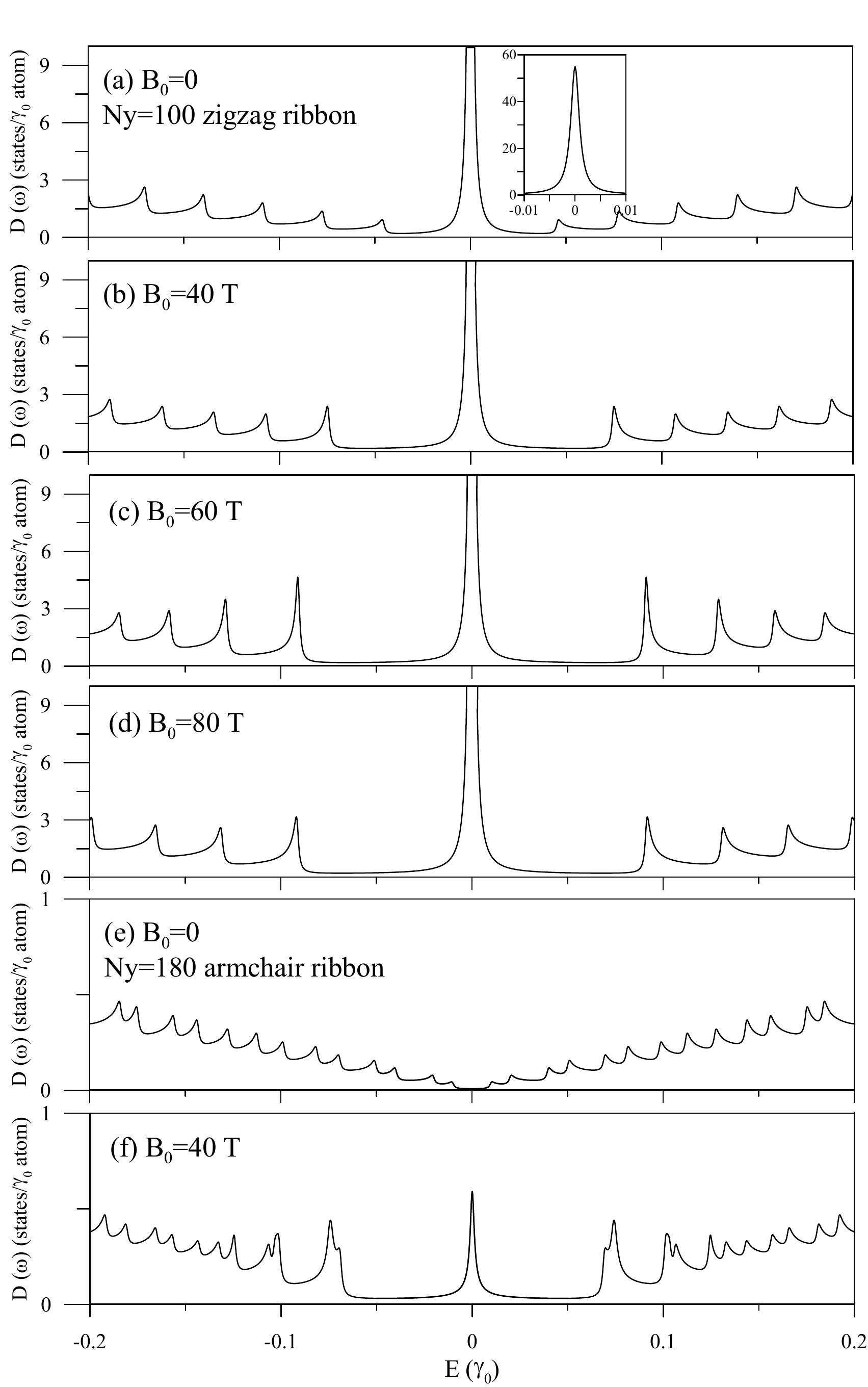}
\caption{Density of states corresponding to electronic structures in Fig. 42.}
\label{fig:graph}
\end{figure}

\begin{figure}
\centering
\includegraphics[width=0.9\linewidth]{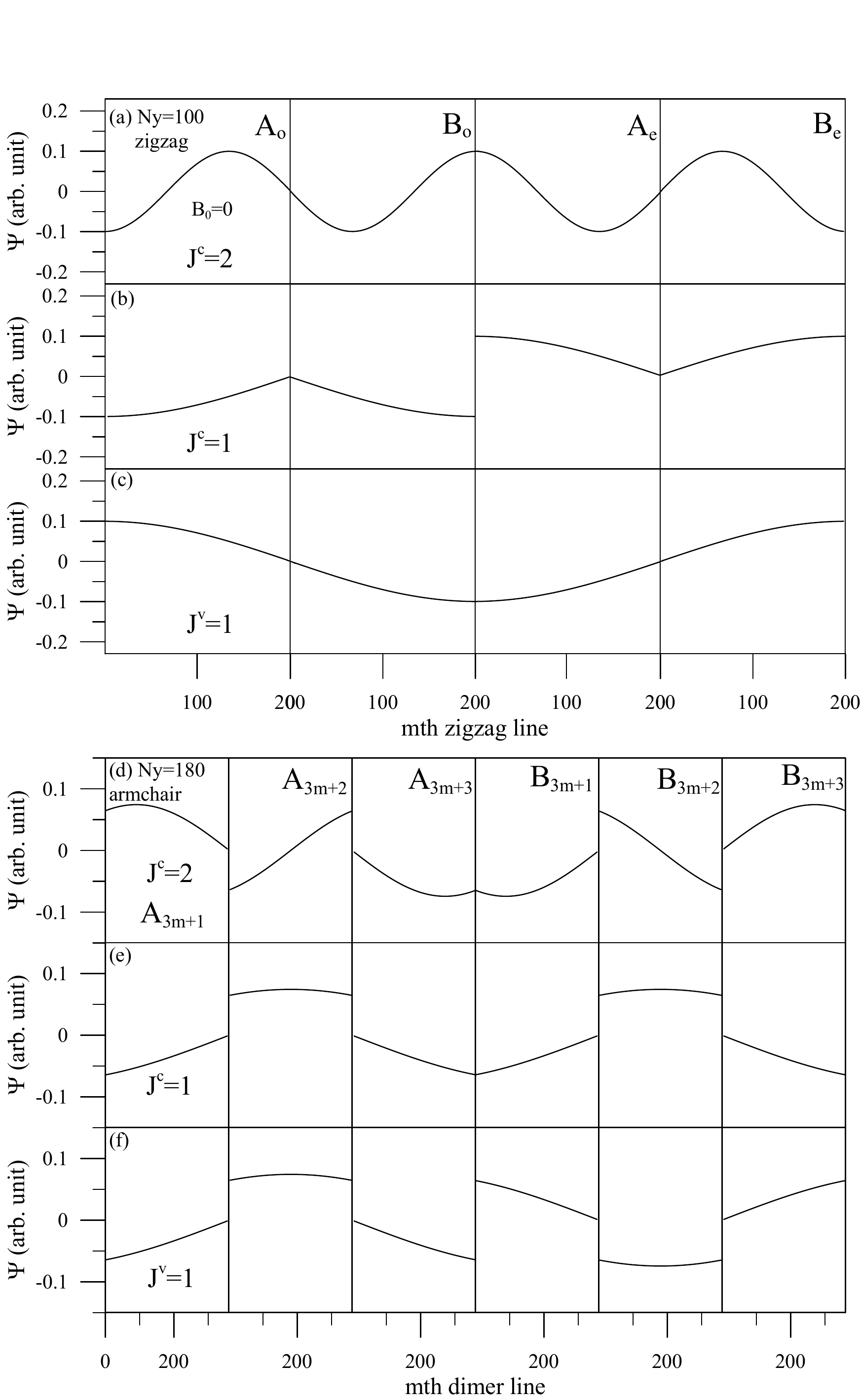}
\caption{The subenvelope functions of a ${N_y=100}$ zigzag nanoribbon for (a) ${J^c=2}$, (b) ${J^c=1}$; (c) ${J^v=1}$, and those of a ${N_y=180}$ armchair system for (d) ${J^c=2}$, (e) ${J^c=1}$; (f) ${J^v=1}$.}
\label{fig:graph}
\end{figure}

\begin{figure}
\centering
\includegraphics[width=0.9\linewidth]{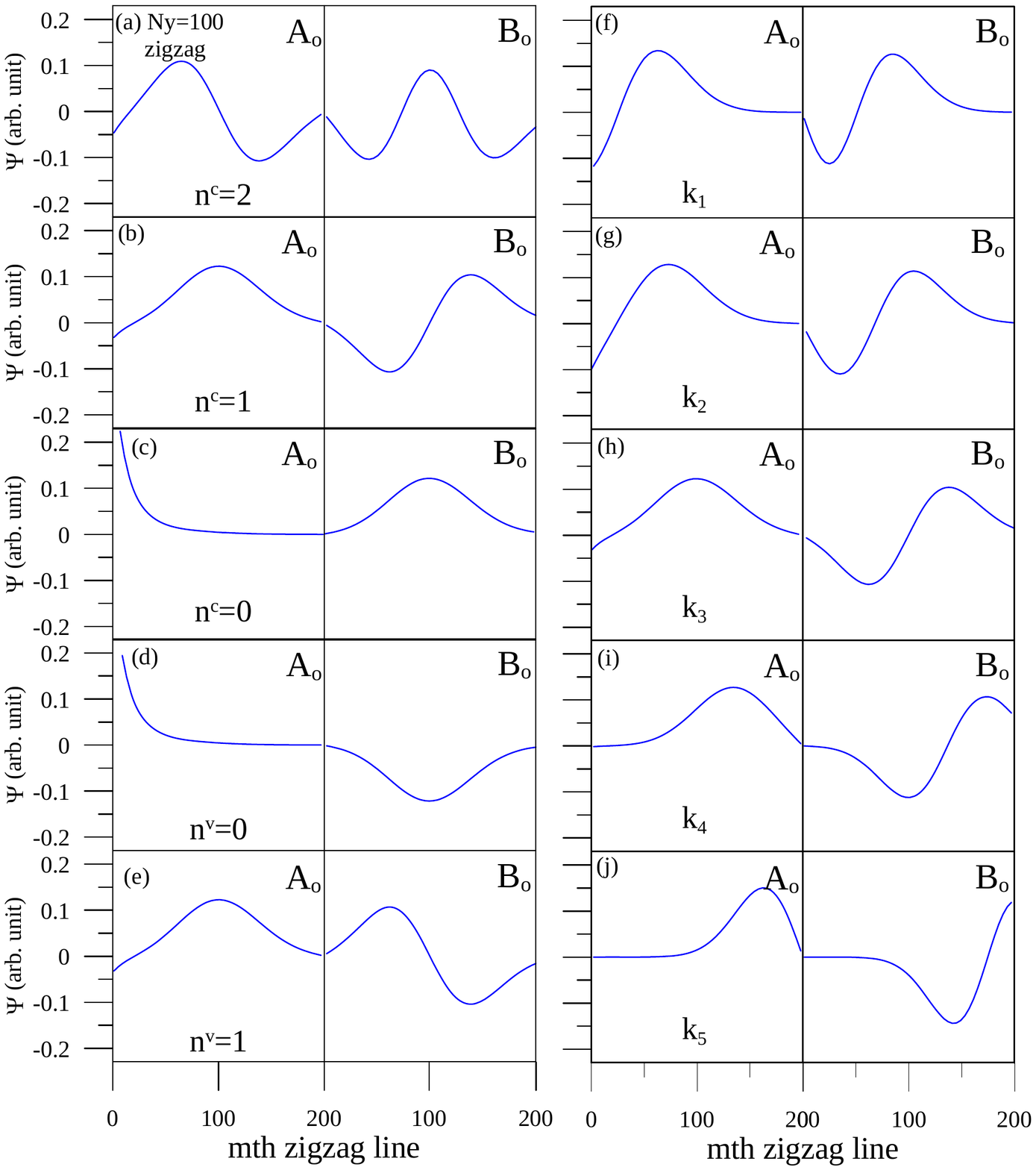}
\caption{The magnetic subenvelope functions of a ${N_y=100}$ zigzag nanoribbon at ${B_0=40}$ T and ${k_x=2/3}$ for (a)-(c) ${n^c=2-0}$
$\&$ (e)-(f) ${n^v=0-1}$. Also shown are those of ${n^c=1}$ at (f)-(j) various $k_x$s (triangles in Fig. 42(c)).}
\label{fig:graph}
\end{figure}

\begin{figure}
\centering
\includegraphics[width=0.9\linewidth]{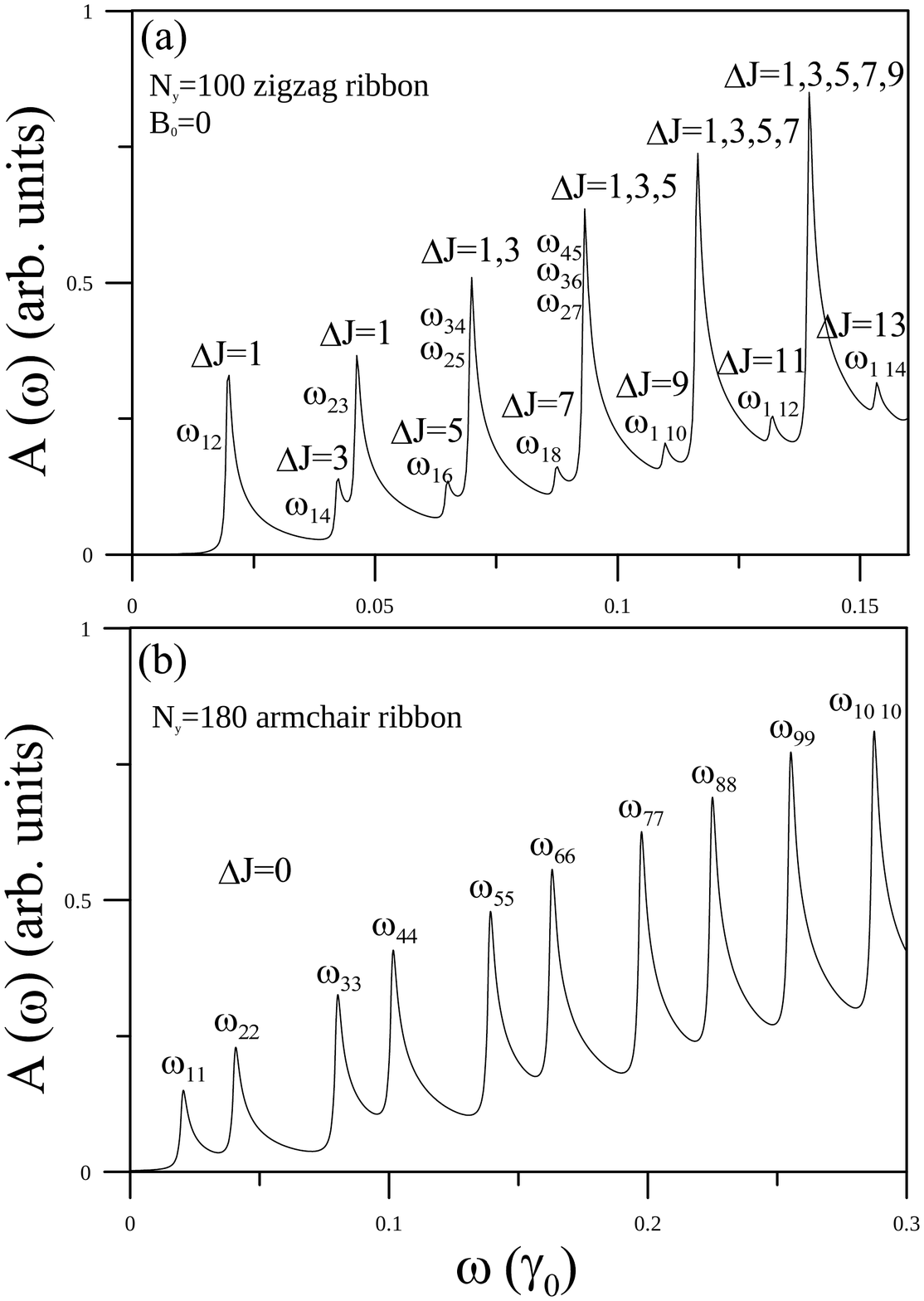}
\caption{Optical absorption spectra of (a) ${N_y=100}$ zigzag and (b) ${N_y=180}$ armchair nanoribbons. Two subscripts in absorption peak frequency represents indexes of the initial valence and final conduction bands.}
\label{fig:graph}
\end{figure}

\begin{figure}
\centering
\includegraphics[width=0.9\linewidth]{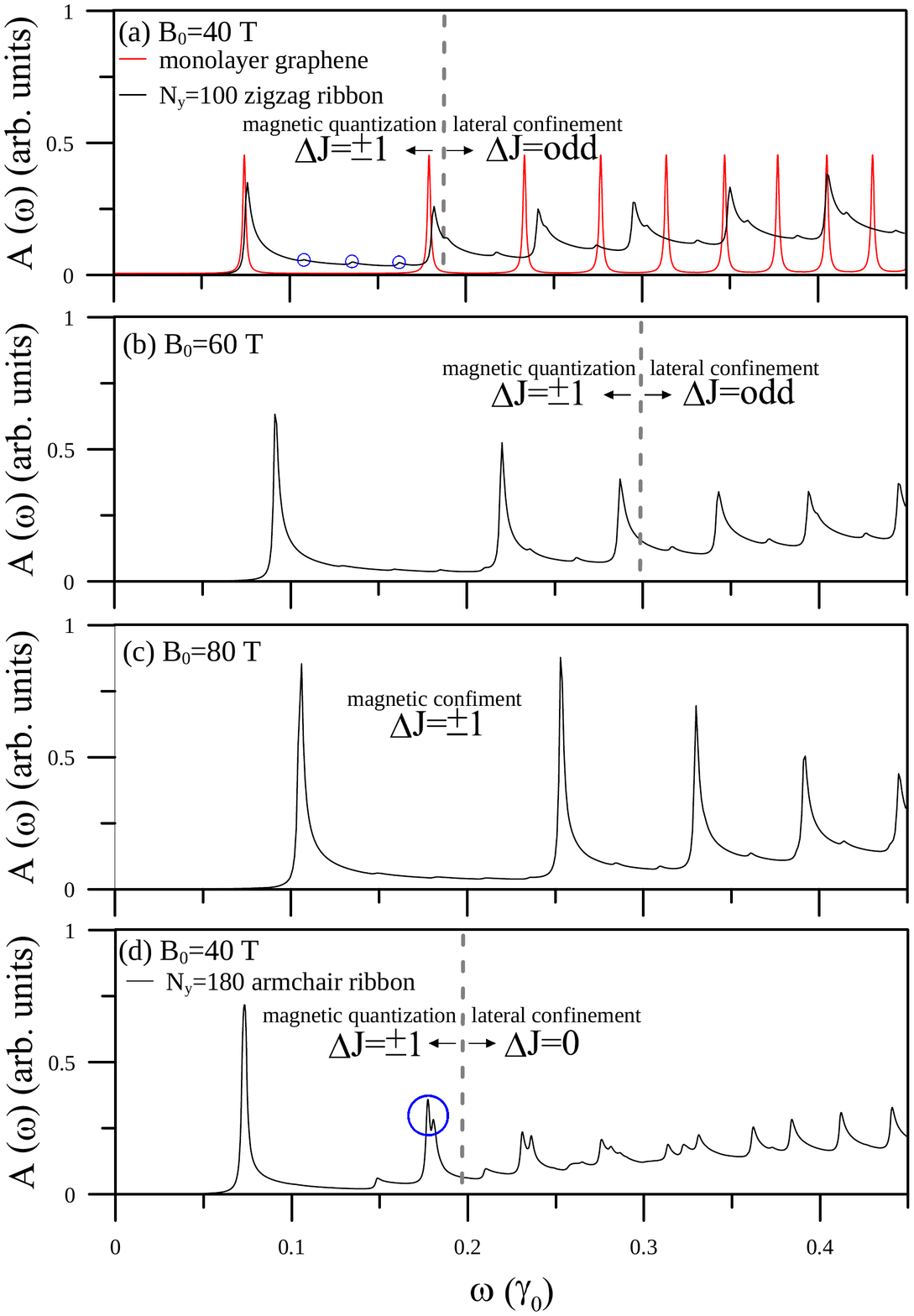}
\caption{Magneto-optical absorption spectra for a ${N_y=100}$ zigzag
nanoribbon at (a) ${B_0=40}$ T, (b) ${60}$ T $\&$ (c) 80 T and a
${N_y=180}$ armchair nanoribbon at (d) ${B_0}$=40 T. Also shown in
(a) is that of monolayer graphene (red curve).}
\label{fig:graph}
\end{figure}

\begin{figure}
\centering
\includegraphics[width=0.9\linewidth]{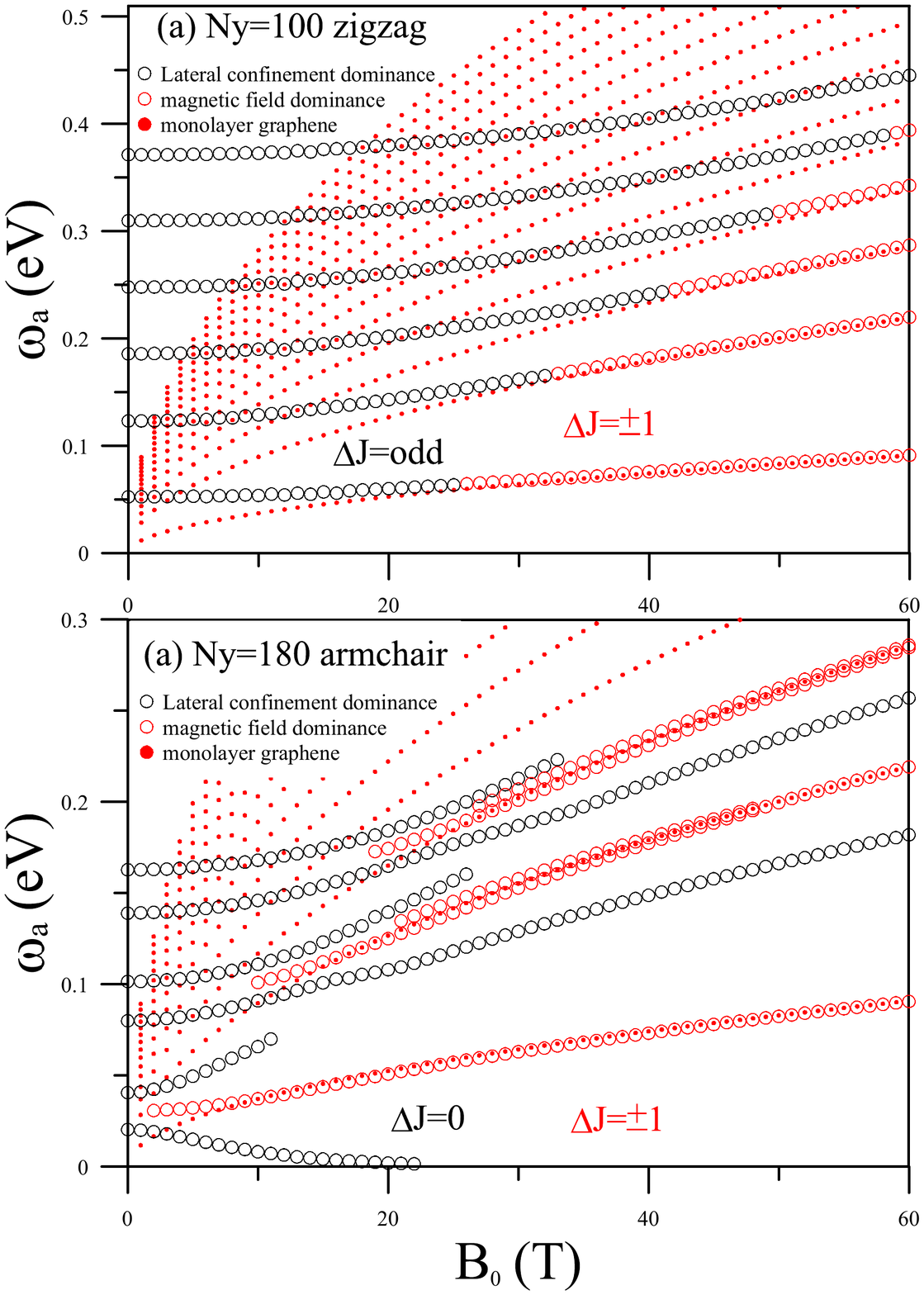}
\caption{Magneto-absorption frequencies of the initial six prominent peaks for (a) ${N_y=100}$ zigzag and ${N_y=180}$ armchair graphene nanoribbons.
The red dots are inter-LL optical excitation frequencies of monolayer graphene.}
\label{fig:graph}
\end{figure}

\end{document}